\documentclass[aps,pra,floatfix,twocolumn,showpacs,amsmath,amssymb,letterpaper]{revtex4-2}
\usepackage{graphicx}
\usepackage{subfigure}
\usepackage{array}
\usepackage{bigstrut}
\usepackage{longtable}
\usepackage{rotating}
\usepackage{diagbox}
\usepackage[mode=text]{siunitx}
\usepackage{bm}
\usepackage{float}
\usepackage{lipsum}
\usepackage{warpcol}
\usepackage{color}
\usepackage{amsthm,amsmath}
\usepackage{mathrsfs}
\usepackage{appendix}
\usepackage{multirow}
\usepackage{threeparttable}
\usepackage{soul}
\usepackage[colorlinks=true, linkcolor=blue, citecolor=blue, urlcolor=blue]{hyperref}
\usepackage{nameref}
\usepackage{csquotes}
\usepackage{tabularx}
\usepackage{booktabs} 
\usepackage{dcolumn}
\allowdisplaybreaks

\newcolumntype{d}[1]{D{.}{.}{#1}}

\begin{document}

	\title{Characterizing resonances in positron-sodium scattering}

	\author{Ning-Ning Gao$^{1, 2}$}
	\author{Hui-Li Han$^{1}$}
	\email{huilihan@wipm.ac.cn}
	\author{Ting-Yun Shi$^{1}$}
	\author{Li-Yan Tang$^{1}$}
	
	\affiliation{
		$^{1}$
		Innovation Academy for Precision Measurement Science and Technology,
		Chinese Academy of Sciences, Wuhan 430071, People's Republic of China
	}
	
	\affiliation{
		$^{2}$University of Chinese Academy of Sciences, Beijing 100049, People's Republic of China
	}
	
	\date{\today}
	
	\begin{abstract}
We investigate resonances in positron-sodium scattering using the $R$-matrix propagation method formulated in hyperspherical coordinates. The interaction between the sodium core and the valence electron is described by analytical model potentials. High partial-wave resonances are calculated for collision energies up to the Na($4f$) threshold. Several resonant states of debated character are identified, and their behavior is analyzed through phase-variation studies, the associated structures in the calculated cross sections, and the characteristic patterns observed in the stability plots. The calculated dipole series of resonances, supported by the ion-dipole interaction between Na$^{\scriptscriptstyle+}$ and Ps($n=2$), shows good agreement with recent complex-scaling calculations. In addition, a sequence of quasi-dipole resonances is found to arise from the near degeneracy of the Na($4d$) and Na($4f$) states in the e$^{+}$-Na system, which accumulate geometrically toward the Na($4d$) threshold.
	\end{abstract}
	
	
	\maketitle
	
	\section{Introduction}
	
	Over the past few decades, atomic resonances involving positrons have attracted considerable attention\,\cite{Bhatia1990,HO1983,Ho1987Apr,Ho1996May,Igarashi1997,Yan2003,Igarashi2004Jul,Ren2011feb,Ward1989nov,Utpal2002apr,Kar2005Jul,Kubota2008feb,Han2008Jan,Han2008Oct,Jiao2012Feb,Umair2015Jul,Umair2016May,Umair2016dec}. Resonant states in positronic atoms generally arise from two main configurations: the excited atom (A) + $e^{\scriptscriptstyle+}$ configuration and the A$^{\scriptscriptstyle+}$ + excited Ps configuration. Both mechanisms are strongly influenced by the degeneracy of the target atom and the Ps atom, which gives rise to an effective dipole interaction capable of supporting an infinite series of resonances. A well known example is provided by the Ps($n=2$) + A$^{\scriptscriptstyle+}$ channels (A = H, He$^{\scriptscriptstyle+}$, Na, and Li), which can support dipole series resonances. The positron--hydrogen system is particularly noteworthy, since both H and Ps possess degenerate spectra, giving rise to dipole series resonances associated with the H($N=4$) and Ps($N=3$) thresholds\,\cite{Ho1988Dec}.
	
	There is now solid experimental evidence that positrons can form bound states with a wide range of molecules. The positron annihilation cross sections for molecules such as C$_3$H$_8$ and C$_6$H$_{14}$ exhibit distinct features attributed to Feshbach resonances, arising from the temporary trapping of positrons in vibrationally excited molecular states\,\cite{Barnes2003Mar}. However, no direct experimental evidence has yet been found for positron-atom bound states. A possible experimental signature would be the observation of resonant structures associated with atomic excited states in positron-atom scattering spectra. However, despite extensive experimental efforts, no clear evidence for such resonant features has yet been reported\,\cite{ratnavelu2019jun}.
	
	Alkali atoms provide particularly suitable systems for systematic studies of positronic resonances owing to their simple electronic structure, characterized by a single weakly bound valence electron\,\cite{Ward1989nov,Utpal2002apr,Kar2005Jul,Kubota2008feb,Han2008Jan,Han2008Oct,Jiao2012Feb,Umair2015Jul,Umair2016May,Umair2016dec}. Positron attachment to low-lying excited states of alkali atoms can lead to Feshbach resonances that manifest as enhancements in low energy annihilation or scattering cross sections.
	
	The first investigation of positronic resonances in alkali systems was conducted by Ward \textit{et al.}\,\cite{Ward1989nov}, who used the eigenphase-sum method combined with a model potential to describe the interaction between the active electron and the ionic core. Later, Kar and Ho\,\cite{Kar2005Jul} examined $S$-wave resonances in the positron-sodium system using the stabilization method with Hylleraas-type basis functions, obtaining resonance energies and widths for states below the Ps($n=2$) threshold. Han \textit{et al.}\,\cite{Han2008Jan} subsequently applied the stabilization method in the hyperspherical coordinate framework to study $S$-wave resonances in e$^{\scriptscriptstyle+}$-Na scattering, achieving results consistent with those of Kar and Ho\,\cite{Kar2005Jul}. More recently, Jiao \textit{et al.}\,\cite{Jiao2012Feb} investigated higher partial-wave resonances using the momentum-space coupled-channel optical method, which incorporates both target continuum and Ps-formation channels through an optical potential. The most recent work by Umair \textit{et al.}\,\cite{Umair2015Jul,Umair2017} employed the complex-scaling method to study natural and unnatural parity states, identifying $S$-, $P$-, and $D$-wave Ps($n=2$) dipole resonance series in the e$^{\scriptscriptstyle+}$-Na and e$^{\scriptscriptstyle+}$-Li systems.
	
	Despite these theoretical advances, discrepancies remain among different approaches regarding the predicted resonance energies and widths. In particular, some resonances that appear as flat regions in stabilization plots are absent in complex-scaling calculations, rendering their physical interpretation ambiguous.
	
	In the present work, we investigate resonances in positron-sodium scattering for total angular momenta $J=0-4$ with natural parity. Resonances are identified using the eigenphase-sum method, where they manifest as sharp peaks in the energy dependence of the time delay. To further clarify the nature of states exhibiting small or ambiguous phase shifts, we also apply the stabilization method within the hyperspherical coordinate framework and analyze their signatures in the corresponding stability plots.
	
	The paper is organized as follows.
	In Sec.\,\ref{sec:method}, our calculation method and the use of our model potentials.
	In Sec.\,\ref{sec:Results}, we discuss the results.
	Finally, we provide a brief summary.
	Atomic units are applied throughout the paper unless stated otherwise. \\
	\section{Theoretical Method}   
	\label{sec:method}
	In this study, the sodium atom is assumed to consist of a single valence electron and a frozen core, thus we treat e$^{\scriptscriptstyle+}$-Na system as a three-body system that consists of a core, an electron, and a positron, their masses are denoted by $m_{1}$, $m_{2}$, and $m_{3}$, respectively. We employ Delves's hyperspherical coordinates and introduce the mass-scaled Jacobi coordinates. The first Jacobi vector $\vec{\rho}_{1}$ is chosen to be the vector from Na$^{\scriptscriptstyle+}$ core to e$^{\scriptscriptstyle-}$, with reduced mass $\mu_{1}$, and the second Jacobi vector $\vec{\rho}_{2}$ goes from the diatom center of mass to e$^{\scriptscriptstyle+}$, with reduced mass $\mu_{2}$. The angle between $\vec{\rho}_{1}$ and $\vec{\rho}_{2}$ is denoted by $\theta$. The hyperradius $R$ and hyperangle $\phi$ are defined as\\
	\begin{equation}
		\label{1}
		\mu R^{2}=\mu_{1}\rho_{1}^{2}+\mu_{2}\rho_{2}^{2}\,,
	\end{equation}
	and\\
	\begin{equation}
		\label{2}
		\tan\phi=\sqrt{\frac{\mu_{2}}{\mu_{1}}}\frac{\rho_{2}}{\rho_{1}},\;\; 0 \leq\phi\leq\frac{\pi}{2}\,,
	\end{equation}
	respectively, where $R$ is the only coordinate with the dimension of length and represents the overall size of the three-body system. The rotation of the plane containing the three particles is described collectively by $\Omega$ $[\Omega \equiv (\theta, \phi, \alpha, \beta, \gamma)]$, which includes $\theta$, $\phi$, and three Euler angles $(\alpha, \beta, \gamma)$. The parameter $\mu$ is an arbitrary scaling factor, and we choose $\mu=\sqrt{\mu_{1}\mu_{2}}$ for our calculations. \\
	
	The Schr$\mathrm{\ddot{o}}$dinger equation in hyperspherical coordinates can be written after rescaling the three-body wave function $\Psi$ as $\psi(R;\theta,\phi)=\Psi(R;\theta,\phi) R^{5/2} \sin\phi \cos\phi$:\\
	\begin{equation}
		\begin{aligned}
			\label{3}
			&\left[ -\frac{1}{2\mu}\frac{d^{2}}{dR^{2}}+\left( \frac{\Lambda^{2}-\frac{1}{4}}{2\mu R^{2}}+V(R;\theta,\phi)\right) \right] \psi_{\upsilon'}(R;\Omega)\\
			&=E\psi_{\upsilon'}(R;\Omega)\,,
		\end{aligned}
	\end{equation}
	where $\Lambda^{2}$ is the squared ``grand angular momentum operator'', and its expression is as given in Ref.\;\cite{lin1995}. The three-body interaction potential $V(R;\theta,\phi)$ is expressed as:\\
	\begin{equation}
		\label{4}
		V(R;\theta,\phi)=v_{\scriptscriptstyle-}(r_{12})+v_{\scriptscriptstyle+}(r_{13})+v(\vec{r}_{12},\vec{r}_{13})\,,
	\end{equation}
	where $r_{12}$, $r_{13}$, and $r_{23}$ is the electron-core distance, the positron-core distance, and the electron-positron distance, respectively.\\
	
	For the e$^{\scriptscriptstyle+}$Na system, the model potential for the valence electron and the core is expressed in the form of \\
	\begin{equation}
		\begin{aligned}
			\label{5}
			v_{\scriptscriptstyle-}(r_{12})=&-\frac{1}{r_{12}}\left[Z_{c}+(Z-Z_{c})e^{-a_{1}r_{12}}+a_{2}re^{-a_{3}r_{12}}\right]\\
			&-\frac{a_{c}}{2r^{4}}\left[W_{3}\left(\frac{r}{r_{c}}\right)\right]^{2 }\,,
		\end{aligned}
	\end{equation}
	where
	\begin{equation}
		\label{6}
		W_{n}(r)=1-e^{-r^{n}}\,,
	\end{equation}
	is the cutoff function used to assure the correct behavior at the origin. The second term in Eq.\,(\ref{5}) describes the polarization of the core, where $a_{c}$=0.9457 a.u. is the Na$^{\scriptscriptstyle+}$ polarizability\,\cite{Johnson1983}. The nuclear charge is $Z=11$ and the charge of the Na$^{\scriptscriptstyle+}$ core is $Z_{c}=1$. The remaining parameters, which are $a_{1}=3.32442452,\, a_{2}=0.71372798,\, a_{3}=1.83281815$, and $r_{c}=0.52450638$, are fitted by Ref.\,\cite{Liu1999} using the least-squares method to reproduce the experimentally measured energy levels of the Na atom. For the positron-core interaction $v_{\scriptscriptstyle+}(r_{13})$, the first term is same as in $v_{\scriptscriptstyle-}(r_{12})$ but with opposite sign, and the polarization potential is chosen to the same as in $v_{\scriptscriptstyle-}(r_{12})$. The interaction between the positron and electron $v(\vec{r}_{12},\vec{r}_{13})$ is taken the form\\
	\begin{equation}
		\label{7}
		v(\vec{r}_{12},\vec{r}_{13})=-\frac{1}{r_{23}}+\frac{a_{c}}{r_{12}^{2}r_{13}^{2}}\cos{\theta} W_{3}\left(\frac{r_{12}}{r_{c}}\right)W_{3}\left(\frac{r_{13}}{r_{c}}\right)\,,
	\end{equation}
	
	The adiabatic potentials $U_{\nu}(R)$ and channel functions $\Phi_{\nu}(R;\Omega)$ at fixed $R$ can be obtained by solving the following adiabatic eigenvalue equation:\\
	\begin{equation}
		\label{8}
		\left( \frac{\Lambda^{2}-\frac{1}{4}}{2\mu R^{2}}+V(R;\theta,\phi)\right) \Phi_{\nu}(R;\Omega)=U_{\nu}(R) \Phi_{\nu}(R;\Omega)\,.
	\end{equation}
	The channel function is further expanded on Wigner rotation matrices $D_{KM}^{J}$ as\\
	\begin{align}
		\label{9}
		\Phi_{\nu}^{J\Pi M}(R;\Omega)=\sum\limits_{K=0}^{J}u_{\nu K}(R;\theta,\phi)\overline{D}_{KM}^{J\Pi}(\alpha,\beta,\gamma)\,.
	\end{align}
	\begin{align}
		\label{10}
		\overline{D}_{KM}^{J\Pi}=\frac{1}{4\pi}\sqrt{2J+1}[D_{KM}^{J}+(-1)^{K+J}\Pi D_{-KM}^{J}]\,,
	\end{align}
	where $J$ is the total nuclear orbital angular momentum, $M$ is its projection onto the laboratory-fixed axis, and $\Pi$ is the parity with respect to the inversion of the nuclear coordinates. The quantum number $K$ denotes the projection of $J$ onto the body-frame $z$ axis and takes the values $K=J,J-2,\ldots,-(J-2),-J$ for the ``parity-favored" case, $\Pi=(-1)^{ J}$, and $K=J-1,J-3,\ldots,-(J-3),-(J-1)$ for the ``parity-unfavored" case, $\Pi=(-1)^{J+1}$. $u_{\nu K}(R;\theta,\phi)$ is expanded with $B$-spline basis sets:\\
	
	\begin{align}
		\label{11}
		u_{\nu K}(R;\theta,\phi)=\sum\limits_{i}^{N_{\phi}} \sum\limits_{j}^{N_{\theta}}c_{i,j}B_{i}(\phi)B_{j}(\theta)\,,
	\end{align}
	where $N_{\theta}$ and $N_{\phi}$ are the sizes of the basis sets in the $\theta$ and $\phi$ directions, respectively. Using the $B$-splines as a basis function has multiple advantages, including high localization, flexibility, completeness, and numerical stability\;\cite{Bachau2001,Kang2006}, which enable us to obtain accurate potential curves and channel functions by employing these advantages. A detailed investigation of the knot distribution suitable for calculations of the positron alkali-atom bound states can be found in our previous work\;\cite{hanbindingenergy2008}.\\
	
	The goal of our scattering study is to determine the scattering matrix $\underline{\mathcal{S}}$ from the solutions of Eq.\;(\ref{3}). We first calculate the $\underline{\mathcal{R}}$ matrix, which is defined as
	\begin{align}
		\label{12}
		\underline{\mathcal{R}}(R)=\underline{\textsl{F}}(R)[\widetilde{\underline{\textsl{F}}}(R)]^{ -1}\,,
	\end{align}
	where matrices $\underline{\textsl{F}}$ and $\widetilde{\underline{\textsl{F}}}$ can be calculated from the solution of Eqs.\;(\ref{3}) and (\ref{9}) by evaluating the integrals:
	\begin{align}
		\label{13}
		F_{\nu,\upsilon'}(R)=\int d\Omega \Phi_{\nu}(R;\Omega)^{\scriptscriptstyle *}\psi_{ \upsilon'}(R;\Omega)\,,
	\end{align}
	\begin{align}
		\label{14}
		\widetilde{F}_{\nu,\upsilon'}(R)=\int d\Omega \Phi_{\nu}(R;\Omega)^{\scriptscriptstyle *}\frac{\partial}{\partial R}\psi_{\upsilon'}(R;\Omega)\,.
	\end{align}
	The hyperradius $R$ is divided into $(N-1)$ intervals using a set of grid points $R_{ 1}<R_{ 2}<\cdots R_{ N}$. At short distances, we utilize the SVD method to solve Eq.\;(\ref{3}) in the interval $[R_{ i},R_{i+1}]$. In this method, we express the total wave function $\psi_{\upsilon'}(R;\Omega)$ in terms of the discrete variable representation (DVR) basis $\pi_{i}$ and the channel functions $\Phi_{\nu}(R;\Omega)$ as follows:
	\begin{align}
		\label{15}
		\psi_{\upsilon'}(R;\Omega)=\sum^{ N_{ DVR}}_{i}\sum^{ N_{chan}}_{\nu}C^{ \upsilon'}_{ i\nu}\pi_{ i}(R)\Phi_{\nu}(R_{i};\Omega)\,,
	\end{align}
	where $N_{DVR}$ represents the number of DVR basis functions and $N_{chan}$ is the number of included channel functions.
	By inserting $\psi_{ \upsilon'}(R;\Omega)$ into the three-body Schr$\mathrm{\ddot{o}}$dinger equation given by Eq.\;(\ref{3}), we arrive at a standard algebraic problem for the coefficients $C^{\upsilon'}_{i\nu}$:
	\begin{align}
		\label{16}
		\sum^{N_{DVR}}_{j}\sum^{ N_{chan}}_{\mu}\mathcal{T}_{ij}
		\mathcal{O}_{i\nu,j\mu}C^{\upsilon'}_{ j\mu}+U_{\nu} (R_{i})C^{\upsilon'}_{i\nu}=E^{ \upsilon'}C^{ \upsilon'}_{i\nu}\,,
	\end{align}
	where
	\begin{align}
		\label{17}
		\mathcal{T}_{ij}=\frac{1}{2\mu}\int^{R_{i+1}}_{ R_{ i}}\frac{d}{dR}\pi_{i}(R)\frac{d}{dR}\pi_{j}(R)dR\,,
	\end{align}
	are the kinetic-energy matrix elements, with ${R_{i}}$ and ${R_{i+1}}$ being the boundaries of the calculation box, and
	\begin{align}
		\label{18}
		\mathcal{O}_{i\nu,j\mu}=\langle\Phi_{\nu}(R_{i};\Omega)|\Phi_{ \mu}(R_{j};\Omega)\rangle
	\end{align}
	are the overlap matrix elements between the adiabatic channels defined at different quadrature points.
	
	At large distances, the traditional adiabatic hyperspherical method is used to solve Eq.\;(\ref{3}).
	When substituting the wave functions $\psi_{\upsilon'}(R;\Omega)$ into Eq.\;(\ref{3}), one obtains a set of coupled ordinary differential equations:
	\begin{equation}
		\begin{aligned}
			\label{19}
			&	\left[-\frac{1}{2\mu}\frac{d^{2}}{dR^{2}}+U_{\nu}(R)- E\right]F_{\nu,\upsilon'}(R)\\
			&-\frac{1}{2\mu}\sum_{\mu}\left[2P_{\mu\nu}(R)\frac{d}{dR}+Q_{\mu\nu}(R)\right]F_{\mu \upsilon'}(R)=0\,,
		\end{aligned}
	\end{equation}
	where
	\begin{align}
		\label{20}
		P_{\mu\nu}(R)=\int d\Omega \Phi_{ \mu}(R;\Omega)^{\scriptscriptstyle *}\frac{\partial}{\partial R}\Phi_{\nu}(R;\Omega)\,,
	\end{align}
	and
	\begin{align}
		\label{21}
		Q_{\mu\nu}(R) = \int d\Omega \Phi_{ \mu}(R;\Omega)^{\scriptscriptstyle *}\frac{\partial^{2}}{\partial R^{2}}\Phi_{\nu}(R;\Omega)\,.
	\end{align}
	are the nonadiabatic couplings that control the inelastic transitions as well as the width of the resonance supported by  adiabatic potential $U_{\nu}(R)$ .
	
	Finally, we use the $R$-matrix propagation method. Within an interval $[R_{1},R_{2}]$, given an $\underline{\mathcal{R}}$ matrix at $R_{1}$, we calculate the corresponding $\underline{\mathcal{R}}$ matrix at another point $R = R_{2}$ using
	\begin{align}
		\label{22}
		\underline{\mathcal{R}}(R_{2})=\underline{\mathcal{R}}_{22}
		-\underline{\mathcal{R}}_{21}\left[\underline{\mathcal{R}}_{11}
		+\underline{\mathcal{R}}(R_{1})\right]^{-1}\underline{\mathcal{R}}_{12}\,.
	\end{align}
	where the corresponding matrices give:
	\begin{align}
		\label{23}
		(\underline{\mathcal{R}}_{11})_{\nu\mu} = \sum\limits_{n }\frac{u_{ \nu}^{(n)}(R_{1})u_{ \mu}^{(n)}(R_{1})}{2\mu (\varepsilon_{n}-E)}\,,
	\end{align}
	\begin{align}
		\label{24}
		(\underline{\mathcal{R}}_{12})_{\nu\mu} = \sum\limits_{n }\frac{u_{\nu}^{(n)}(R_{ 1})u_{\mu}^{(n)}(R_{2})}{2\mu (\varepsilon_{n}-E)}\,,
	\end{align}
	\begin{align}
		\label{25}
		(\underline{\mathcal{R}}_{21})_{\nu\mu} = \sum\limits_{n }\frac{u_{\nu}^{(n)}(R_{2})u_{ \mu}^{(n)}(R_{1})}{2\mu (\varepsilon_{n}-E)}\,,
	\end{align}
	\begin{align}
		\label{26}
		(\underline{\mathcal{R}}_{22})_{\nu\mu} = \sum\limits_{n }\frac{u_{\nu}^{(n)}(R_{2})u_{\mu}^{(n)}(R_{2})}{2\mu (\varepsilon_{n}-E)}\,,
	\end{align}
	where $\nu$ and $\mu$ denote different channels, indices 1 and 2 do not label the channel, and more details can be found in Ref.\;\cite{WangJia2011}.
	
	The $\underline{\mathcal{K}}$ matrix can be expressed as the following matrix equation:
	\begin{align}
		\label{27}
		\underline{\mathcal{K}}=
		(\underline{f}-\underline{f}'\underline{\mathcal{R}})
		(\underline{g}-\underline{g}'\mathcal{R})^{-1}\,,
	\end{align}
	where $f_{\nu \nu'}=\sqrt{\frac{2\mu k_{\nu}}{\pi}} R j_{l_{\nu}}(k_{\nu} R)\delta_{\nu \nu'}$ and $g_{\nu \nu'}=\sqrt{\frac{2\mu k_{\nu}}{\pi}} R n_{l_{\nu}}(k_{ \nu} R)\delta_{\nu \nu'}$ are the diagonal matrices of energy-normalized spherical Bessel and Neumann functions.
	For the bound-state channel, $l_{\nu}$ is the angular momentum of the third atom relative to the dimer and $k_{\nu}$ is given by $k_{\nu}=\sqrt{2 \mu\left(E-E_{2 b}\right)}$.
	For the continuous channel, $l_{\nu}=\lambda_{\nu}+3 / 2$, and $k_{\nu}=\sqrt{2 \mu E}$. The scattering matrix $\underline{\mathcal{S}}$ is related to $\underline{\mathcal{K}}$ as follows:
	\begin{align}
		\label{28}
		\underline{\mathcal{S}}=(\underline{1}+i\underline{\mathcal{K}})(\underline{1}-i\underline{\mathcal{K}})^{-1}\,.
	\end{align}
	
	To analyze the resonance energies $E_{R}$ and width $\Gamma$, we mainly employ the eigenphase sum method. The eigenphase shifts $\delta(E)$ are obtained by diagonalizing the $\mathcal{K}$-matrix (Eq.\;(\ref{27})) followed by taking the arctan. Consequently, the total eigenphase shift is expressed as\;\cite{Gao1989,Greene1987}:\\
	\begin{equation}
		\label{29}
		\delta_{tot}(E)=\sum\limits_{i=1}^{N_{o}}\delta_{i}(E)=\sum\limits_{i=1}^{N_{o}}arctan(\lambda_{i})\,.
	\end{equation}
	where $\lambda_{i}$ is the $i$-th eigenvalue of $\mathcal{K}$-matrix, $E$ is the collision energy, and $N_o$ is the number of open channels. The resonant position is the point at which, the time delay $\tau = \frac{d \delta_{tot}(E)}{dE}$ is maximal.
	
	\section{Results and Discussions}
	\label{sec:Results}
	We focus on studying resonances for total angular momentum $J=0-4$ in e$^{\scriptscriptstyle+}$-Na system using the the eigenphase sum and stabilization method. In Table\,\ref{t1} we present the resonance energies and widths of the e$^{\scriptscriptstyle+}$-Na system below the Na($5s$) threshold for $S$-, $P$-, and $D$-waves, together with a comparison to previous theoretical results. In these calculations, the basis sets $N_{\theta}=76$ and $N_{\phi}=218$ are chosen, and the potential curves converge to at least six significant digits. \\
	
	\subsection{Partial-wave resonances associated with excited thresholds of $\text{Na}$ atom}
\begin{figure*}[htbp]
	\centering
	\label{fig1}
	\renewcommand{\thesubfigure}{(a\arabic{subfigure})}
	\setcounter{subfigure}{0}
	\subfigure{
		\includegraphics[scale=0.28]{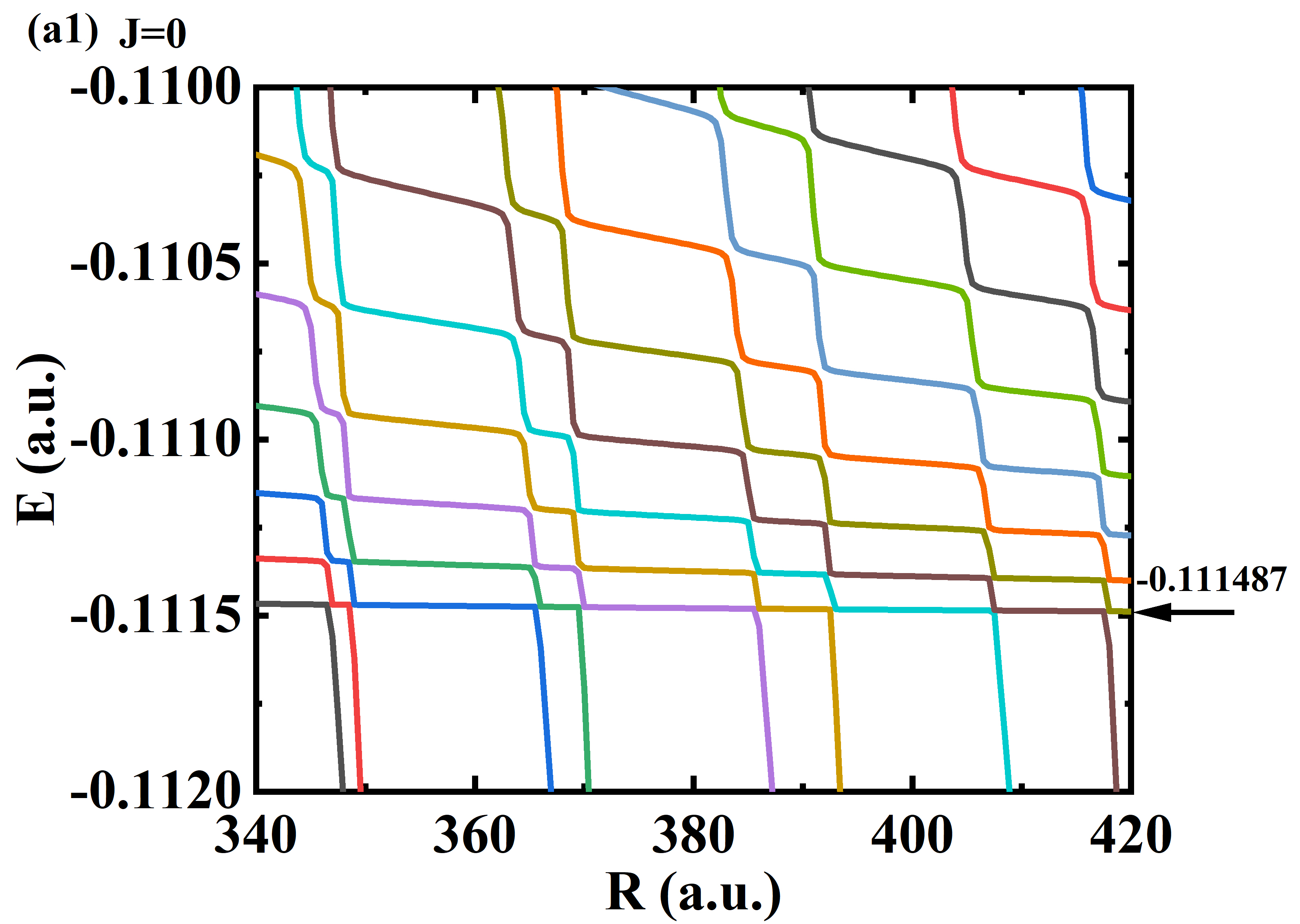}
		\label{fig1a1}
	}
	\subfigure{
		\includegraphics[scale=0.28]{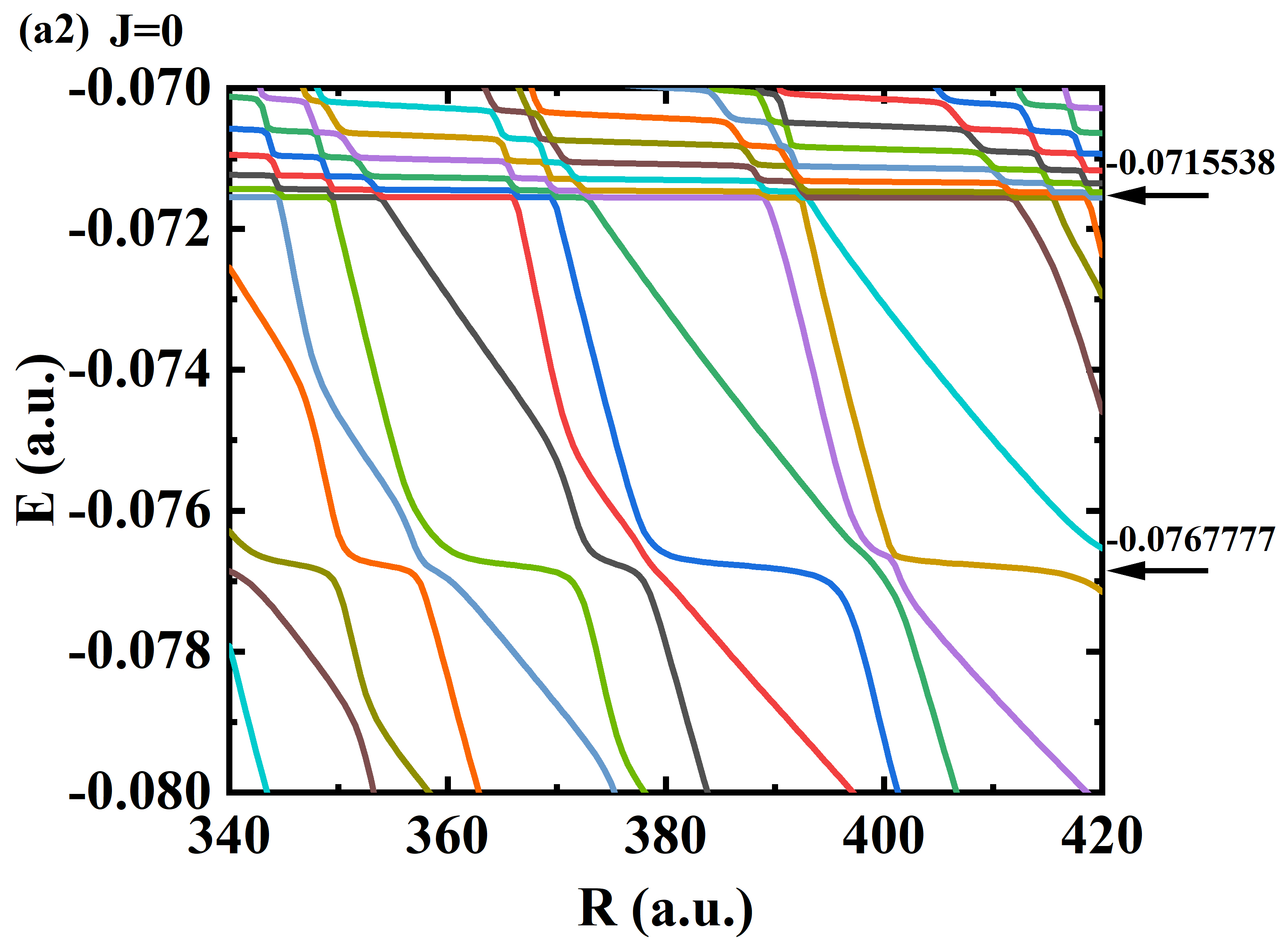}
		\label{fig1a2}
	}
	\renewcommand{\thesubfigure}{(b\arabic{subfigure})}
	\setcounter{subfigure}{0}
	\subfigure{
		\includegraphics[scale=0.28]{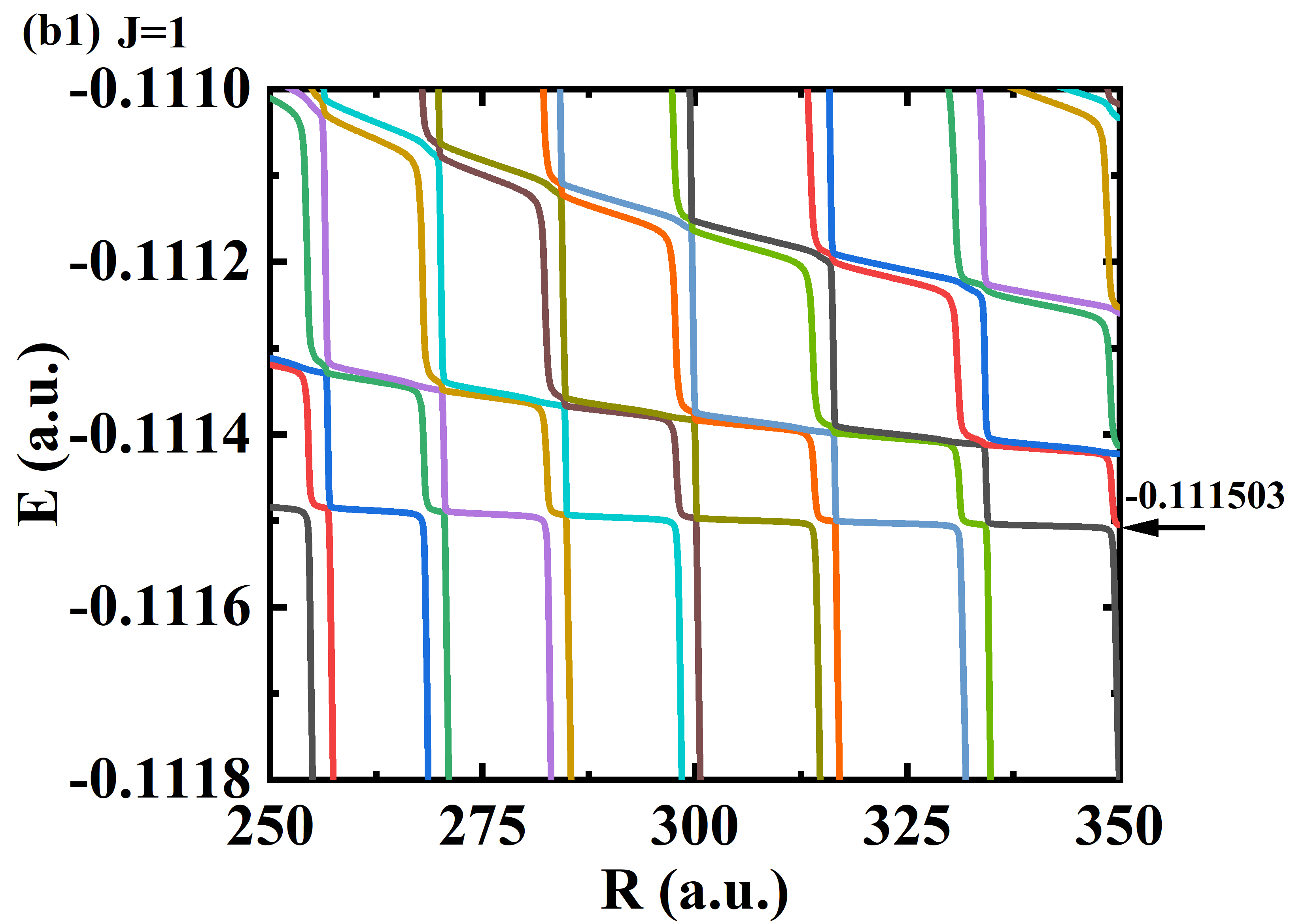}
		\label{fig1b1}
	}
	\subfigure{
		\includegraphics[scale=0.285]{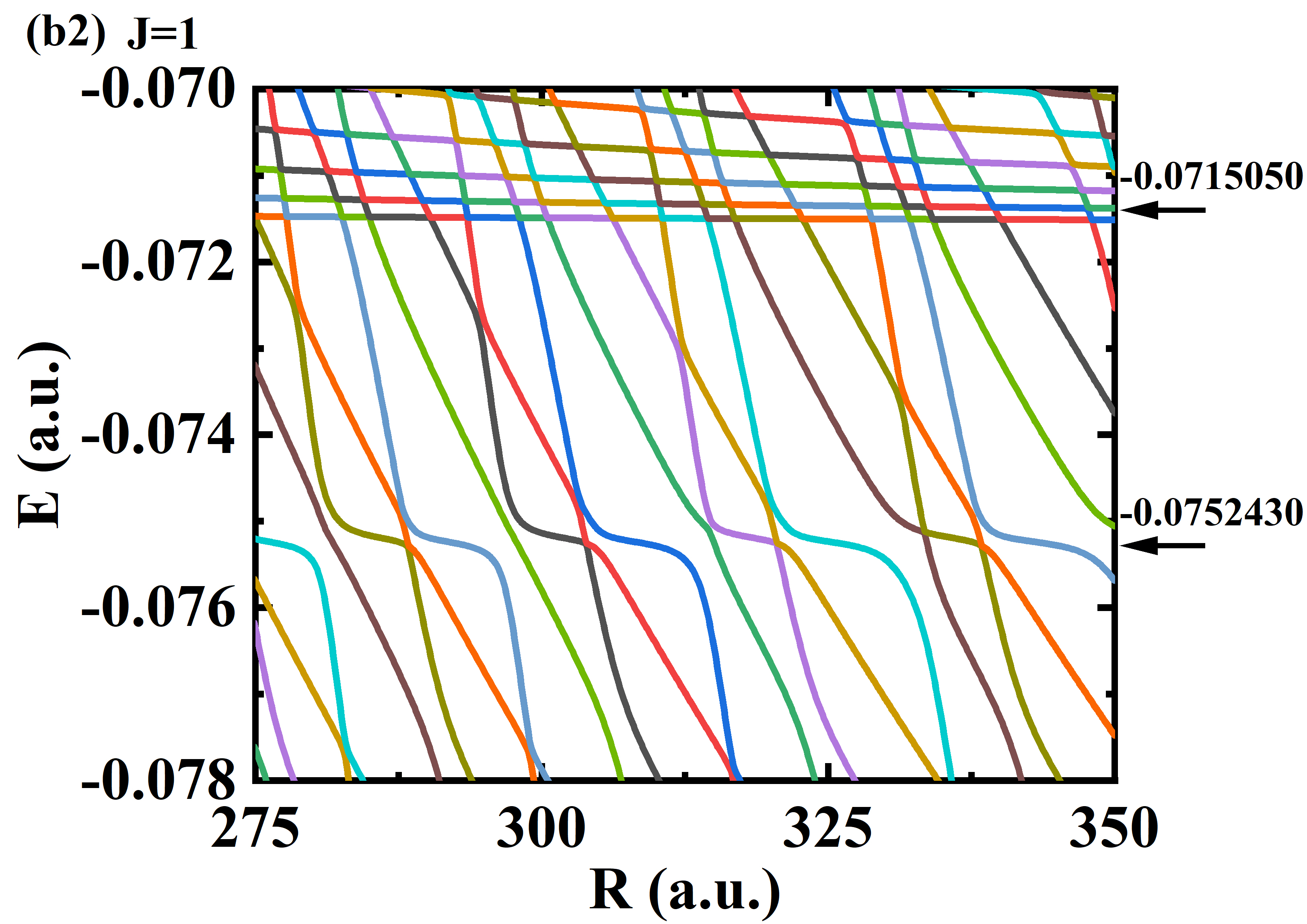}
		\label{fig1b2}
	}
	\renewcommand{\thesubfigure}{(c\arabic{subfigure})}
	\setcounter{subfigure}{0}
	\subfigure{
		\includegraphics[scale=0.28]{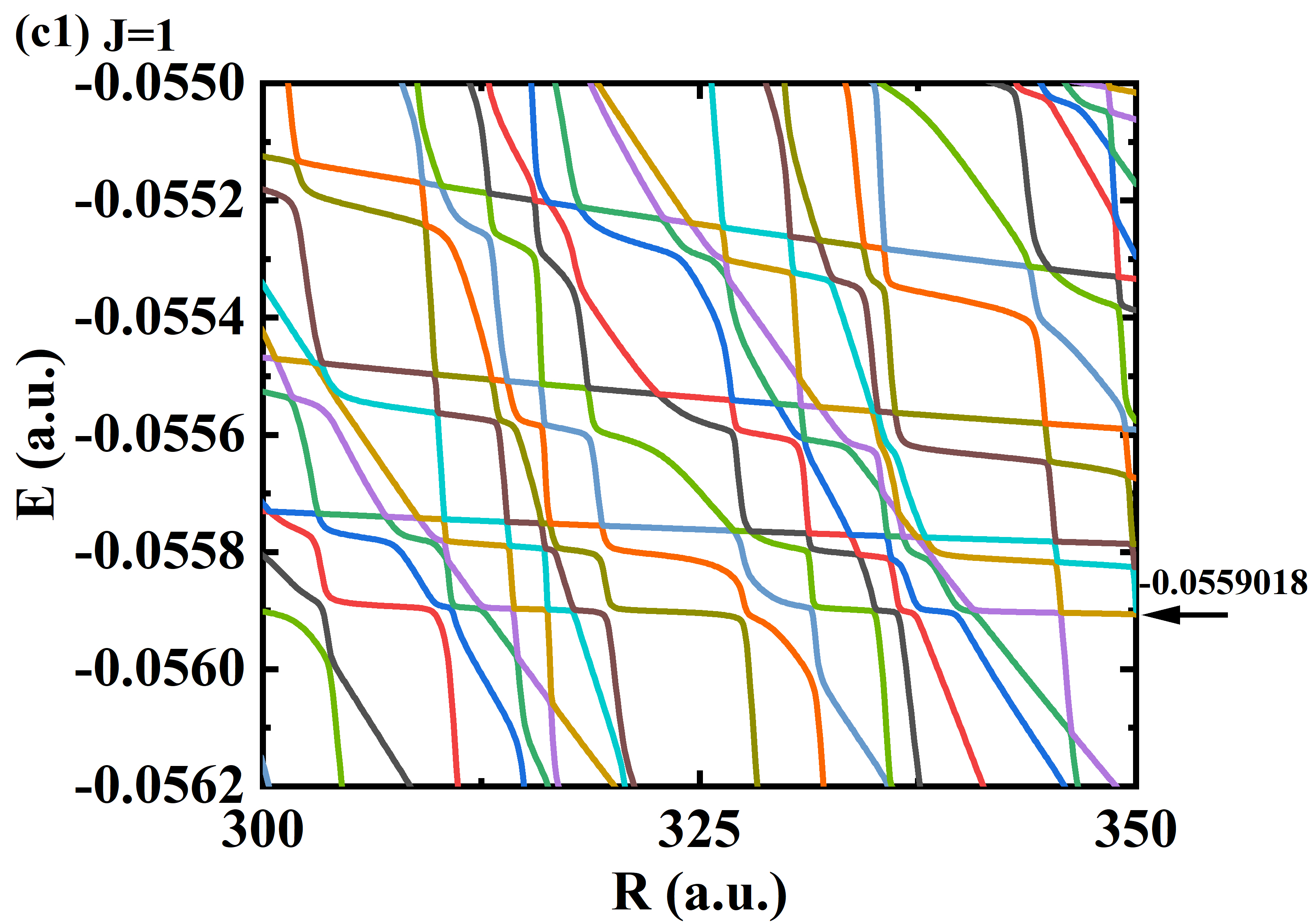}
		\label{fig1c1}
	}
	\subfigure{
		\includegraphics[scale=0.285]{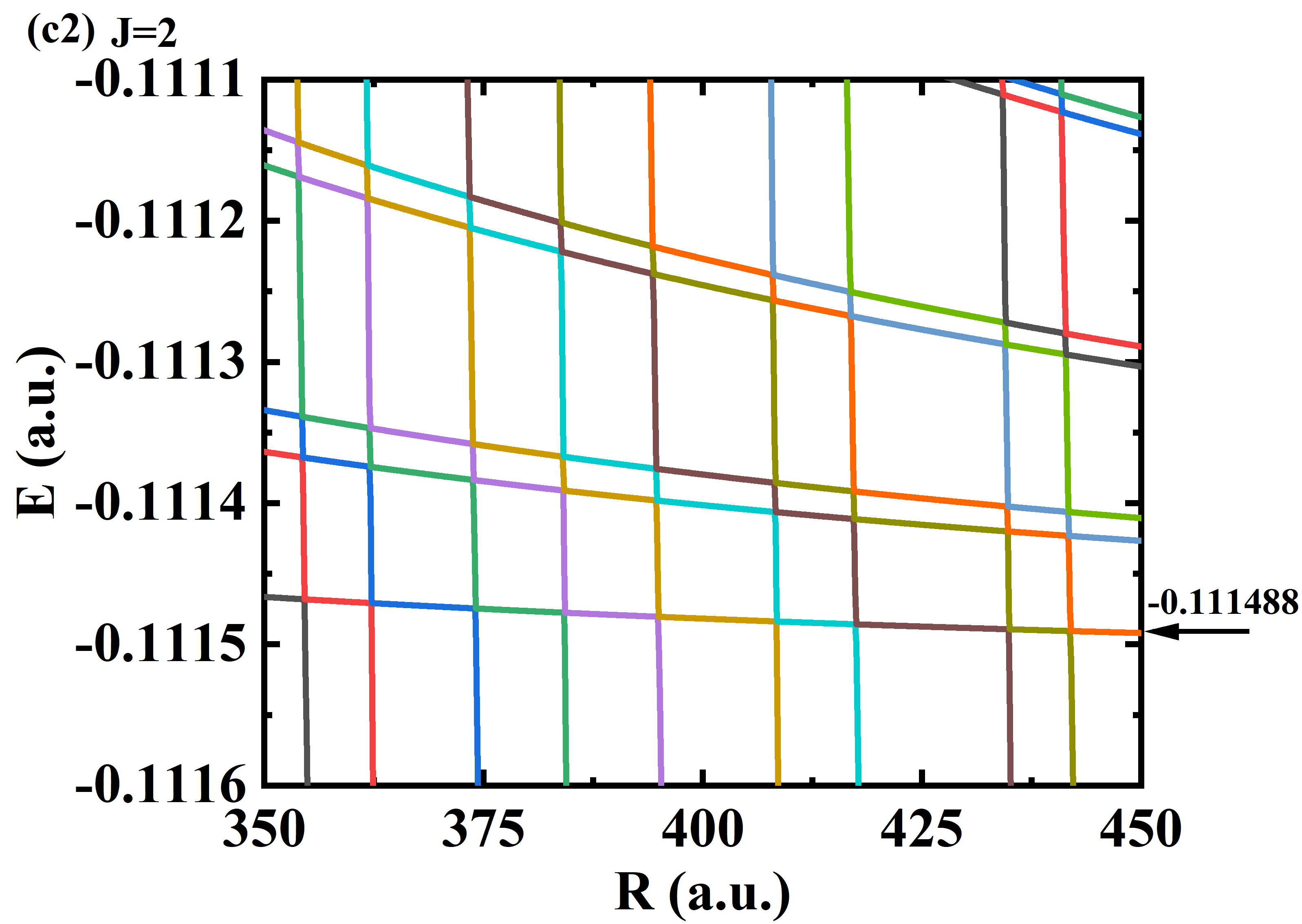}
		\label{fig1c2}
	}
	\renewcommand{\thesubfigure}{(d\arabic{subfigure})}
	\setcounter{subfigure}{0}
	\subfigure{
		\includegraphics[scale=0.29]{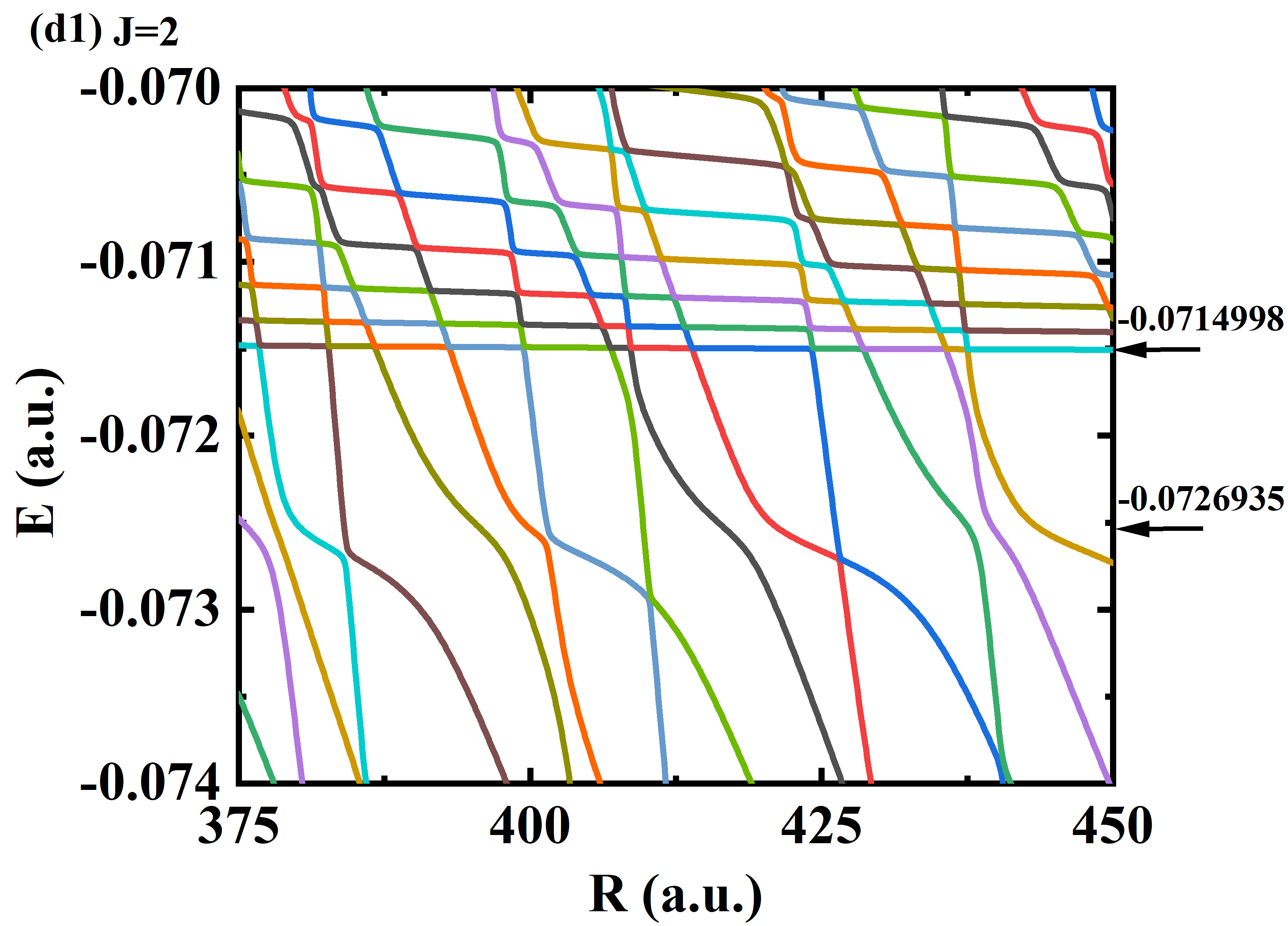}
		\label{fig1d1}
	}
	\subfigure{
		\includegraphics[scale=0.295]{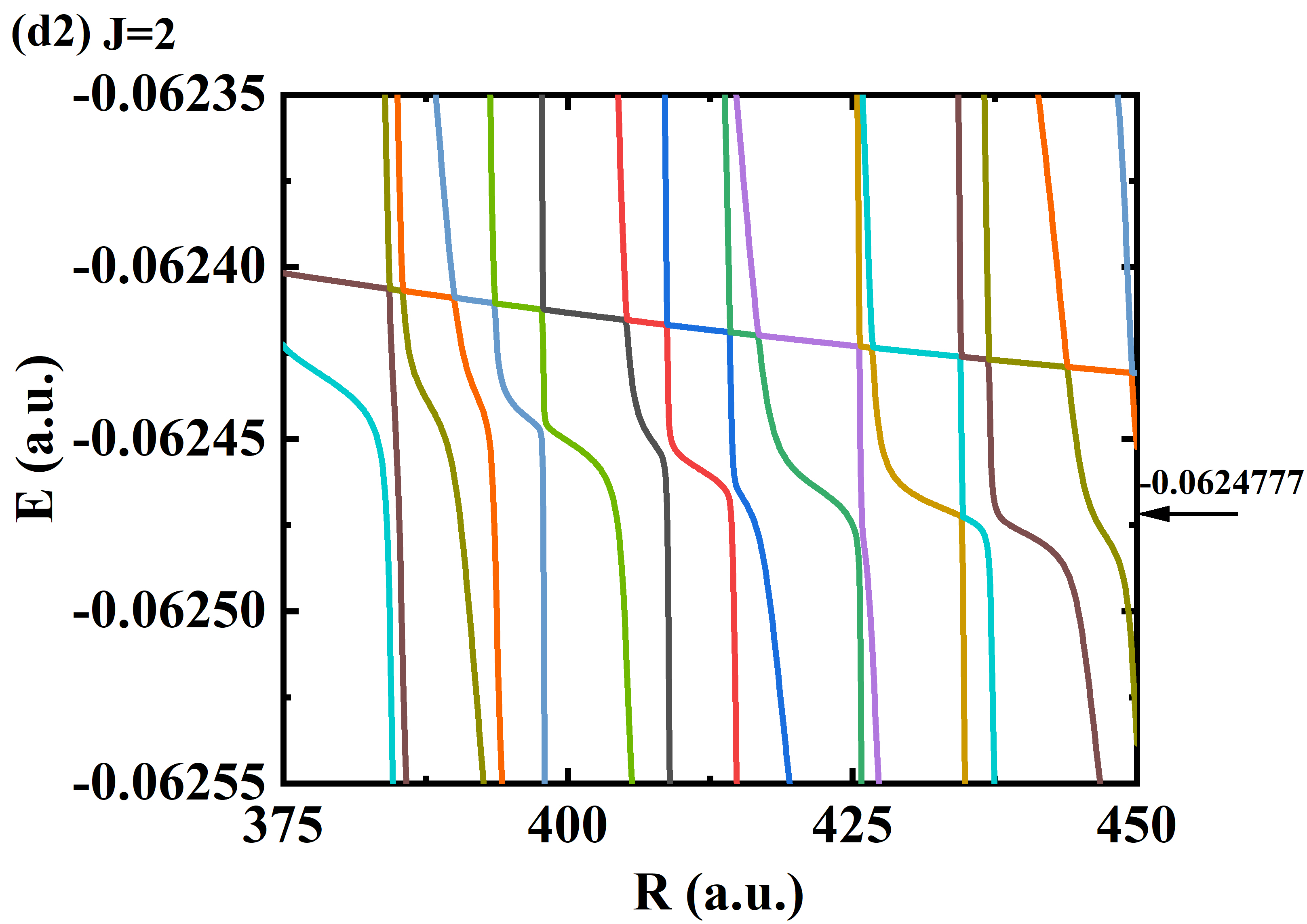}
		\label{fig1d2}
	}
	\caption{(Color online) Stabilization plot of energy $E$ versus box size $R$ in the e$^{\scriptscriptstyle+}$ + Na system with $J=0-2$. The arrow marks the resonance position. }	
\end{figure*}

\begin{figure*}[htbp]
	\centering
	\label{fig2}
	\renewcommand{\thesubfigure}{(a\arabic{subfigure})}
	\setcounter{subfigure}{0}
	\subfigure{
		\includegraphics[scale=0.28]{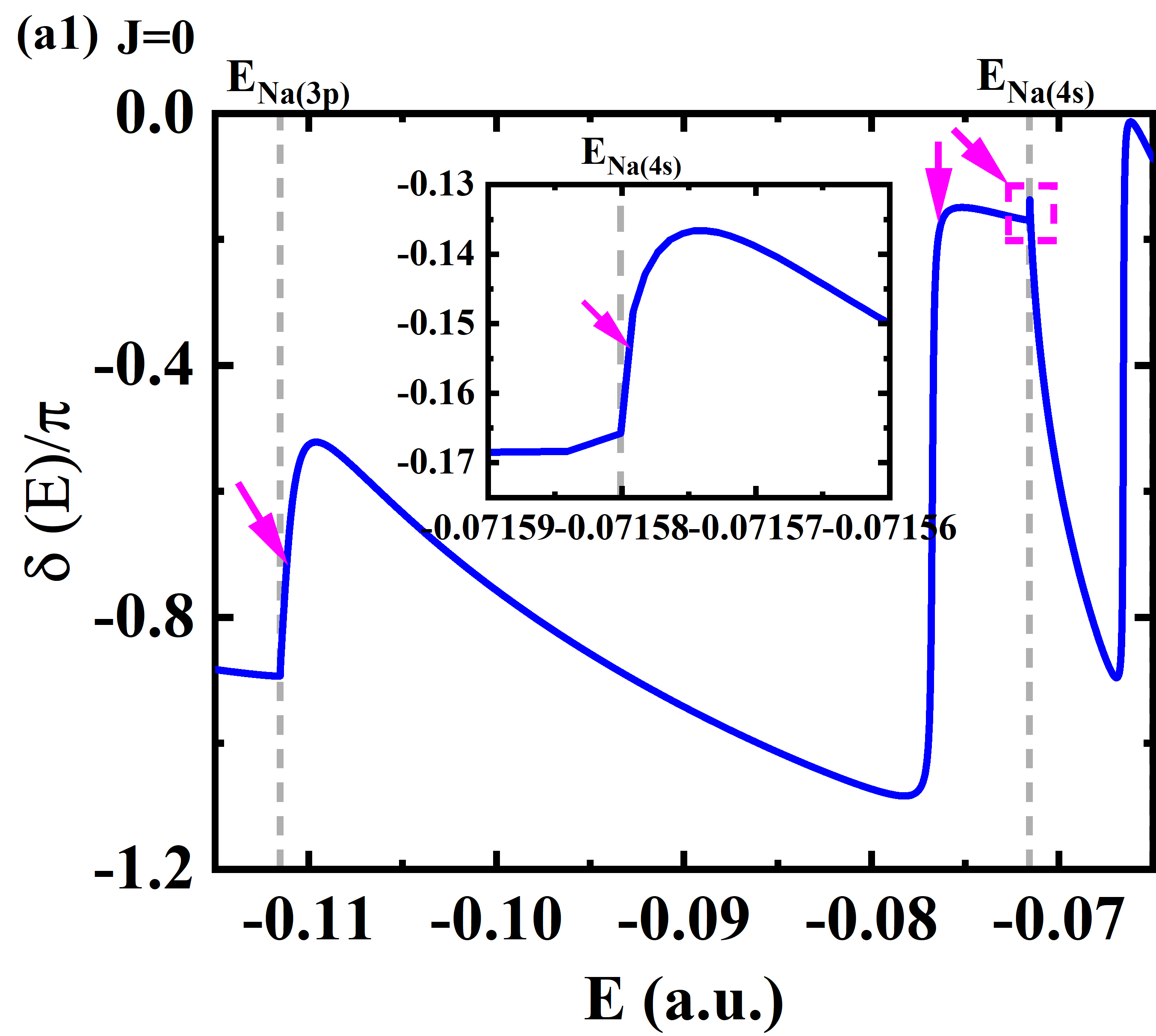}
		\label{fig2a1}
	}
	\subfigure{
		\includegraphics[scale=0.28]{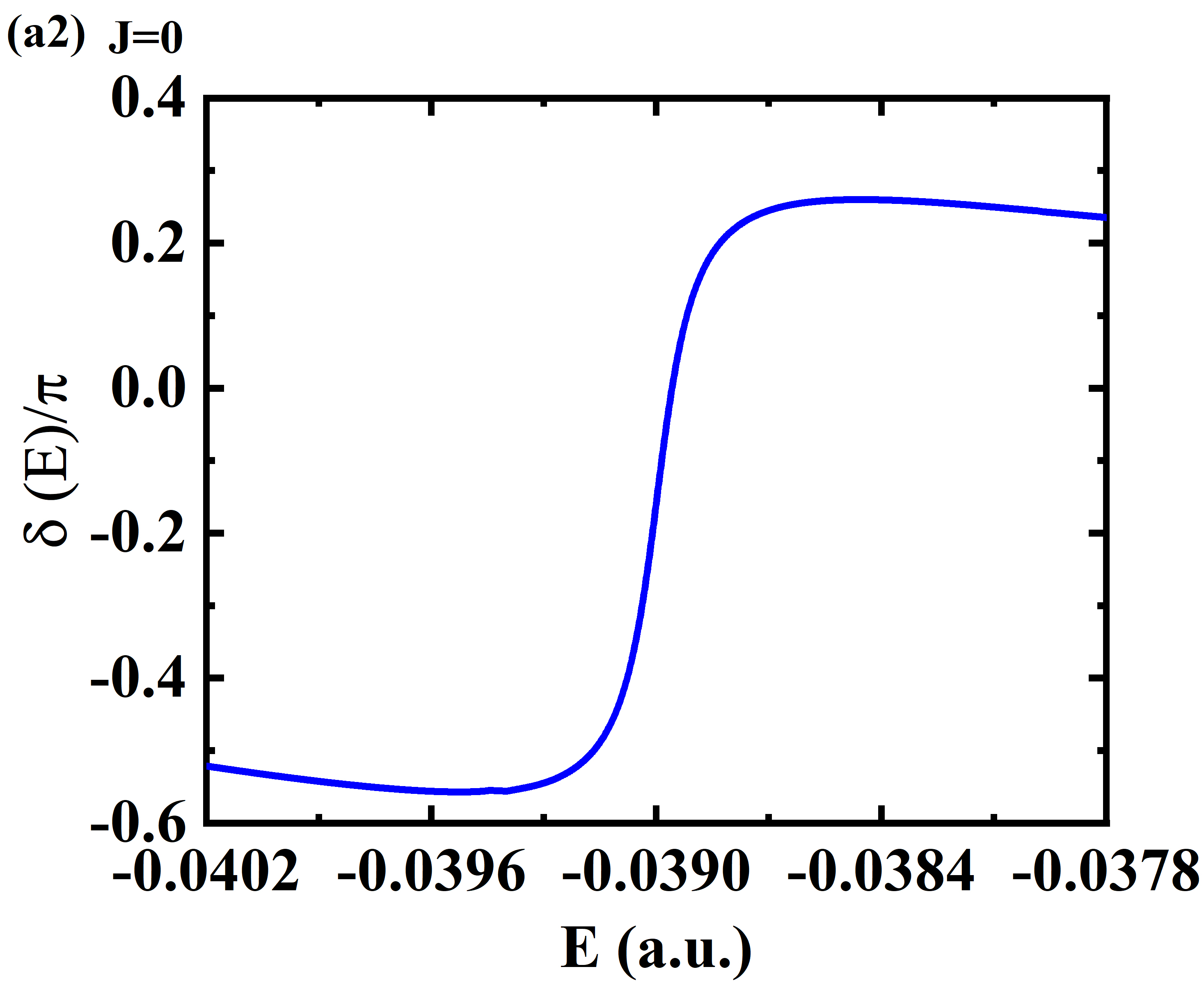}
		\label{fig2a2}
	}
	\renewcommand{\thesubfigure}{(b\arabic{subfigure})}
	\setcounter{subfigure}{0}
	\subfigure{
		\includegraphics[scale=0.28]{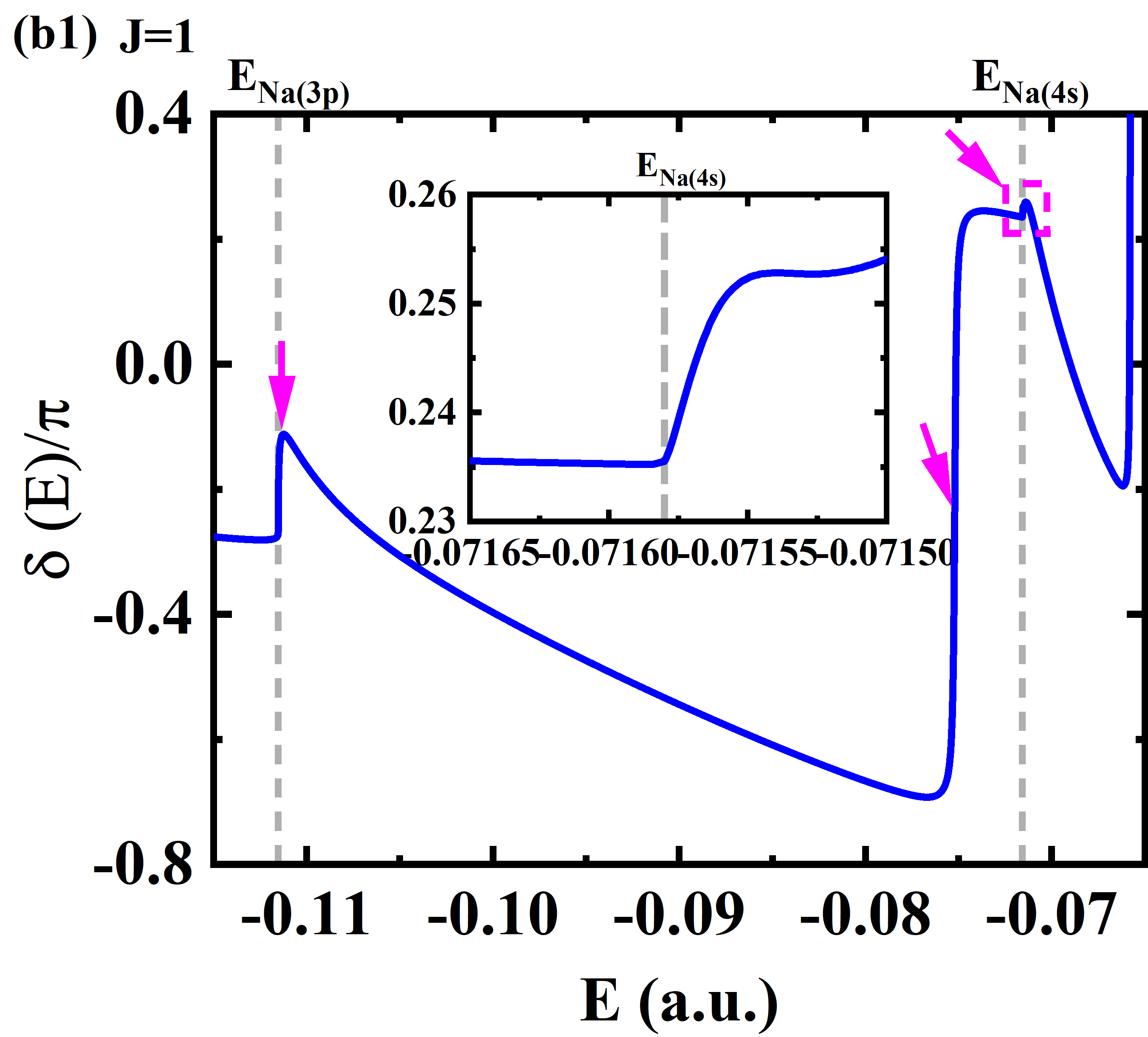}
		\label{fig2b1}
	}
	\subfigure{
		\includegraphics[scale=0.28]{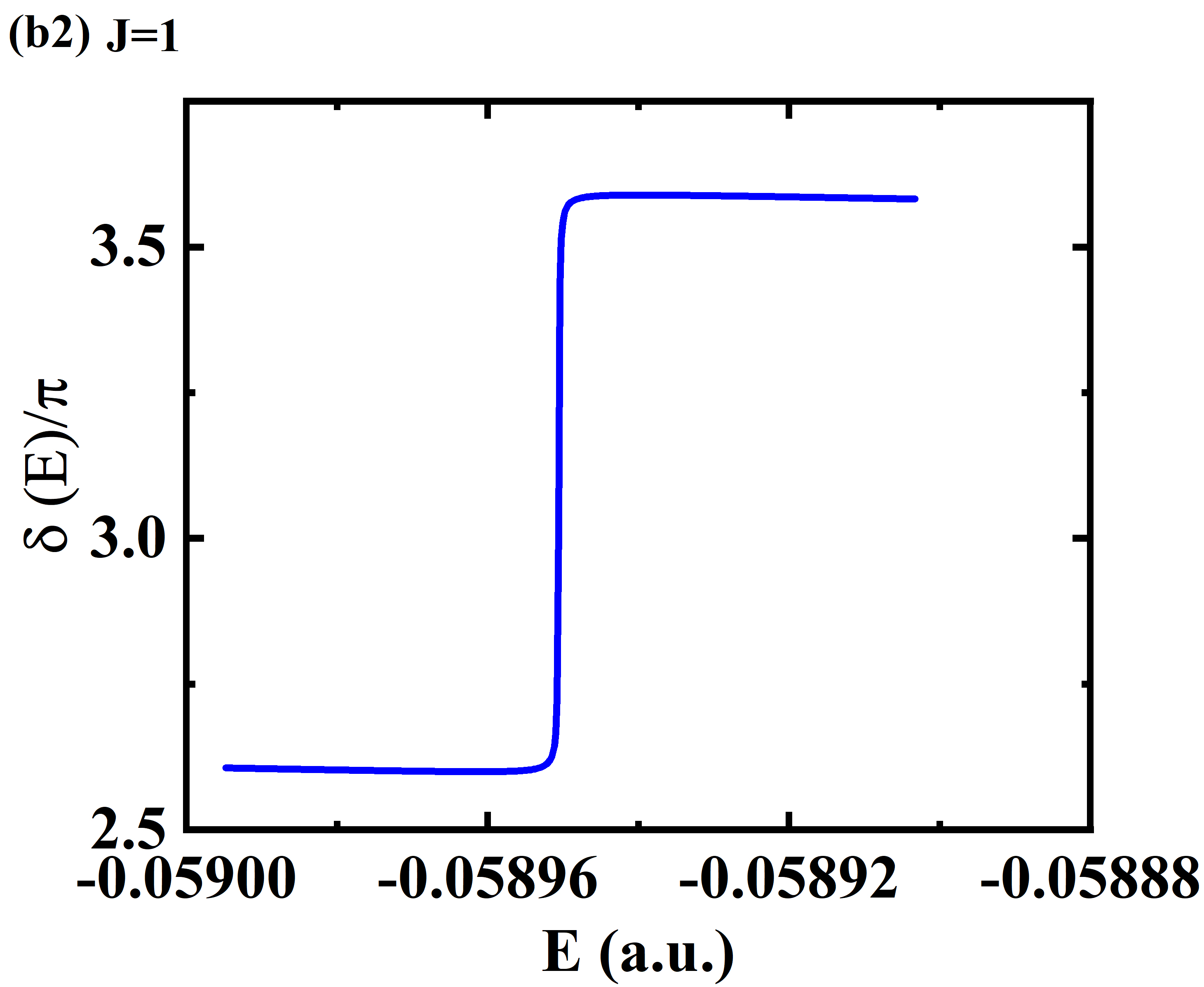}
		\label{fig2b2}
	}
	\renewcommand{\thesubfigure}{(c\arabic{subfigure})}
	\setcounter{subfigure}{0}
	\subfigure{
		\includegraphics[scale=0.28]{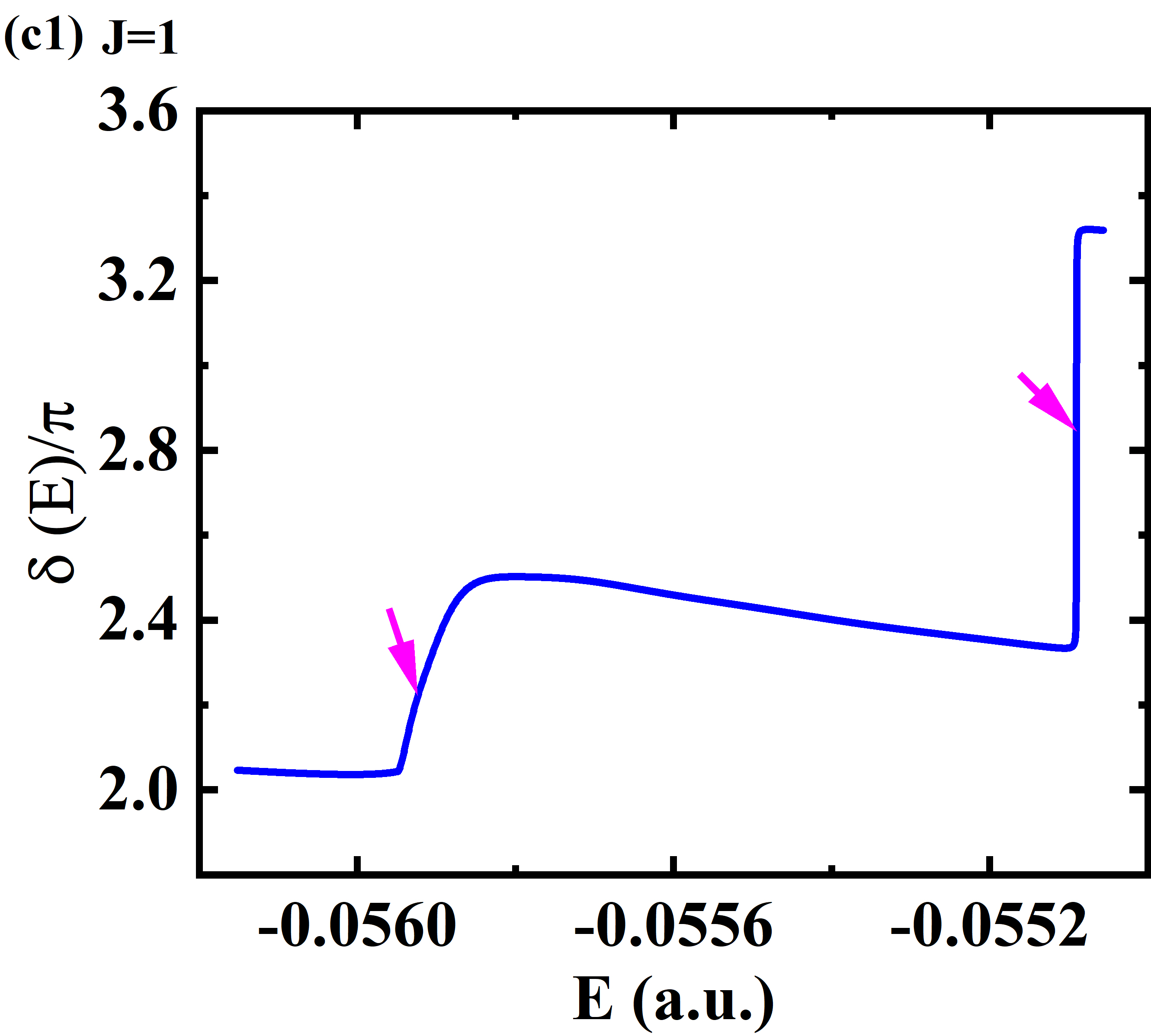}
		\label{fig2c1}
	}
	\subfigure{
		\includegraphics[scale=0.28]{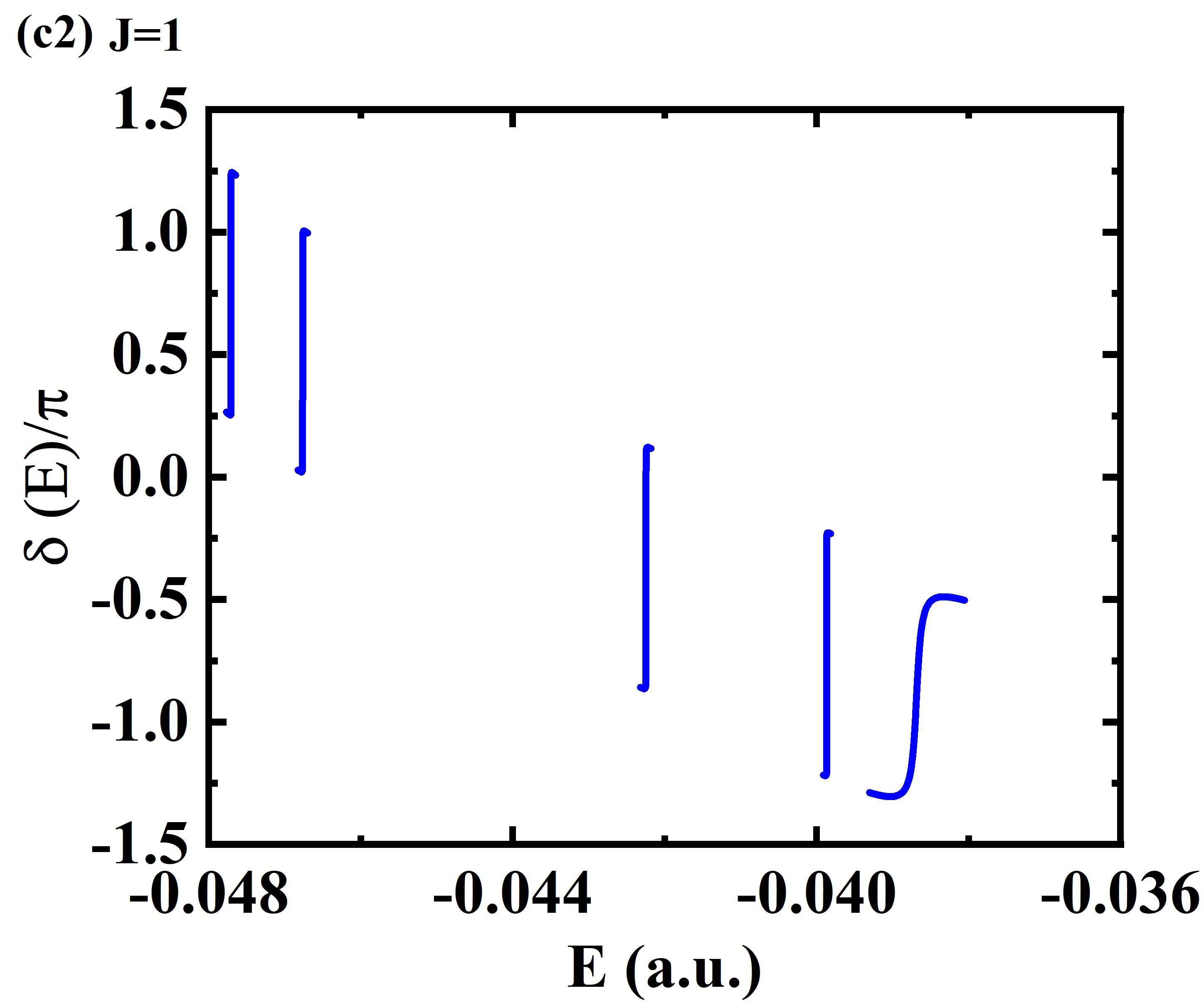}
		\label{fig2c2}
	}
	\renewcommand{\thesubfigure}{(d\arabic{subfigure})}
	\setcounter{subfigure}{0}
	\subfigure{
		\includegraphics[scale=0.29]{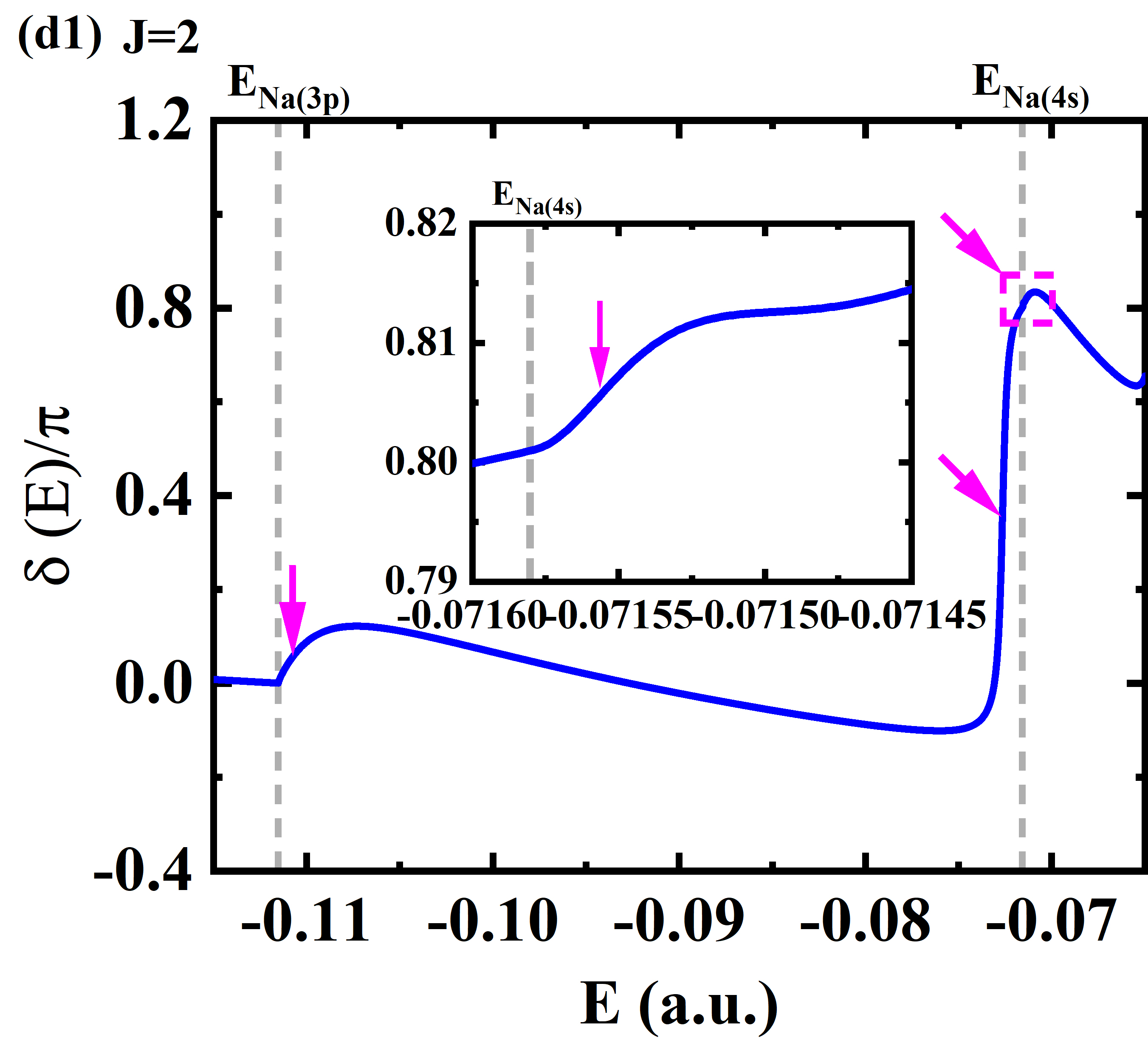}
		\label{fig2d1}
	}
	\subfigure{
		\includegraphics[scale=0.29]{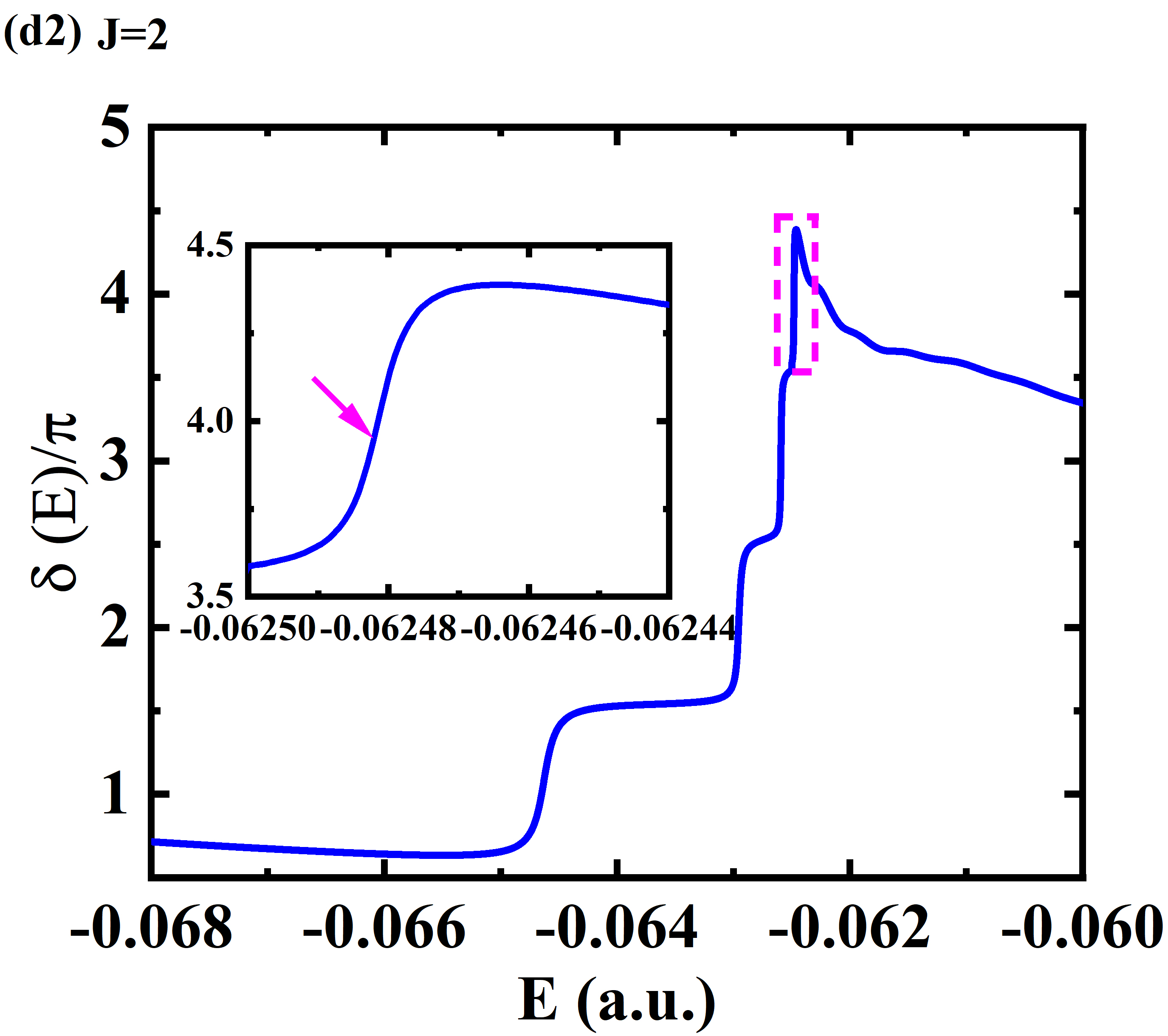}
		\label{fig2d2}
	}
	\caption{(Color online) The eigenphase sums in the e$^{\scriptscriptstyle+}$-Na system with $J=0-2$. Arrows indicate the resonance positions.}
\end{figure*}

\begin{figure*}[htbp]
	\centering
	\label{fig3}
	\renewcommand{\thesubfigure}{(a\arabic{subfigure})}
	\setcounter{subfigure}{0}
	\subfigure{
		\includegraphics[scale=0.275]{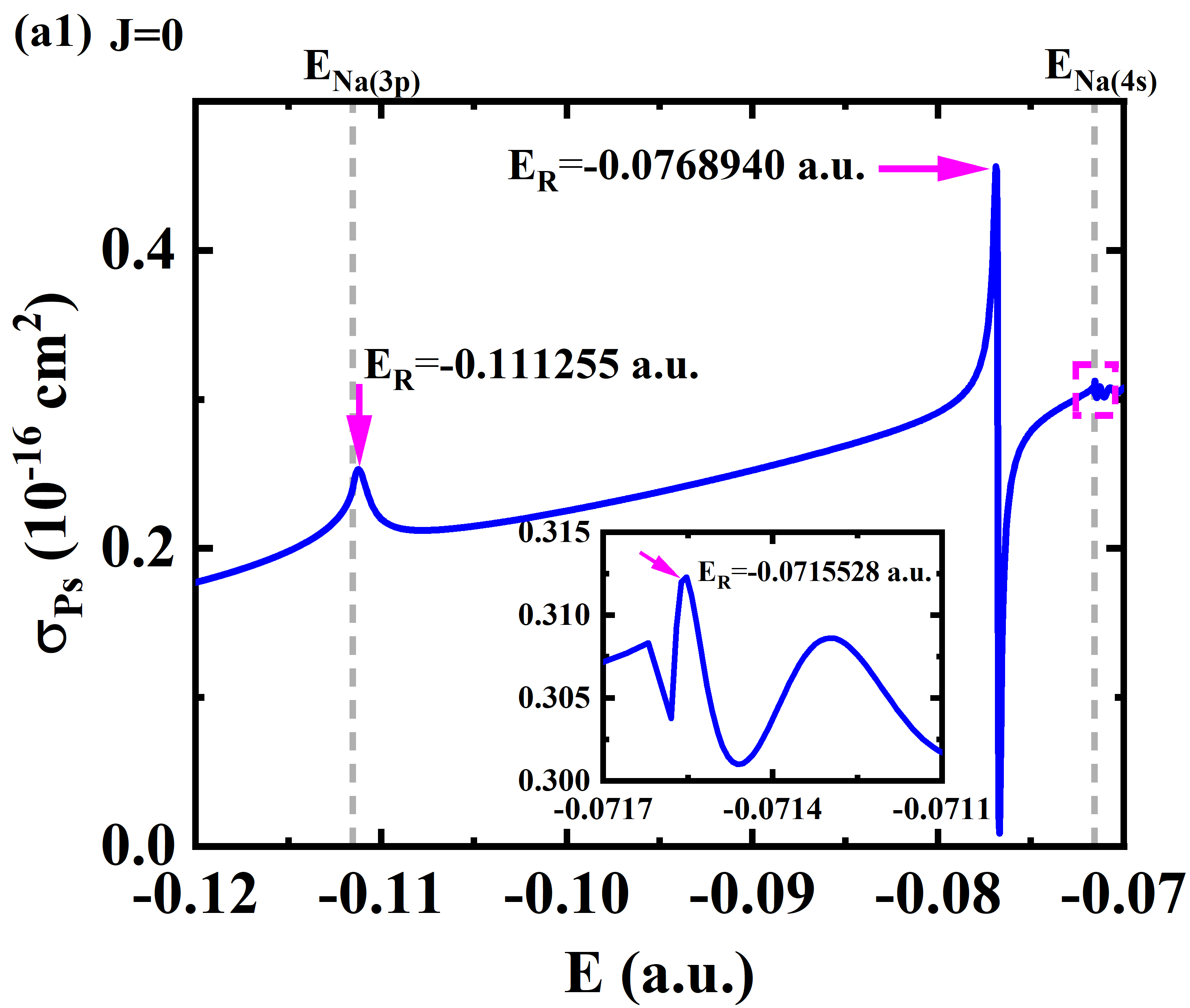}
		\label{fig3a1}
	}
	\subfigure{
		\includegraphics[scale=0.28]{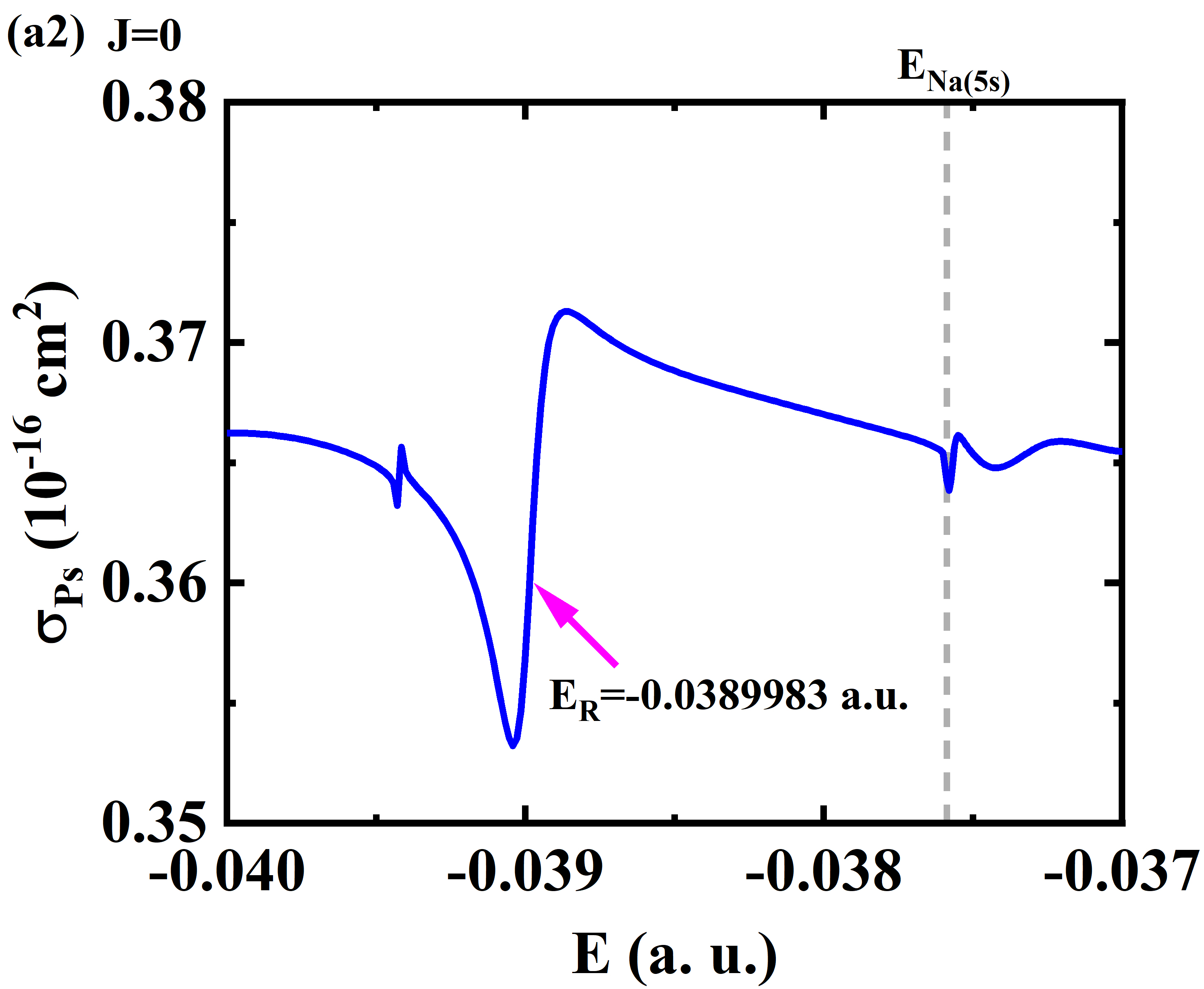}
		\label{fig3a2}
	}
	\renewcommand{\thesubfigure}{(b\arabic{subfigure})}
	\setcounter{subfigure}{0}
	\subfigure{
		\includegraphics[scale=0.275]{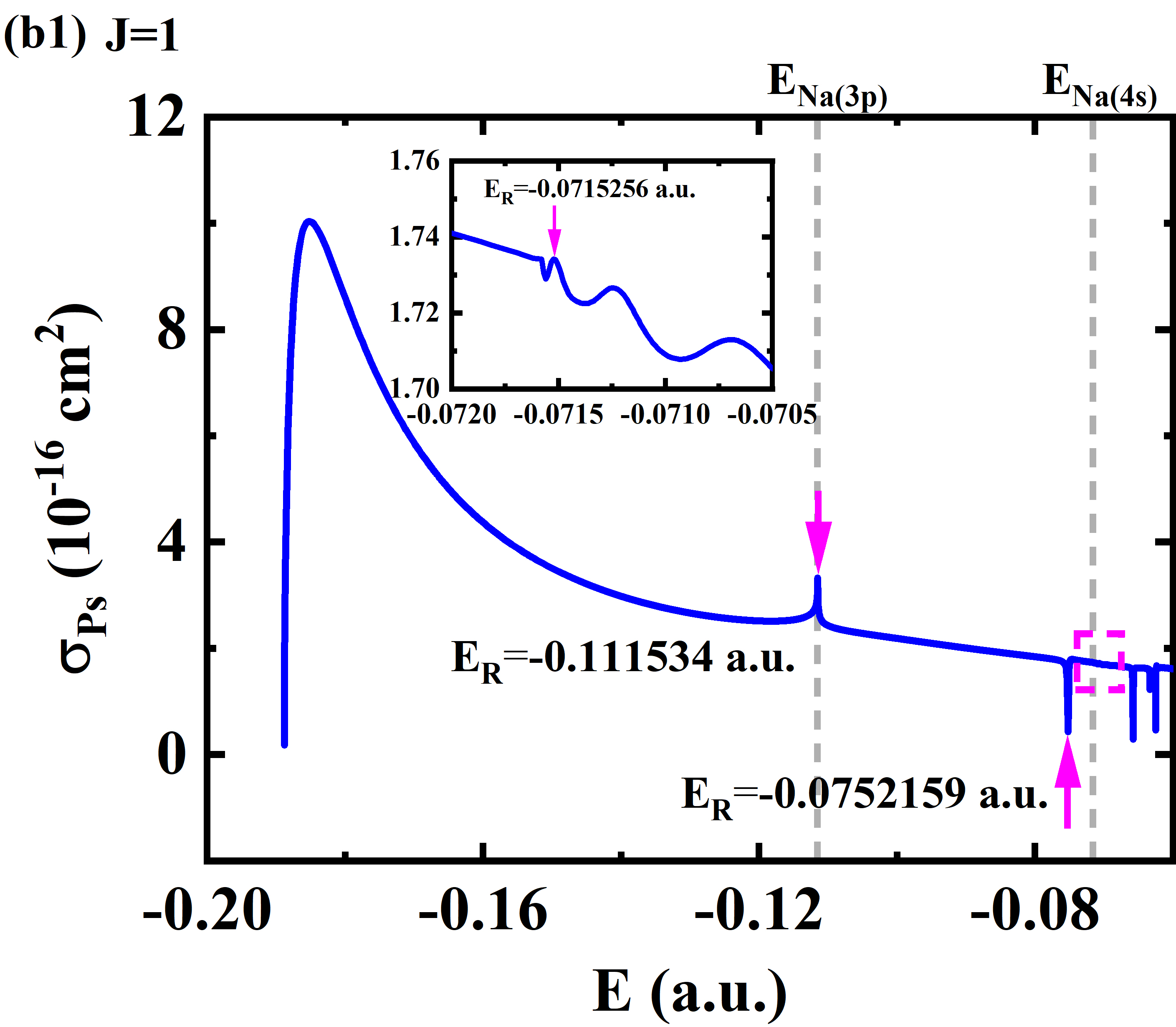}
		\label{fig3b1}
	}
	\subfigure{
		\includegraphics[scale=0.275]{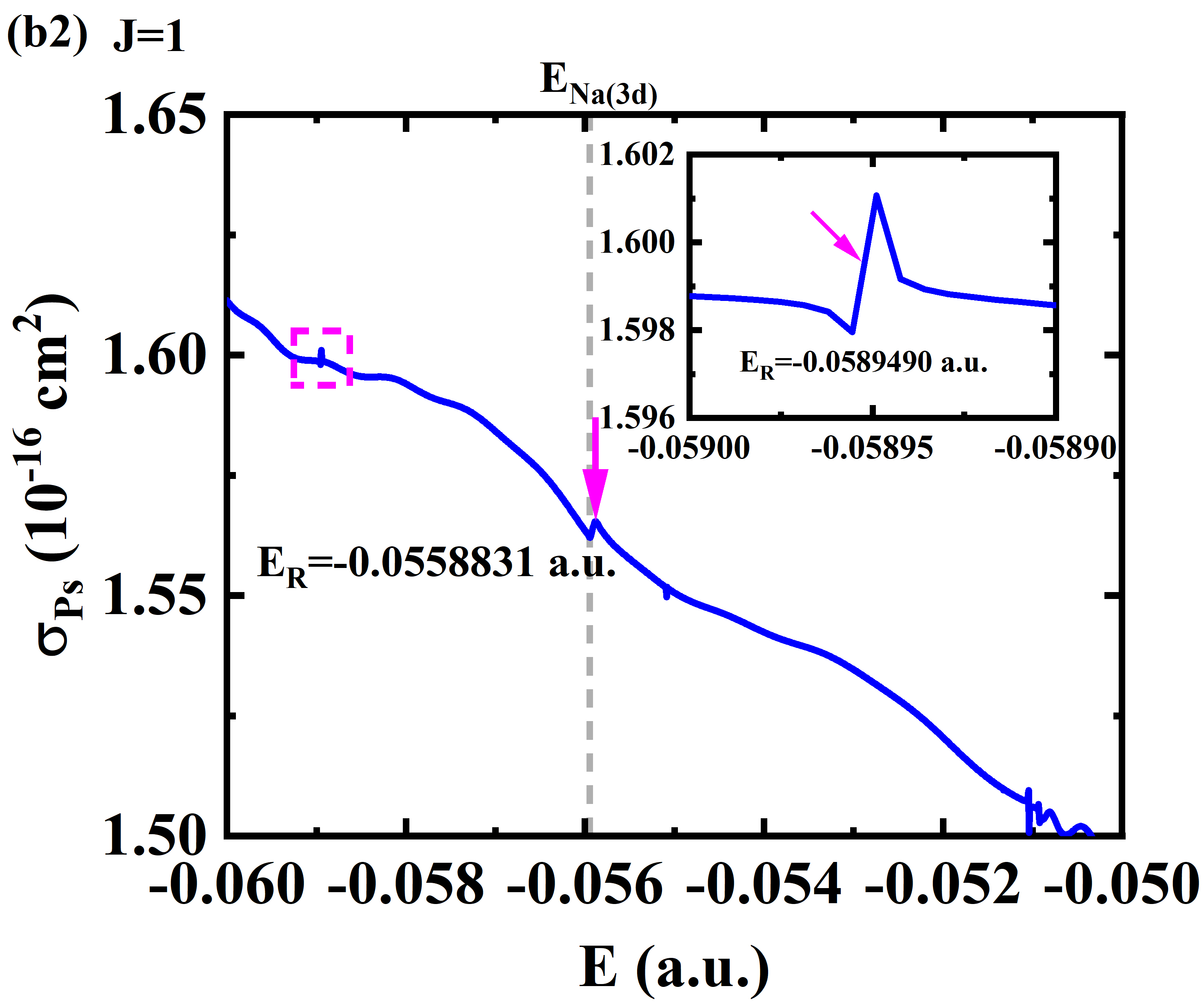}
		\label{fig3b2}
	}
	\renewcommand{\thesubfigure}{(c\arabic{subfigure})}
	\setcounter{subfigure}{0}
	\subfigure{
		\includegraphics[scale=0.275]{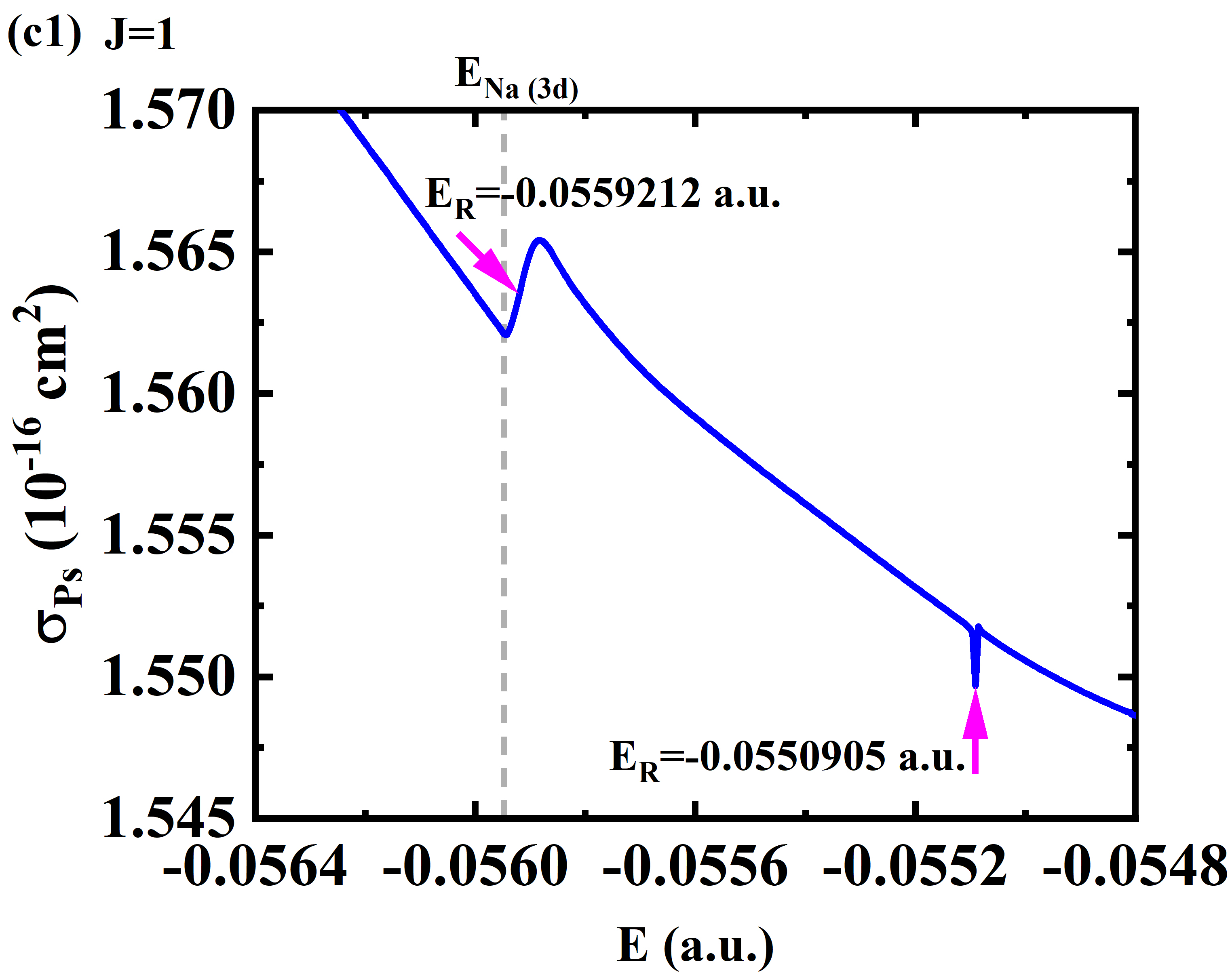}
		\label{fig3c1}
	}
	\subfigure{
		\includegraphics[scale=0.275]{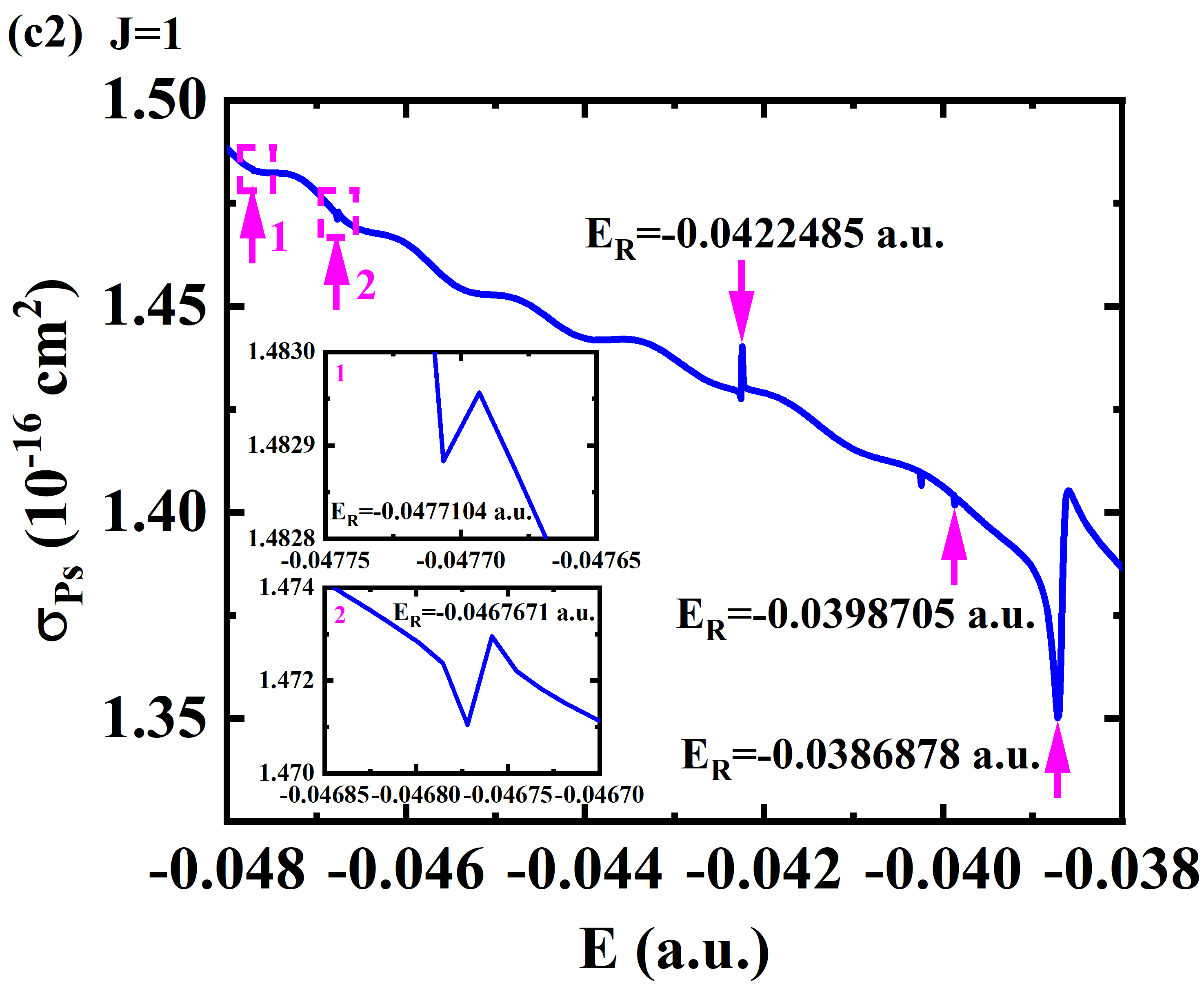}
		\label{fig3c2}
	}
	\renewcommand{\thesubfigure}{(d\arabic{subfigure})}
	\setcounter{subfigure}{0}
	\subfigure{
		\includegraphics[scale=0.285]{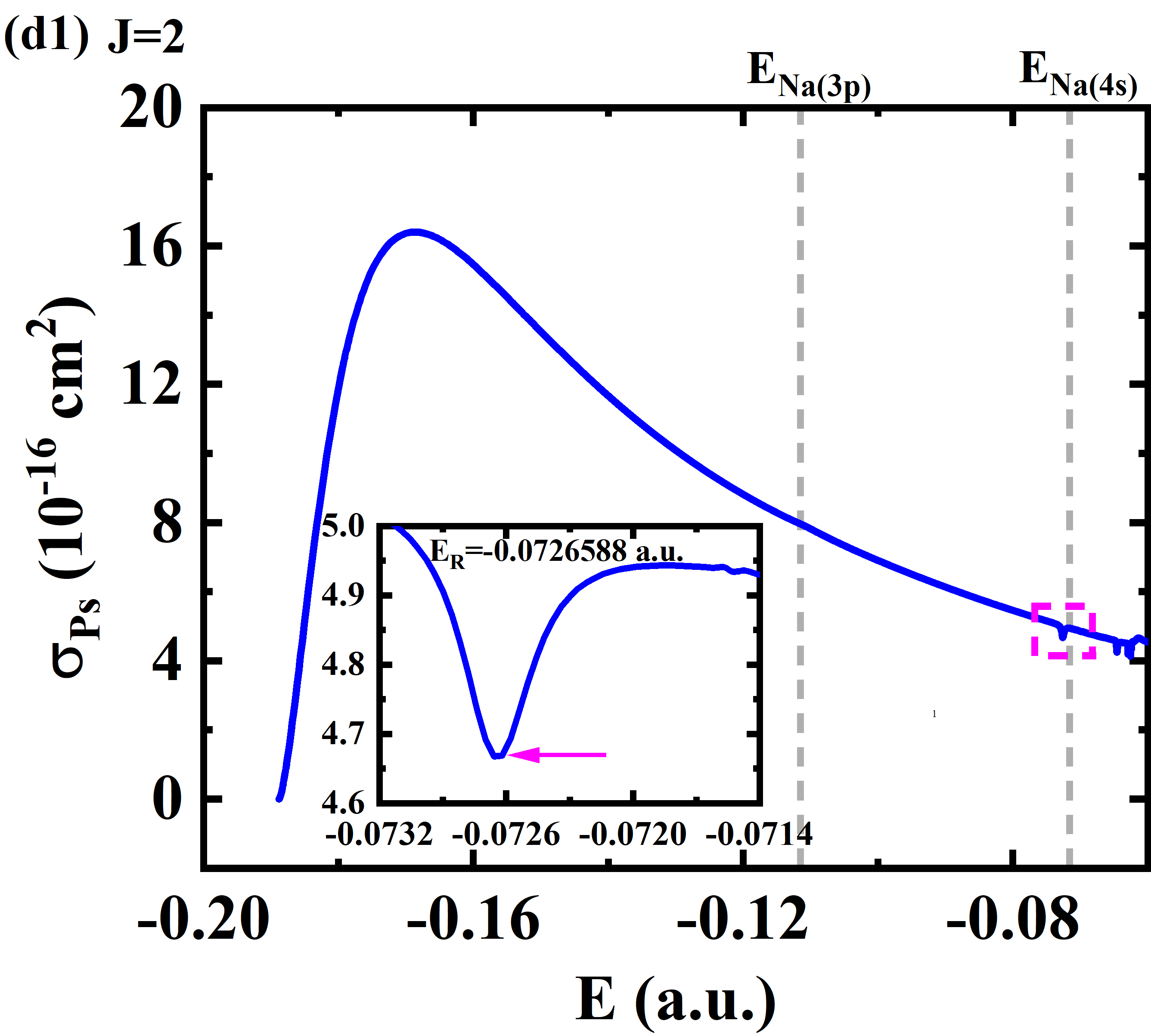}
		\label{fig3d1}
	}
	\subfigure{
		\includegraphics[scale=0.285]{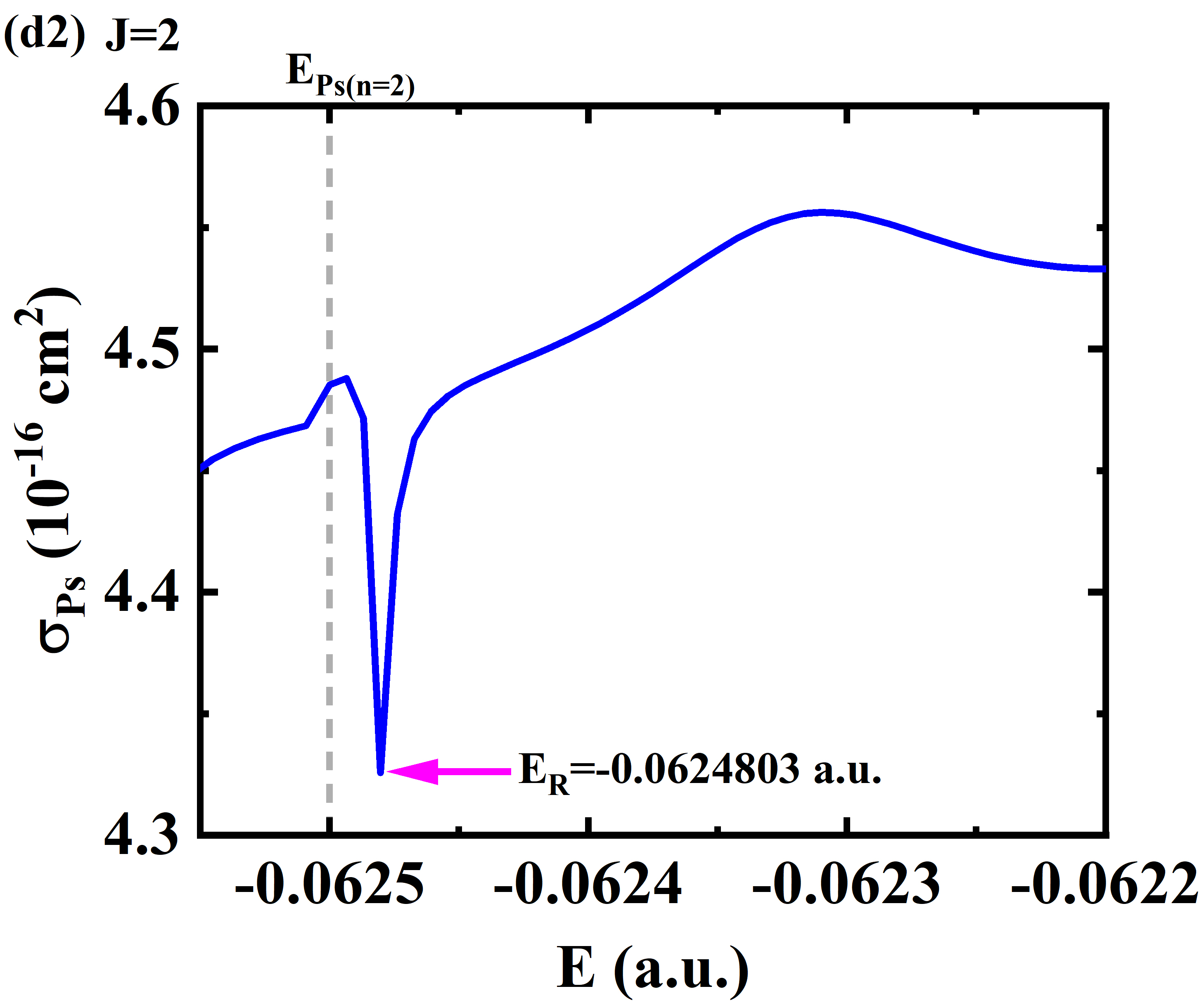}
		\label{fig3d2}
	}
	\caption{(Color online) The Ps($n=1,\,2$) formation cross sections for the e$^{\scriptscriptstyle+}$ + Na($3s$) $\rightarrow$ Ps($n=1,\,2$) + Na$^{\scriptscriptstyle+}$ process with $J=0-2$. Arrows indicate the resonance positions.}
\end{figure*}

For the resonance states associated with the Na($3p$) threshold, both the $S$- and $P$-wave resonances have been reported in previous coupled-channel optical and eigenphase-sum calculations, whereas the complex-coordinate-rotation method shows no corresponding resonance ~\cite{Umair2015Jul}. Figures~\ref{fig1a1}, \ref{fig1b1}, and \ref{fig1d1} display the stabilization plots for the $S$-, $P$-, and $D$-waves of the e$^{\scriptscriptstyle+}$--Na system. Nearly flat regions appear near $E_{R}=-0.111513$~a.u. for the $S$- and $P$-waves, while the $D$-wave exhibits a slight slope. The corresponding partial eigenphase sums shown in Figs.~\ref{fig2a1}, \ref{fig2b1}, and \ref{fig2d1} reveal a phase variation at $E_{R}=-0.111513$~a.u. for the $S$-wave, which becomes progressively smaller for $P$-wave; the phase variation in the $D$-wave is only about $0.2\pi$. The $S$- and $P$-waves Ps($n=1,\,2$) formation cross sections presented in Figs.~\ref{fig3a1}, \ref{fig3b1}, and \ref{fig3d1} for the e$^{\scriptscriptstyle+}$ + Na($3s$) $\rightarrow$ Ps($n=1,\,2$) + Na$^{\scriptscriptstyle+}$ process exhibit distinct resonance structures near $E_{R}=-0.111513$~a.u. No discernible structure is observed in the $D$-wave cross section. The narrow width, on the order of $10^{-5}$~a.u., and the small phase variation indicate that this resonance is weakly coupled to the open channels.

Below the Na($4s$) threshold, two pronounced flat regions are observed in the $S$-wave stabilization plots at $E_{R}=-0.0768010$~a.u. and $E_{R}=-0.0715790$~a.u., as shown in Fig.~\ref{fig1a2}. For higher partial waves, these features shift slightly in energy. The partial eigenphase sums in Fig.~\ref{fig2a1} exhibit a clear phase variation near $E_{R}=-0.0767715$~a.u., while the variation at $E_{R}=-0.0715790$~a.u., which lies close to the Na($4s$) threshold, is much smaller and nearly vanishes for higher partial waves. In the partial wave Ps($n=1,\,2$) formation cross sections for the e$^{\scriptscriptstyle+}$ + Na($3s$) $\rightarrow$ Ps($n=1,\,2$) + Na$^{\scriptscriptstyle+}$ process, shown in Figs.~\ref{fig3a1}, \ref{fig3b1}, and \ref{fig3d1}, a distinct resonance structure appears at $E_{R}=-0.0767715$~a.u., whereas only a weak bump is visible at $E_{R}=-0.0715790$~a.u. Ward's calculations~\cite{Ward1989nov} predict a resonance near $E_{R}=-0.0713846$~a.u. with a width of approximately $10^{-6}$~a.u., consistent with the weak or nearly absent signatures seen in both the eigenphase sum and cross section analyses. As summarized in Table~\ref{t1}, the deeper resonance associated with the Na($4s$) threshold agrees well with results from the complex-rotation method~\cite{Umair2017}. The near-threshold resonances at $E_{R}=-0.111513$~a.u. and $E_{R}=-0.0715790$~a.u. are very weakly coupled states, characterized by extremely small imaginary parts and minimal phase jumps in the eigenphase sum or stabilization analysis. Such weakly coupled resonances may therefore be difficult to detect using complex-coordinate-rotation calculations.

Between the Na($4s$) and Ps($n=2$) thresholds, the $S$-wave resonant state $E_{R}=-0.0615316$~a.u. predicted in Ref.~\cite{Jiao2012Feb} is absent in the present calculations, as well as in Ward's eigenphase-sum results~\cite{Ward1989nov} and in the complex-coordinate-rotation approach~\cite{Umair2015Jul}. A pronounced flat region is observed in the $D$-wave stabilization plots at $E_{R}=-0.0624815$~a.u., as shown in Fig.~\ref{fig1d2}. The $D$-wave Ps($n=1,\,2$) formation cross section also exhibits a corresponding structure at this energy (Fig.~\ref{fig3d2}). However, the hyperspherical potential curves indicate that this energy does not coincide with any threshold. As noted by Ward \textit{et al.}~\cite{Ward1989nov}, such structures arise from poles of the scattering matrix rather than threshold cusps, and therefore signify the presence of genuine resonant states.
	
Below the Na($3d$) threshold, the $P$-wave eigenphase sums shown in Fig.~\ref{fig2b2} exhibit a clear phase variation near $E_{R}=-0.058951$~a.u., with an estimated width on the order of $3\times10^{-7}$~a.u. This narrow resonance manifests as a faint structure in the scattering cross sections, becoming visible only upon magnification of the data, as shown in the enlarged view of the $P$-wave Ps($n=1,\,2$) formation cross section in Fig.~\ref{fig3b2}. For this resonance, no distinct flat region is observed in the $P$-wave stabilization plot, likely because its extremely small width and weak coupling to the open channels make it too narrow in energy to be clearly resolved within the discrete stabilization grid.
	
Below the Na($4p$) threshold, the $P$-wave eigenphase sums in Fig.~\ref{fig2c1} show two clear variations: one near $E_{R}=-0.0559212$~a.u. with an estimated width of about $6\times10^{-5}$~a.u., and another near $E_{R}=-0.0550905$~a.u. with an estimated width of about $3\times10^{-7}$~a.u. The $P$-wave Ps($n=1,\,2$) formation cross sections shown in Fig.~\ref{fig3c1} exhibit corresponding resonance structures at these energies. The stabilization plot in Fig.~\ref{fig1c1} displays a single horizontal line near $E_{R}=-0.0559212$~a.u., whereas the narrow resonance $E_{R}=-0.0550905$~a.u. does not produce a visible flat region. As the incident energy increases, more channels become energetically accessible and the interchannel coupling becomes increasingly complex, making the stabilization plot itself more intricate. This indicates that the stabilization method is most effective for identifying resonances in systems dominated by a single open channel.
	
Below the highly excited Na($5s$) threshold, an $S$-wave resonance is found near $E_{R}=-0.0389983$~a.u., with an estimated width of about $1 \times 10^{-4}$~a.u., as shown in the eigenphase sums of Fig.~\ref{fig2a2}. The $S$-wave Ps($n=1,\,2$) formation cross section in Fig.~\ref{fig3a2} also exhibits a corresponding structure at this energy. In contrast, the $P$-wave eigenphase sums in Fig.~\ref{fig2c2} display five clear variations, the first four correspond to narrow resonances with widths on the order of $10^{-7}$, while the last one has a width on the order of $10^{-4}$. These narrow resonances produce faint structures in the $P$-wave Ps($n=1,\,2$) formation cross section, as shown in Fig.~\ref{fig3c2} and its enlarged view.
	
	\begin{table*}[ht]\footnotesize
		\centering
		\renewcommand{\arraystretch}{1.3}
		\begin{threeparttable}
			\caption{\label{t1} Comparison of the lowest five partial-wave resonance energies $E_{R}$ and widths $\Gamma$ for the e$^{\scriptscriptstyle+}$-Na system with previous calculations. The threshold energies are shown here. The notation x[y] means $10^{-y}$.}
			\begin{ruledtabular}
				\begin{tabular}{ccccccccccc}		
					&\multicolumn{2}{c}{Present\tnote{a}}
					&\multicolumn{2}{c}{Present\tnote{b}}
					&\multicolumn{2}{c}{Ref.~\cite{Ward1989nov}\tnote{c}}
					&\multicolumn{2}{c}{Ref.~\cite{Jiao2012Feb}\tnote{d}}
					&\multicolumn{2}{c}{Ref.~\cite{Umair2017}\tnote{e}}\\
					\cline{2-3} \cline{4-5} \cline{6-7} \cline{8-9} \cline{10-11}
					\multicolumn{1}{c}{Partial wave}&
					\multicolumn{1}{c}{$E_{R}$}&\multicolumn{1}{c}{$\Gamma$}&
					\multicolumn{1}{c}{$E_{R}$}&\multicolumn{1}{c}{$\Gamma$}&\multicolumn{1}{c}{$E_{R}$}&\multicolumn{1}{c}{$\Gamma$}&\multicolumn{1}{c}{$E_{R}$}&\multicolumn{1}{c}{$\Gamma$}&\multicolumn{1}{c}{$E_{R}$} & \multicolumn{1}{c}{$\Gamma$}\\
					\hline					
					$S$ & -0.111513&5.11[5]&-0.111486&2.41[6]&-0.1158699&1.47[5]   &-0.116311&1.31[3] &&\\[1.0ex]	
					$P$ &-0.111530&9.17[6]&-0.111503&3.59[6]& -0.1129385& 2.21[4] &-0.113086 &1.75[3]   &         & \\[1.0ex]	
					$D$ &-0.111518&3.69[5]&-0.111488&1.00[5]\\	
					\multicolumn{11}{c}{Na(3p) threshold ($E_{t}=-0.11153440)$}\\
					\hline	
					$S$ &-0.0767715&1.58[4]&-0.0767777&1.52[4]&& &-0.0814949&5.55[4]&-0.0767885&1.51[4]\\
					&-0.0715790&1.20[6] &-0.0715538&2.20[6]&-0.0713846 &7.35[6]&\\[1.0ex]
					$P$ &	-0.0752191 & 1.49[4]&-0.0752430&1.36[4]& -0.0740047& 1.18[3]&-0.0778282 &3.36[3]   &-0.0751853& 1.00[4]\\
					&-0.0715714&2.04[5]&-0.0715050&2.57[6]\\[1.0ex]	
					$D$ &-0.0726528 & 4.01[4]&-0.0726935 &3.45[4]&    &  &-0.0730758    &6.44[3]   &-0.0725989 & 4.96[4] \\
					&-0.0715546&4.66[5]&-0.0714998&1.37[6]\\
					\multicolumn{11}{c}{Na(4s) threshold ($E_{t}=-0.07158008)$}\\
				\hline	
                     $S$ &&&&&&&-0.0615316&1.11[3]\\[1.0ex]	
                     $D$ &-0.0624815&7.42[6]&-0.0624777&9.53[6]\\
					\multicolumn{11}{c}{Ps($n=2$) threshold ($E_{t}=-0.0625)$}\\
					\hline	
					$P$ &-0.0589506&3.45[7]& &  \\				
					\multicolumn{11}{c}{Na(3d) threshold ($E_{t}=-0.05594829)$}\\
					\hline	
					$S$ &&&&&-0.0557596 &1.47[4] &      -0.0547669 &3.64[4]\\[1.0ex]	
					$P$ &-0.0559212 & 6.18[5]&-0.0559018 &6.34[6]&           &         &-0.0544091 &2.28[4]   &       &\\
					&-0.0550905&3.81[7]	\\
					\multicolumn{11}{c}{Na(4p) threshold ($E_{t}=-0.05093737)$}\\
					\hline	
					$S$ &-0.0389983&1.52[4]&\\[1.0ex]	
					$P$ &-0.0477104&4.19[7]&-0.0508918&1.38[6]\\
					&-0.0467671&5.88[7]&\\
					&-0.0422485&6.73[7]&\\
					&-0.0398705 & 3.91[7]&\\
					&-0.0386878&1.03[4]&-0.0386996&5.01[5]\\
					\multicolumn{11}{c}{Na(5s) threshold ($E_{t}=-0.03758689)$}\\				
				\end{tabular}
			\end{ruledtabular}
			\begin{tablenotes}
				\footnotesize
				\item[a] The eigenphase sum method
				\item[b] The stabilization method
				\item[c] The eigenphase sum method
				\item[d] The momentum-space coulped-channel optical method
				\item[e] The complex scaling method
			\end{tablenotes}
		\end{threeparttable}
	\end{table*}
	
	\subsection{Dipole series resonances}
	\subsubsection{Dipole series resonances with \text{Ps}$(n=1,2)$+\text{Na}$^{\scriptscriptstyle+}$}
	
	The energies of Ps atom are degenerate with respect to the $l$ quantum number, and
	hence, the Ps-A$^{\scriptscriptstyle+}$ interaction give a dipole
	interaction, i.e., an interaction-potential with long-range form
	proportional to $-1/r^2$. This long range potential gives, in
	principle, an infinite sequence of quasibound states clustering
	towards the Ps($n=2$) thresholds.
	The binding energies (and widths) of each sequence follow a fixed ratio\,\cite{Gailitis1963},
	
	\begin{align}
		\label{Evratio}
		\frac{E_{\nu}}{E_{\nu+1}}=e^{\frac{2\pi}{\alpha}}\,.
	\end{align}
	where the subscript $\nu$ denotes different states in the series.
	\begin{align}
		\label{Ev}
		\ln(E_{\nu})=\ln(E_0)-\alpha\nu\,.
	\end{align}
	The resonances within each dipole series can be connected by straight lines fitted using Eq.\,(\ref{Ev}).
	
\begin{figure*}[htbp]
	\centering	
	\label{fig10}
	\renewcommand{\thesubfigure}{(a\arabic{subfigure})}
	\setcounter{subfigure}{0}	
	\subfigure{
		\includegraphics[scale=0.23]{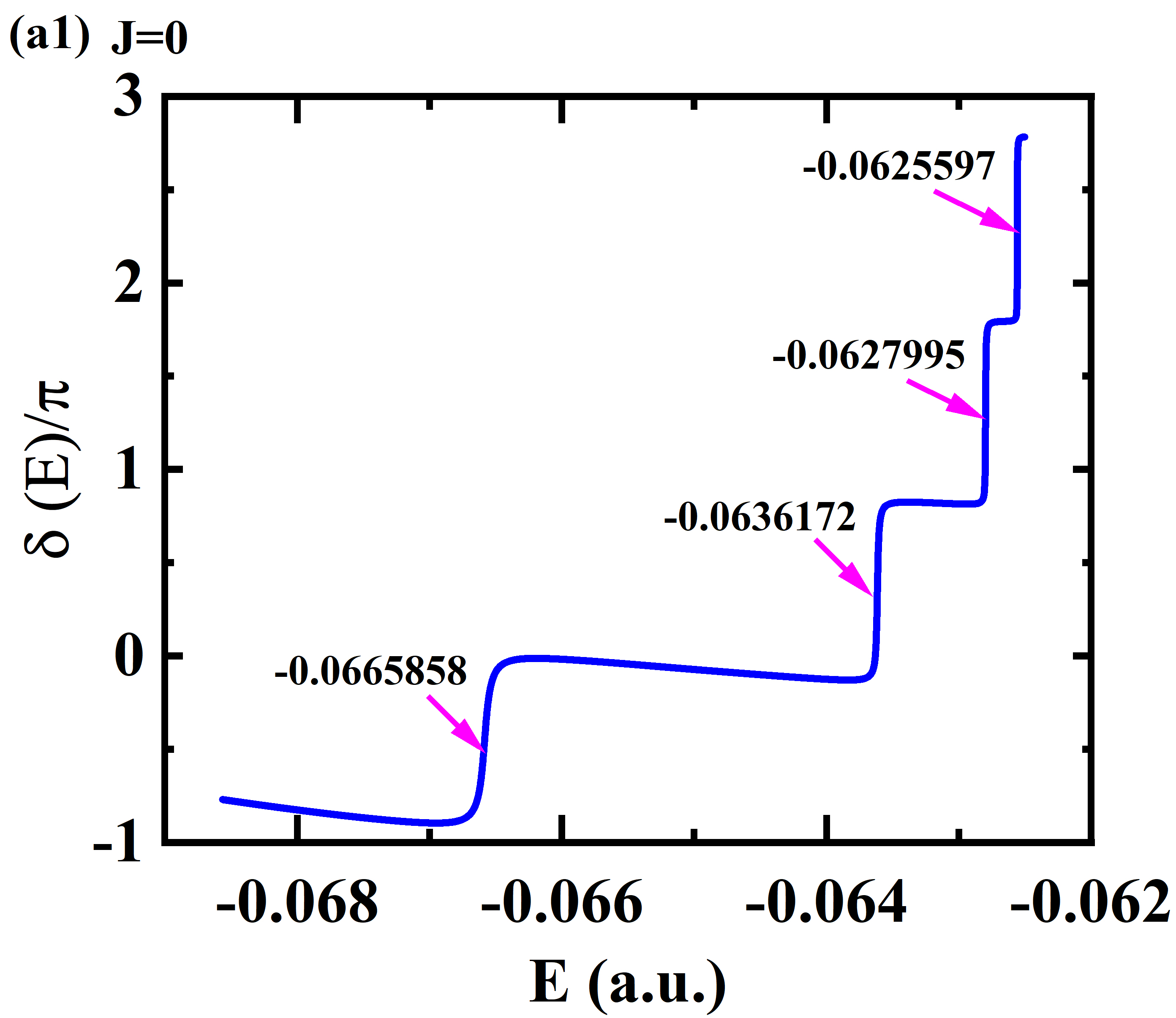}
		\label{fig10a1}
	}
	\subfigure{
		\includegraphics[scale=0.235]{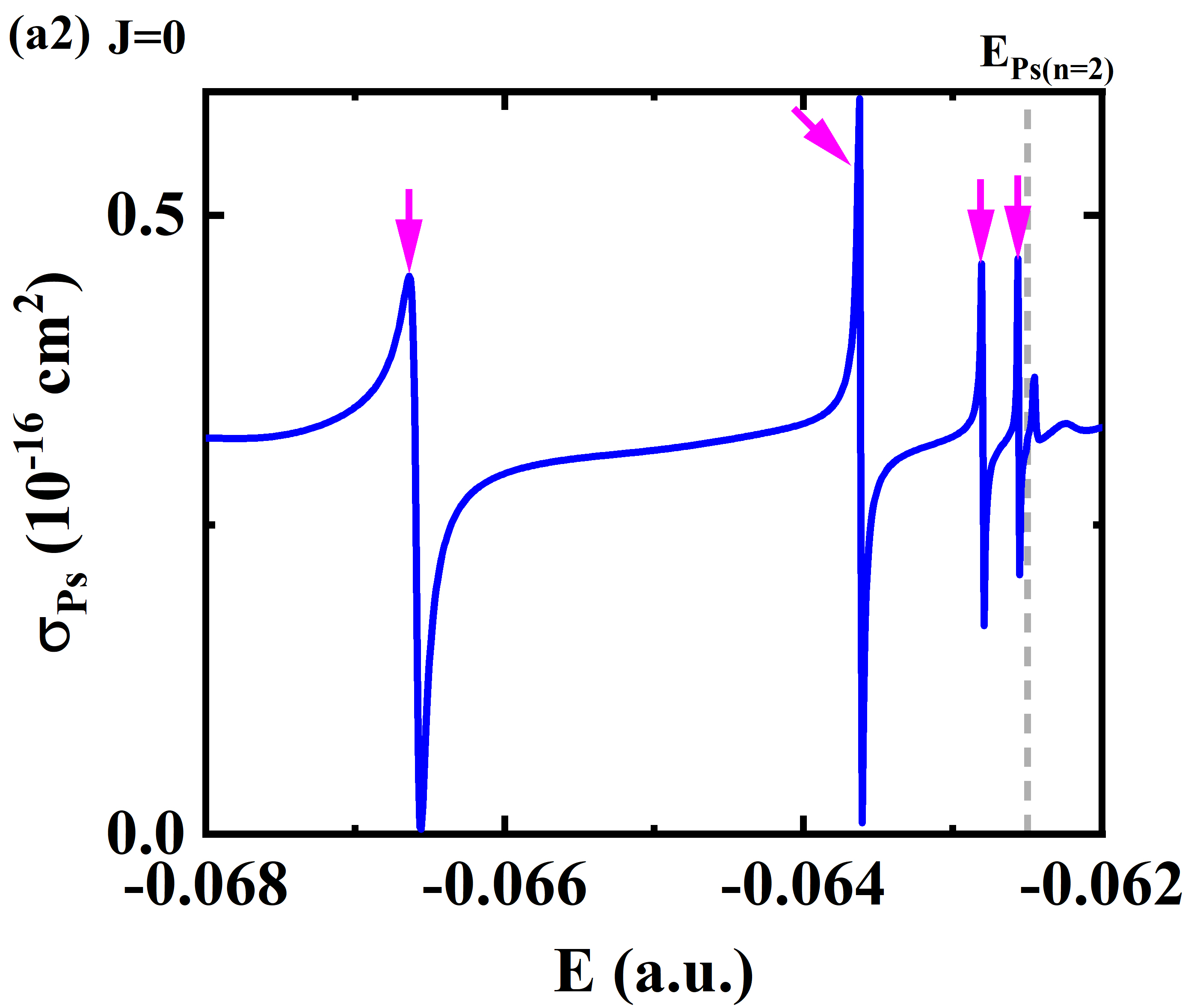}
		\label{fig10a2}
	}
	\subfigure{
		\includegraphics[scale=0.235]{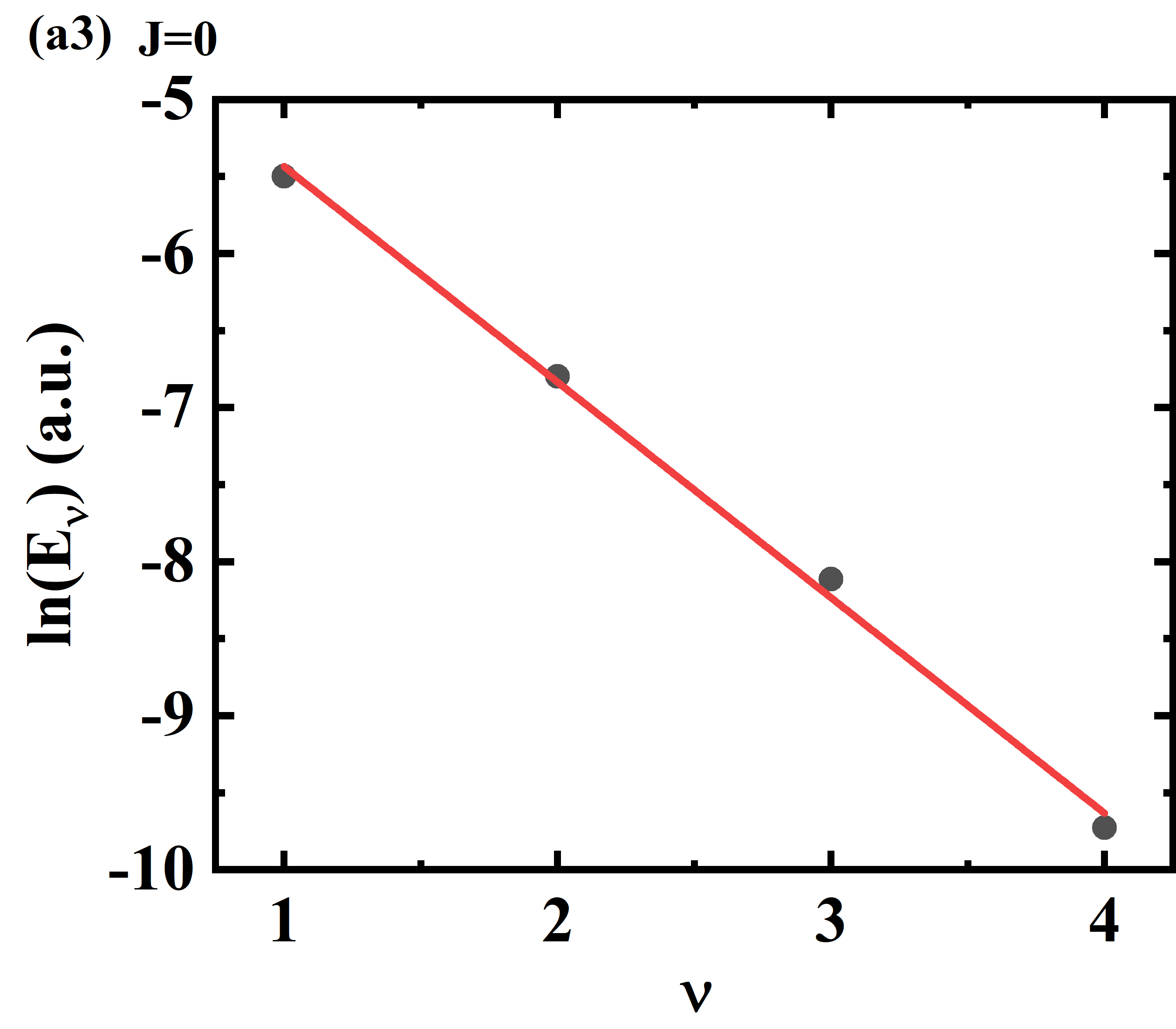}
		\label{fig10a3}
	}
	\renewcommand{\thesubfigure}{(b\arabic{subfigure})}
	\setcounter{subfigure}{0}
	\subfigure{
		\includegraphics[scale=0.23]{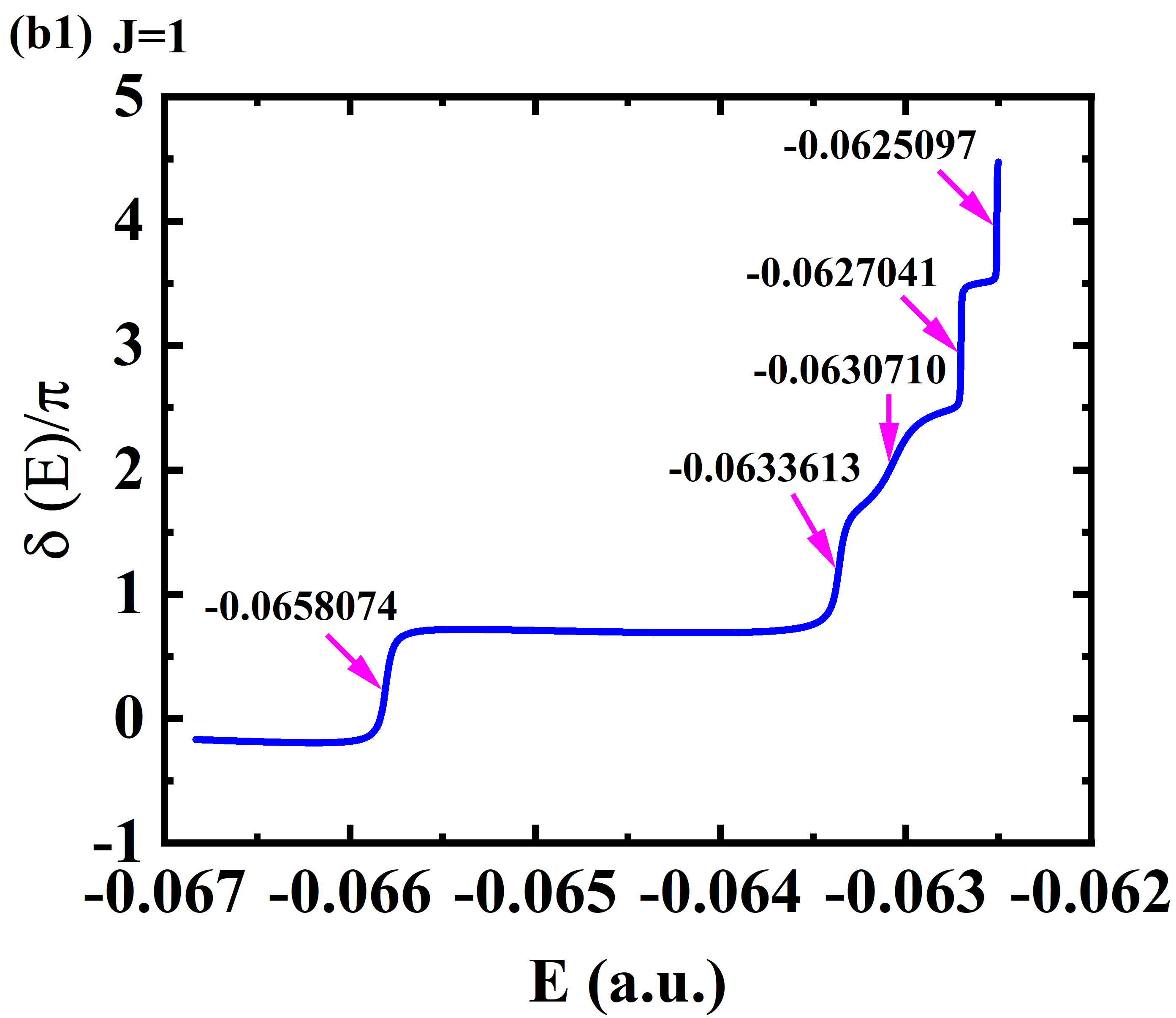}
		\label{fig10b1}
	}
	\subfigure{
		\includegraphics[scale=0.236]{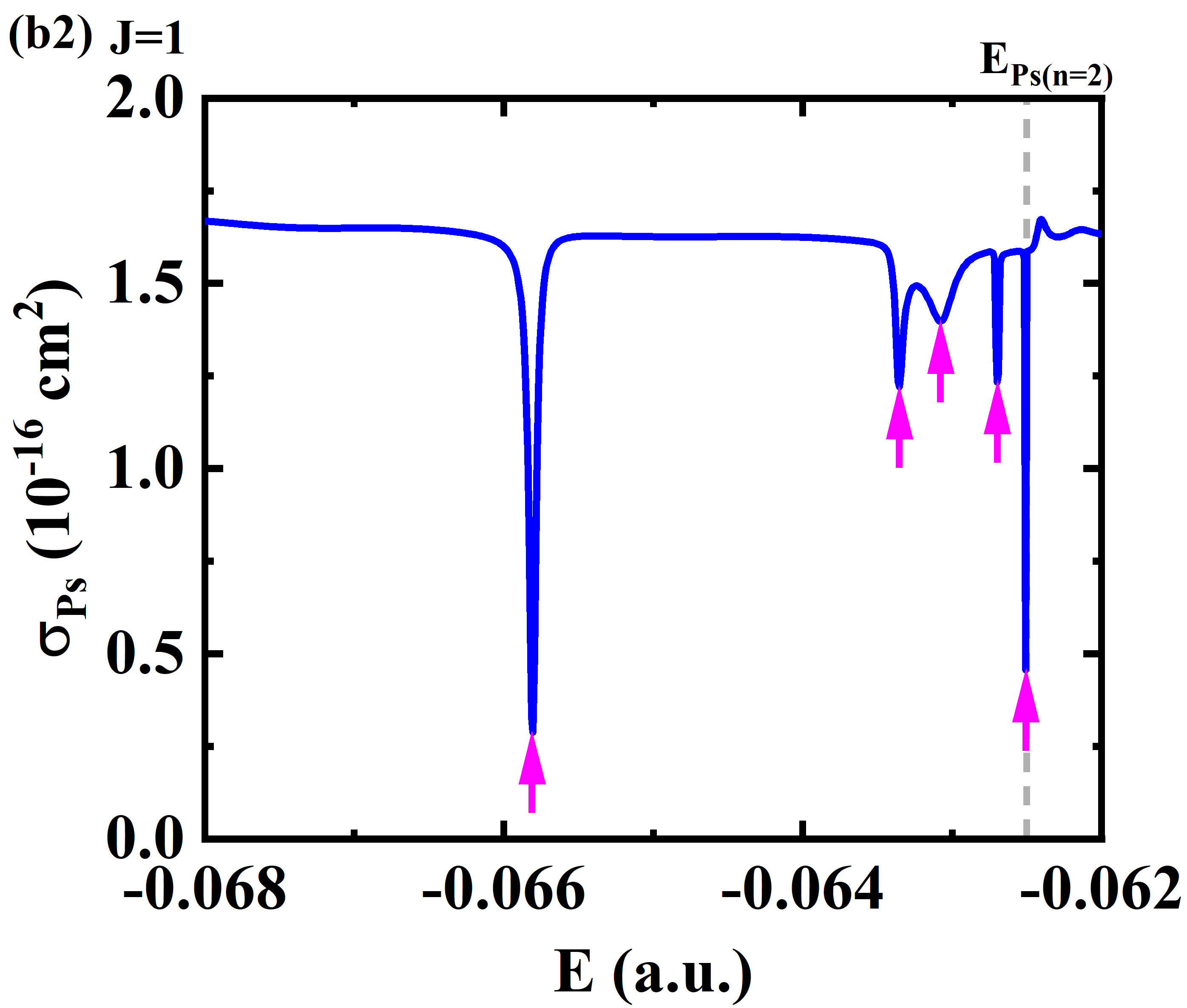}
		\label{fig10b2}
	}
	\subfigure{
		\includegraphics[scale=0.236]{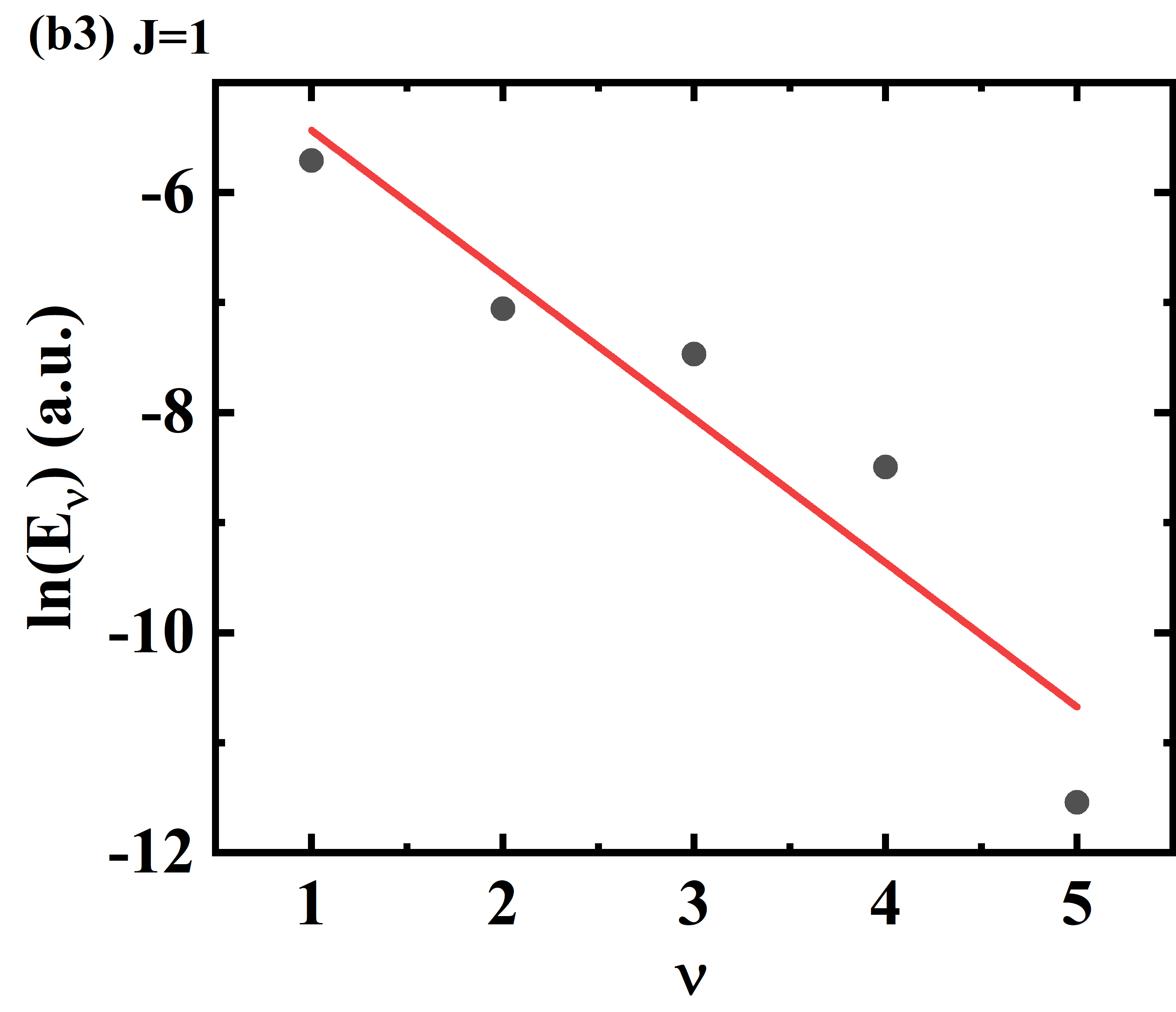}
		\label{fig10b3}
	}
	\renewcommand{\thesubfigure}{(c\arabic{subfigure})}
	\setcounter{subfigure}{0}
	\subfigure{
		\includegraphics[scale=0.236]{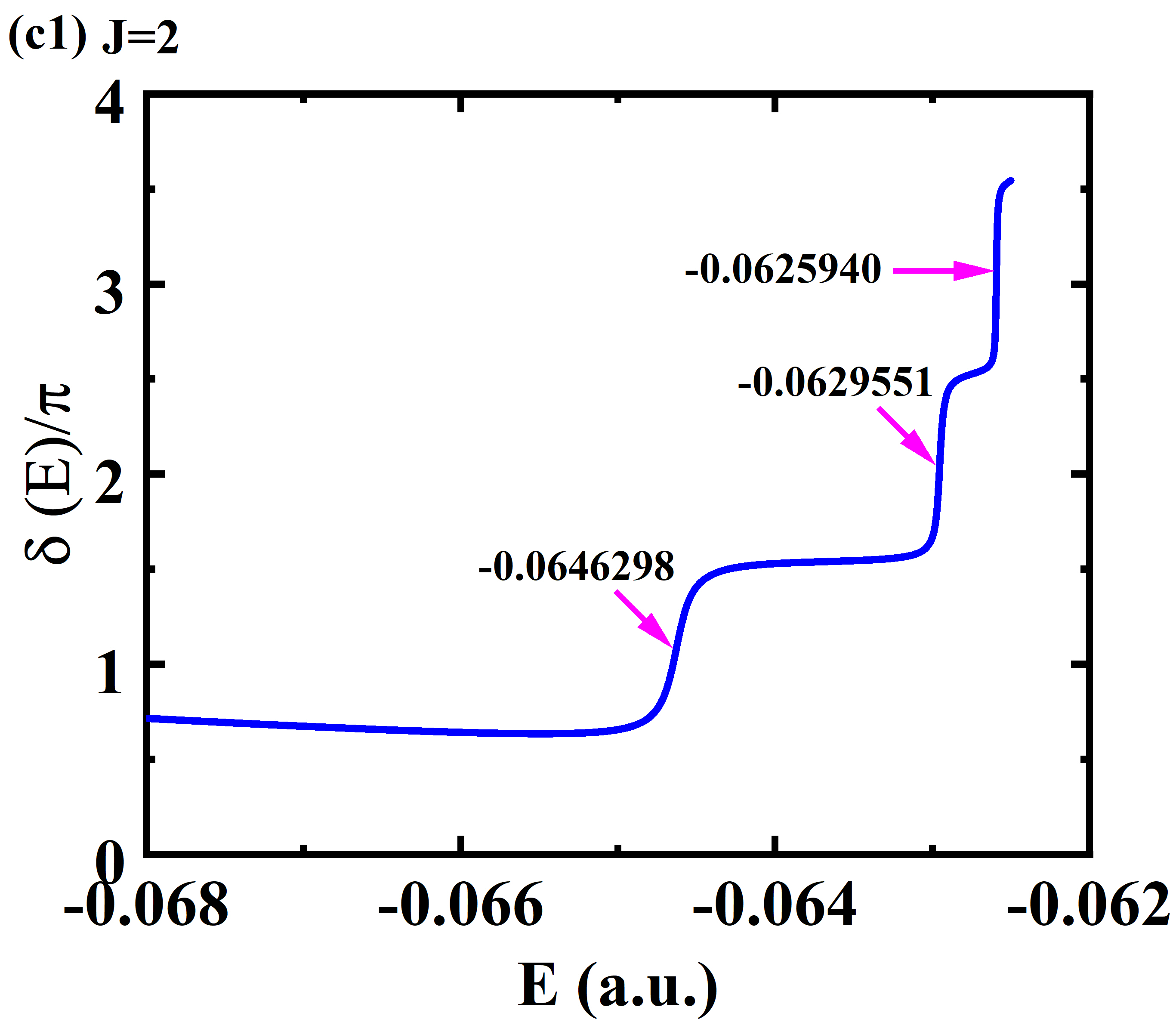}
		\label{fig10c1}
	}
	\subfigure{
		\includegraphics[scale=0.238]{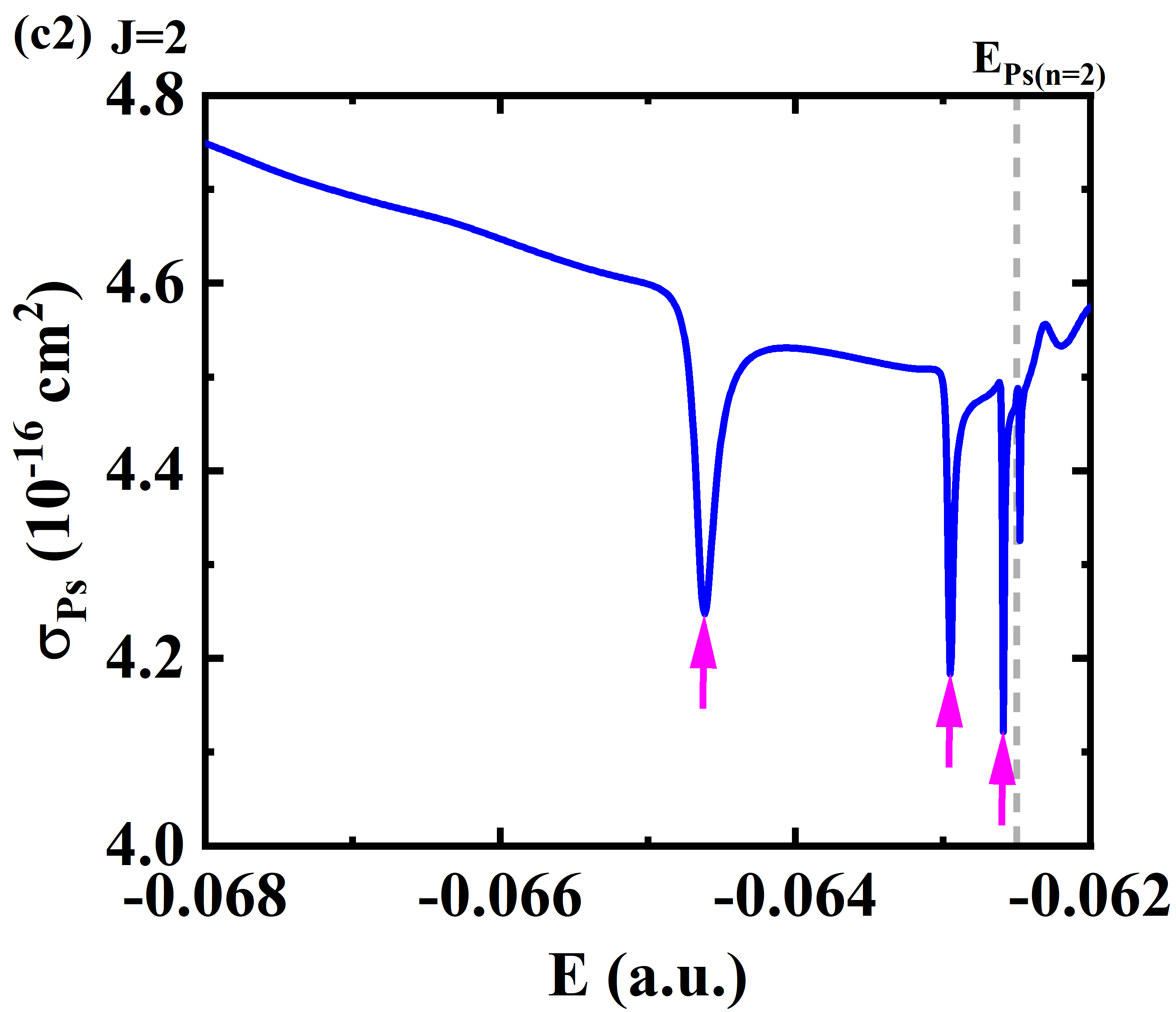}
		\label{fig10c2}
	}
	\subfigure{
		\includegraphics[scale=0.238]{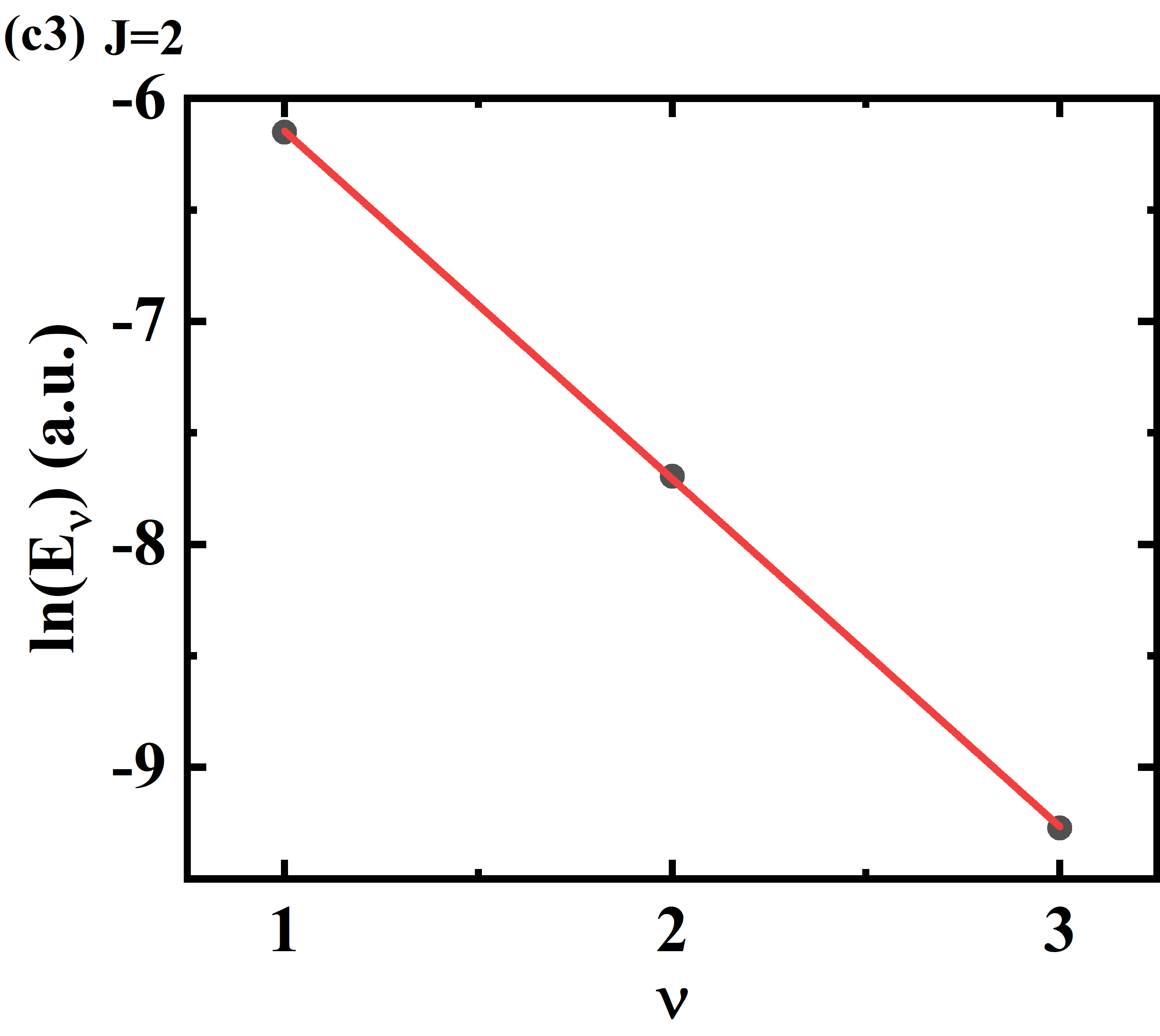}
		\label{fig10c3}
	}
	\renewcommand{\thesubfigure}{(d\arabic{subfigure})}
	\setcounter{subfigure}{0}
	\subfigure{
		\includegraphics[scale=0.236]{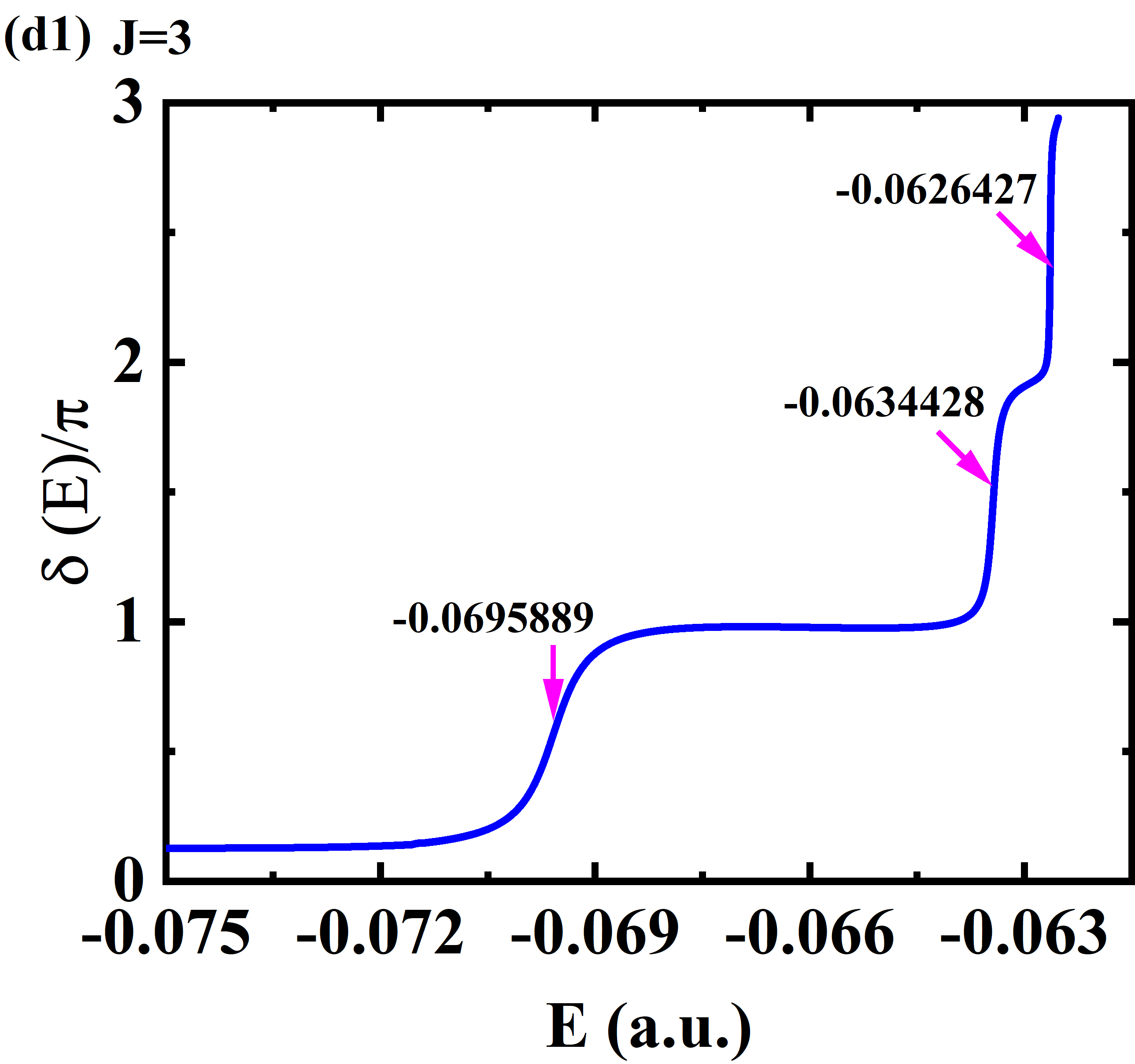}
		\label{fig10d1}
	}
	\subfigure{
		\includegraphics[scale=0.236]{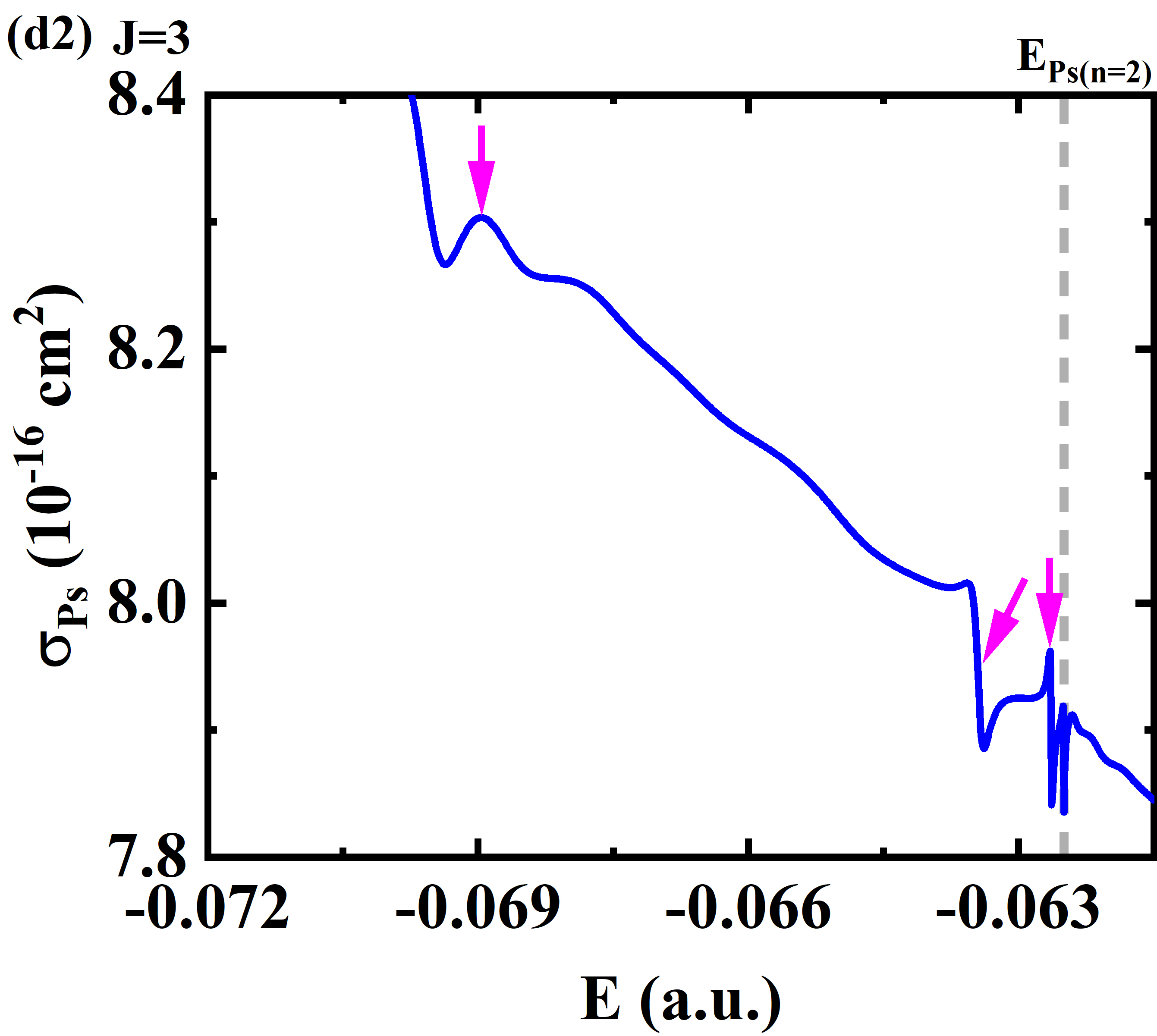}
		\label{fig10d2}
	}
	\subfigure{
		\includegraphics[scale=0.238]{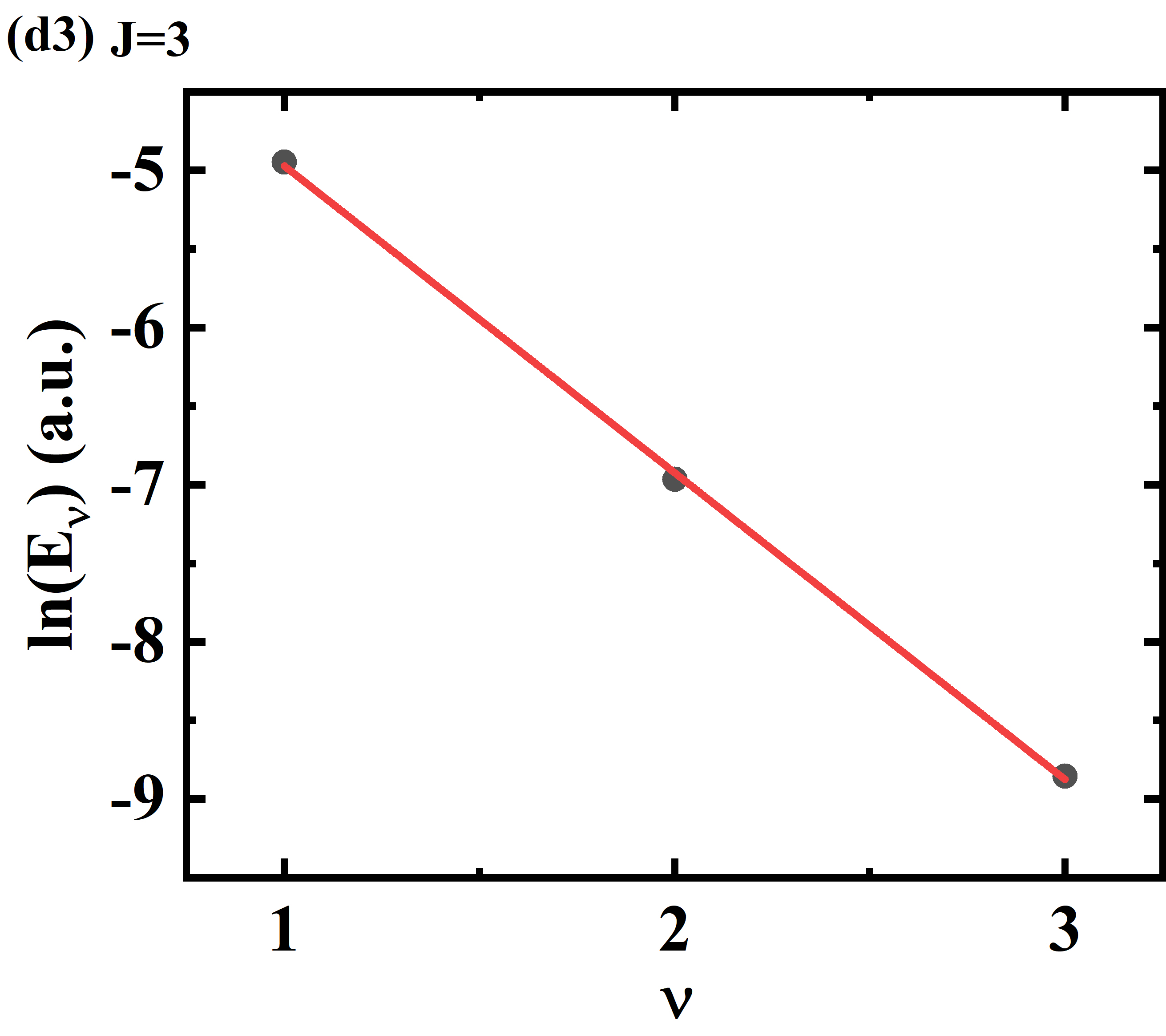}
		\label{fig10d3}
	}
	\caption{(Color online) (a1), (b1), (c1) and (d1) Eigenphase sums; (a2), (b2), (c2) and (d2) Ps-formation cross sections for the e$^{\scriptscriptstyle+}$-Na system with $J=0-3$ at the Ps($n=2$) threshold. Arrows indicate the resonance positions near the Ps($n=2$) threshold. (a3), (b3), (c3) and (d3) Semilogarithmic plots of the resonance positions $E_{R}=E_{th}-E_{\nu}$ for Ps($n=2$)+Na$^{\scriptscriptstyle+}$ below the Ps($n=2$) threshold. Straight lines represent the fits using Eq.\,(\ref{Ev}).}
\end{figure*}

	The $S$-, $P$-, $D$-, and $F$-wave eigenphase sum spectra in the energy region approaching the Ps($n=2$) threshold for the e$^{\scriptscriptstyle+}$-Na system are shown in Figs.~\ref{fig10a1}-\ref{fig10d1}, where the arrows indicate the resonance positions. The successive sudden increases of the eigenphase sums by approximately $\pi$ can be interpreted as clear signatures of resonances. The partial-wave Ps($n=1,\,2$) formation cross sections also exhibit corresponding resonance structures, as shown in Figs.~\ref{fig10a2}-\ref{fig10d2}. Table~\ref{t2} summarizes the dipole resonances below the Ps($n=2$) threshold in the e$^{\scriptscriptstyle+}$-Na system, as reported in the literature, some of which are also visible in the stabilization plot. Our results are in reasonable agreement with those of Ref.~\cite{Umair2017}, obtained using the complex scaling method.
	
	The dipole resonance energies $E_{R}$ near the Ps($n=2$) threshold are plotted on a semi-logarithmic scale in Figs.~\ref{fig10a3}-\ref{fig10d3} and fitted to Eq.\,(\ref{Ev}), as indicated by the straight lines. Table~\ref{t3} lists the energy ratios of successive resonances identified in the present calculations near the Ps($n=2$) threshold of the e$^{\scriptscriptstyle+}$-Na system. For comparison, the universal value of $\alpha$ obtained from the analytic formula of Temkin and Walker\,\cite{Temkin1965Nov} is also given. We find that the ratios for the states closest to the threshold deviate significantly from the analytic prediction, which may be understood as a consequence of strong coupling with other channels.
	
	\begin{table*}[ht]
		\centering
		\renewcommand{\arraystretch}{1.3}
		\begin{threeparttable}
			\caption{\label{t2} Comparison of the lowest five partial-wave resonance energies ($E_{R}$) and widths ($\Gamma$) for the e$^{\scriptscriptstyle +}$-Na system, converging to the Ps($n=2$) threshold, with previous results. Here, x[y] denotes $10^{-y}$.}
			\begin{ruledtabular}
				\begin{tabular}{cccccccccc}		
					&\multicolumn{2}{c}{Present\tnote{a}}
					&\multicolumn{2}{c}{Present\tnote{b}}
					&\multicolumn{2}{c}{Ref.~\cite{Umair2017}\tnote{e}}
					&\multicolumn{2}{c}{Ref.~\cite{Jiao2012Feb}\tnote{d}}\\
					\cline{2-3} \cline{4-5} \cline{6-7} \cline{8-9}
					\multicolumn{1}{c}{Partial wave}&
					\multicolumn{1}{c}{$E_{R}$}&\multicolumn{1}{c}{$\Gamma$}&\multicolumn{1}{c}{$E_{R}$}&\multicolumn{1}{c}{$\Gamma$}&\multicolumn{1}{c}{$E_{R}$}&\multicolumn{1}{c}{$\Gamma$}&\multicolumn{1}{c}{$E_{R}$} & \multicolumn{1}{c}{$\Gamma$}\\
					\hline					
					$S$ &	-0.0665858&6.96[5] &-0.0665606&6.1[5]&-0.06659598&6.91[5]&-0.0663478 &2.01[3]\\
					&-0.0636172&1.62[5] &-0.0635301&4.1[5]&-0.06362215&1.61[5]&           &      \\
					&-0.0627995&4.95[6]&           &       &-0.06280162&4.34[6]&           &      \\
					&-0.0625597&2.34[6]&           &       &-0.06258079&1.16[6]&           &      \\
					&& & & & -0.06252164&3.11[7]& &\\
					&& &          &      &-0.06250580&8.33[8]& &\\[1.5ex]
					$P$ &-0.0658074 & 5.59[5]&-0.0658034&4.01[5]      &-0.06579619& 4.77[5]&            &          \\
					&	-0.0633613 &6.20[5]&-0.0633450&4.11[5]                &-0.06335707& 5.51[5]&&\\
					&-0.0630710&2.21[4]&&       &-0.06306132&1.99[4]&&\\
					&-0.0627041 & 4.74[6]      & -0.0627009           &3.60[6]    &-0.06270311& 4.50[6]&&\\
					&-0.0625097&2.67[6] &          &                   &-0.06255171&8.17[7]&&\\
					&         &                  &            &         & -0.06251308   &1.11[7]   &&\\[1.5ex]
					$D$ &	-0.0646298 & 1.55[4]&-0.0646300             &9.22[5]          &-0.06464101 & 1.70[4]&& \\
					&	-0.0629551 & 3.89[5]&-0.0629496 &3.73[5]         &-0.06295016 & 2.57[5]&            &         \\
					&	-0.0625940 & 1.07[5]&-0.0625949 &1.17[5]         &-0.06259650 & 5.58[6]&            &          \\
					&             & &            &                       &-0.06252115 & 1.23[6]&             &   \\
					&             & &            &                          &-0.06250466 & 2.69[7]&            & \\[1.5ex]
					$F$&-0.0695889&7.51[4]\\
					&-0.0634428&1.53[4]\\
					&-0.0626427&2.30[5]\\[1.5ex]
					$G$	&-0.0656010&1.36[3]\\
					&-0.0626070&4.52[5]\\			
				\end{tabular}
			\end{ruledtabular}
			\begin{tablenotes}
				\footnotesize
				\item[a] The eigenphase sum method
				\item[b] The stabilization method
				\item[d] The momentum-space coulped-channel optical method
				\item[e] The complex scaling method
			\end{tablenotes}
		\end{threeparttable}
	\end{table*}
	
	\begin{table}[ht]
		\caption{\label{t3} The energy ratios of successive resonances located by the present calculations near the Ps($n=2$) threshold in the e$^{\scriptscriptstyle+}$-Na system for $J=0-3$.}
		\renewcommand{\arraystretch}{1.3}
		\begin{ruledtabular}
			\begin{tabular}{cccc}
				\multicolumn{1}{c}{$E_{\nu}/E_{\nu+1}$}
				&\multicolumn{1}{c}{Present}
				&\multicolumn{1}{c}{Calculated\,(Ref.~\cite{Umair2017})}
				&\multicolumn{1}{c}{Analytic\,(Ref.~\cite{Umair2017})}\\
				\hline
				\multicolumn{4}{c}{$S$ wave}\\
				$1/2$ &3.66&3.65&3.73 \\
				$2/3$ &3.73&3.72& \\
				$3/4$ &5.02&3.73& \\
				\multicolumn{4}{c}{$P$ wave}\\
				$1/2$ &3.84&3.82&3.95 \\
				$2/3$ &1.51&4.26& \\
				$3/4$ &2.80&3.95& \\
				$4/5$ &21.04&4.04& \\
				\multicolumn{4}{c}{$D$ wave}\\
				$1/2$ &4.68&4.76&4.54 \\
				$2/3$ &4.84&4.73& \\
				\multicolumn{4}{c}{$F$ wave}\\
				$1/2$ &7.52 &&\\
				$2/3$ &6.61&& \\			
			\end{tabular}
		\end{ruledtabular}
	\end{table}
	
	\subsubsection{Quasi dipole resonances with \text{e}$^{\scriptscriptstyle+}$+\text{Na}($4d$) channel}
	
	It is well known that dipole resonances arise from the degeneracy of excited atomic energy levels. Such resonances have previously been observed in the excited states of positronium and in the hydrogen atom within the e$^{\scriptscriptstyle+}$-H system~\cite{Ho1988Dec}. In the case of Na, the $4d$ ($E_{R}=-0.03144895$ a.u.) and $4f$ ($E_{R}=-0.03126768$ a.u.) levels lie very close to each other. Consequently, we have identified a series of dipole-like resonances just below the Na($4d$) threshold. These resonances are characterized by eigenphase variations of approximately $\pi$, and they give rise to pronounced structures in the Ps($n=1,\,2$) formation cross sections.
	
\begin{figure*}[htbp]
	\centering
	\label{fig11}
	\renewcommand{\thesubfigure}{(a\arabic{subfigure})}
	\setcounter{subfigure}{0}
	\subfigure{
		\includegraphics[scale=0.28]{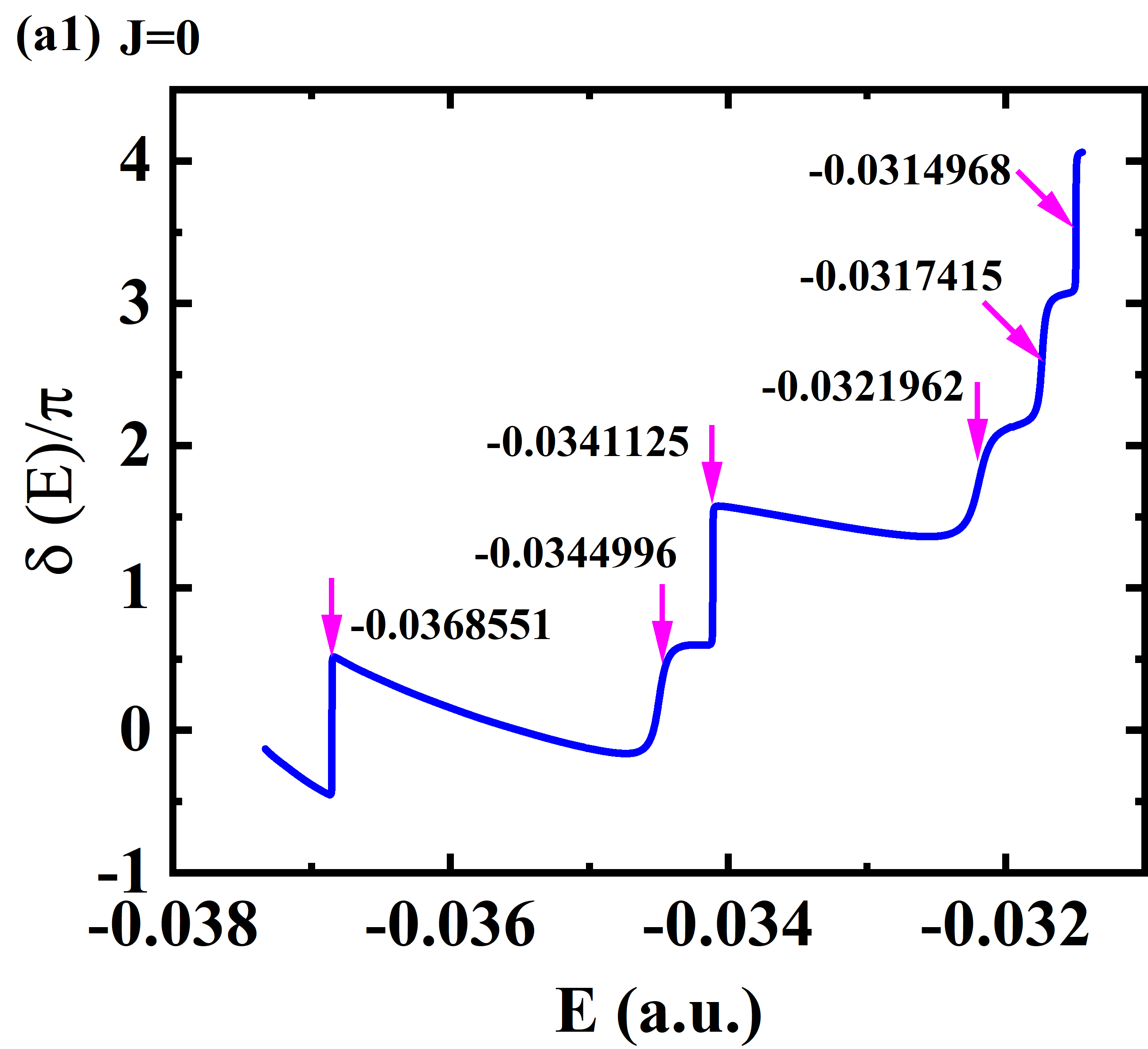}
		\label{fig11a1}
	}
	\subfigure{
		\includegraphics[scale=0.28]{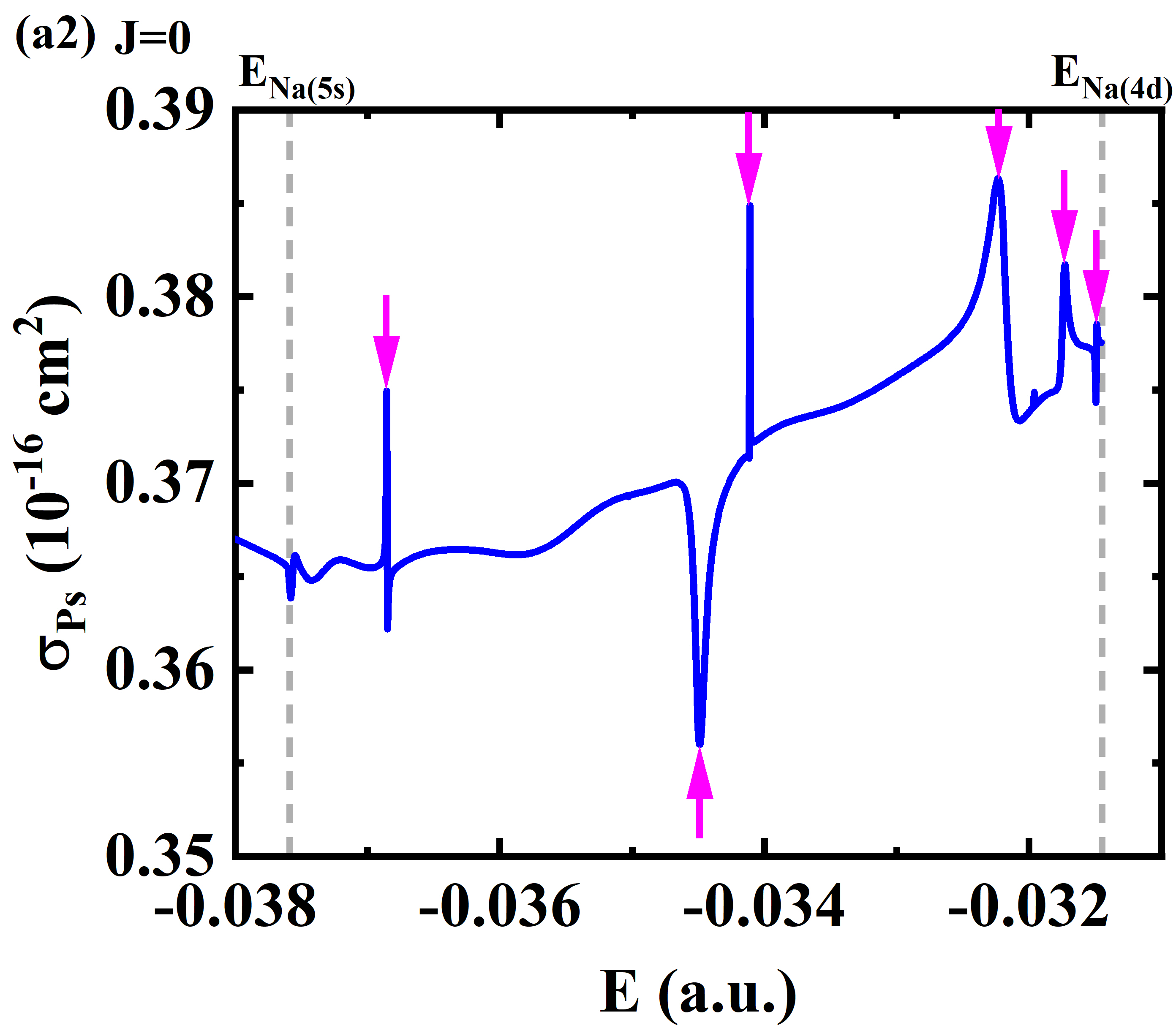}
		\label{fig11a2}
	}\\
	\renewcommand{\thesubfigure}{(b\arabic{subfigure})}
	\setcounter{subfigure}{0}
	\subfigure{
		\includegraphics[scale=0.28]{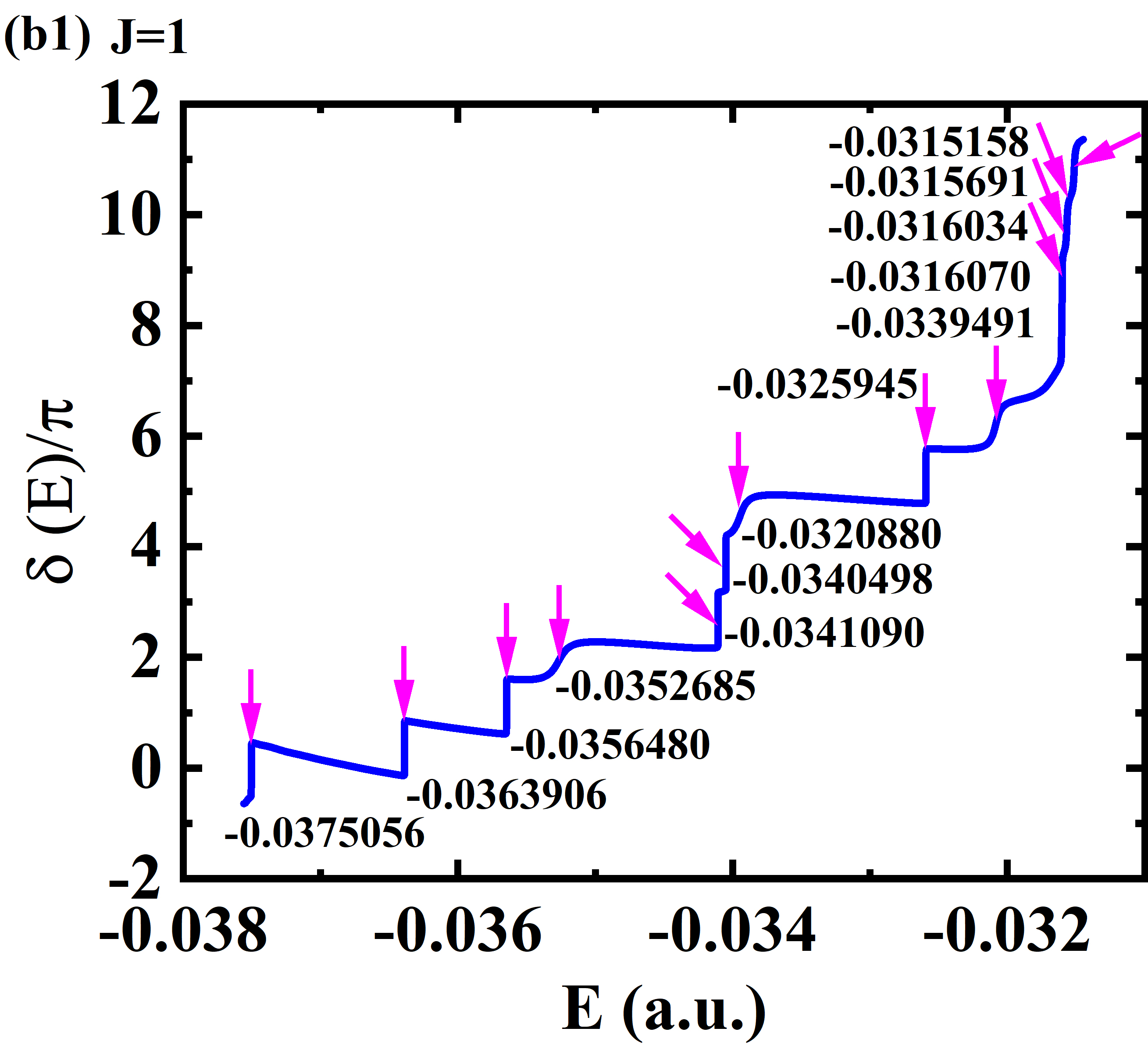}
		\label{fig11b1}
	}
	\subfigure{
		\includegraphics[scale=0.28]{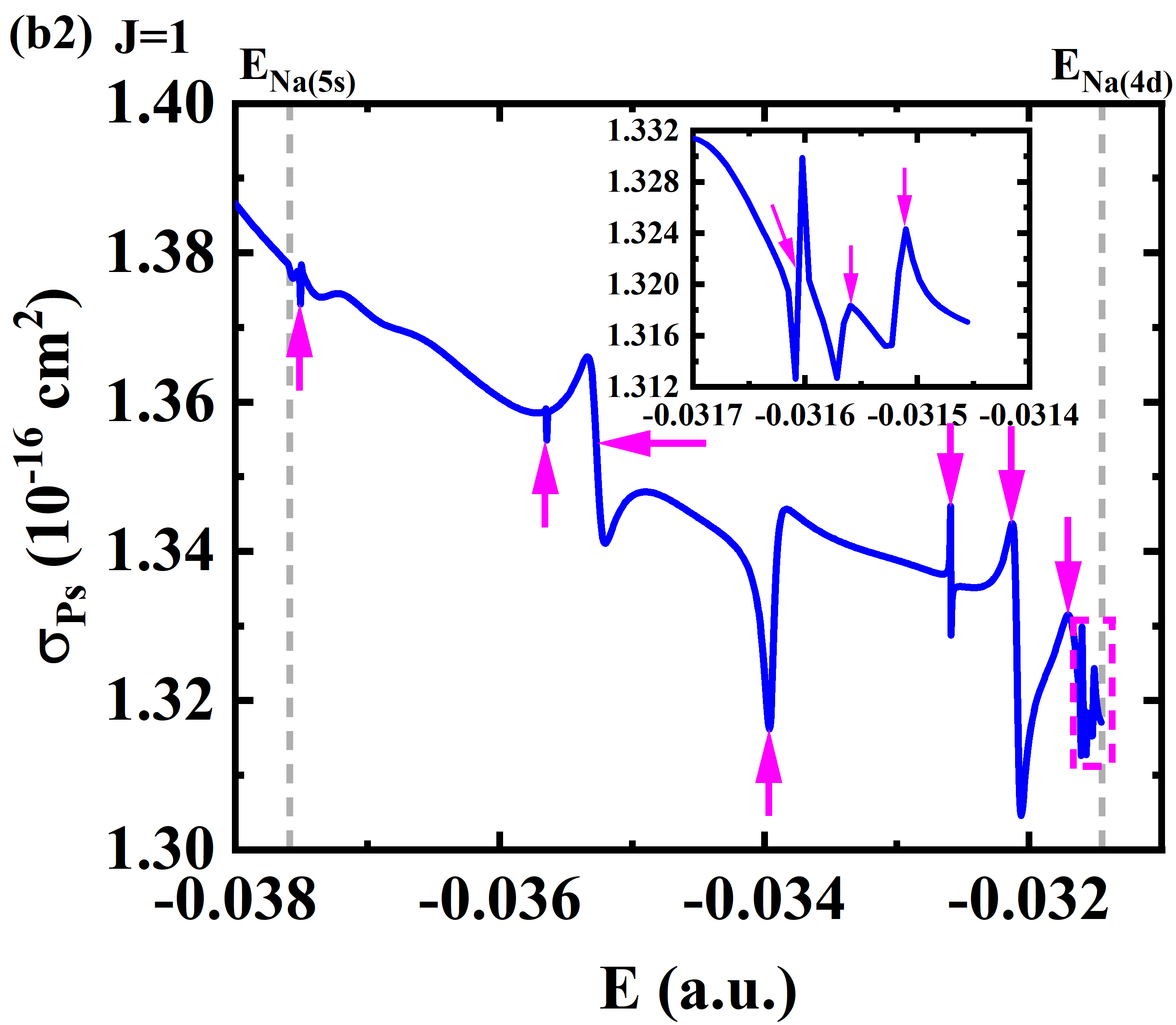}
		\label{fig11b2}
	}\\
	\renewcommand{\thesubfigure}{(c\arabic{subfigure})}
	\setcounter{subfigure}{0}
	\subfigure{
		\includegraphics[scale=0.28]{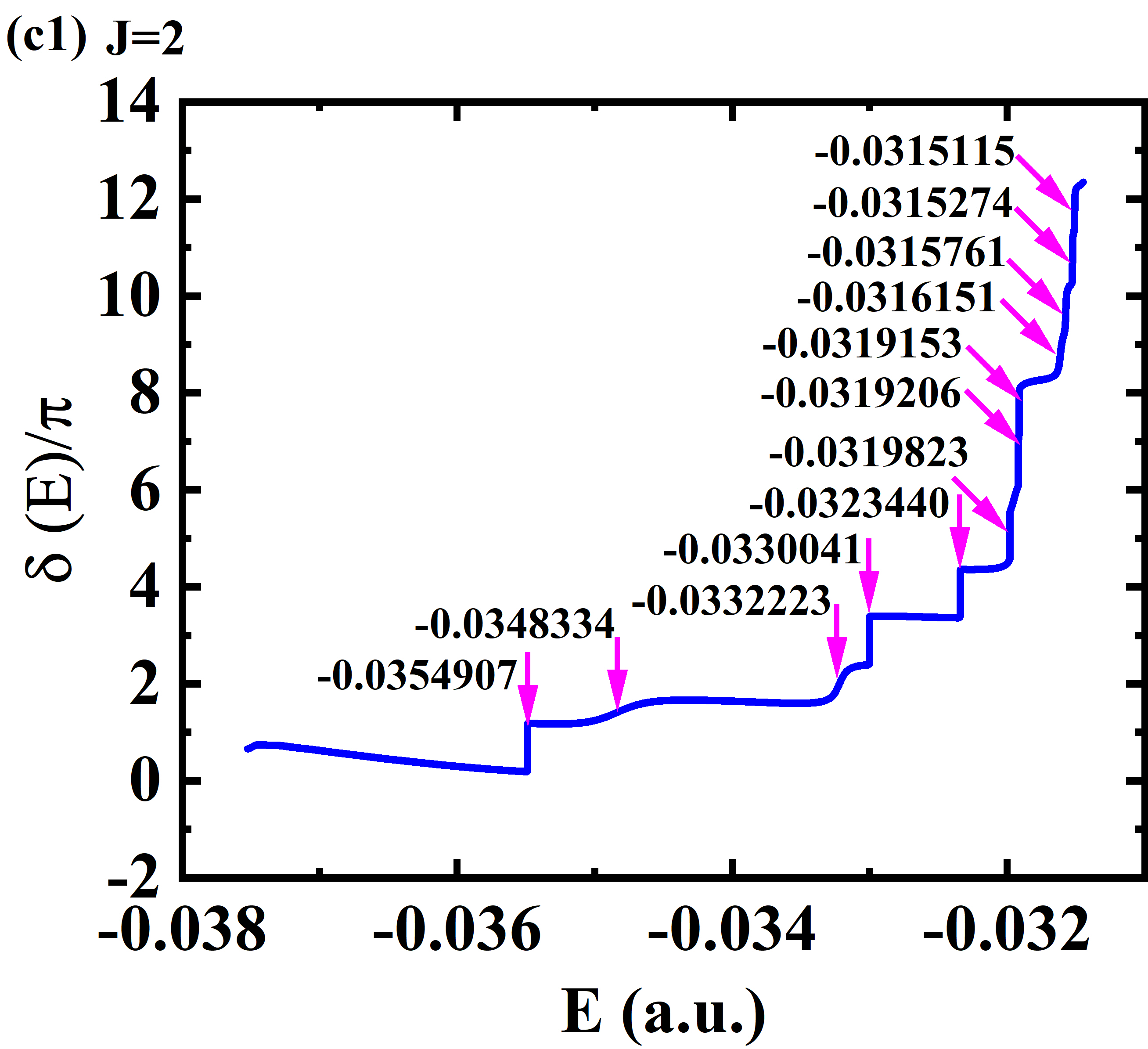}
		\label{fig11c1}
	}
	\subfigure{
		\includegraphics[scale=0.285]{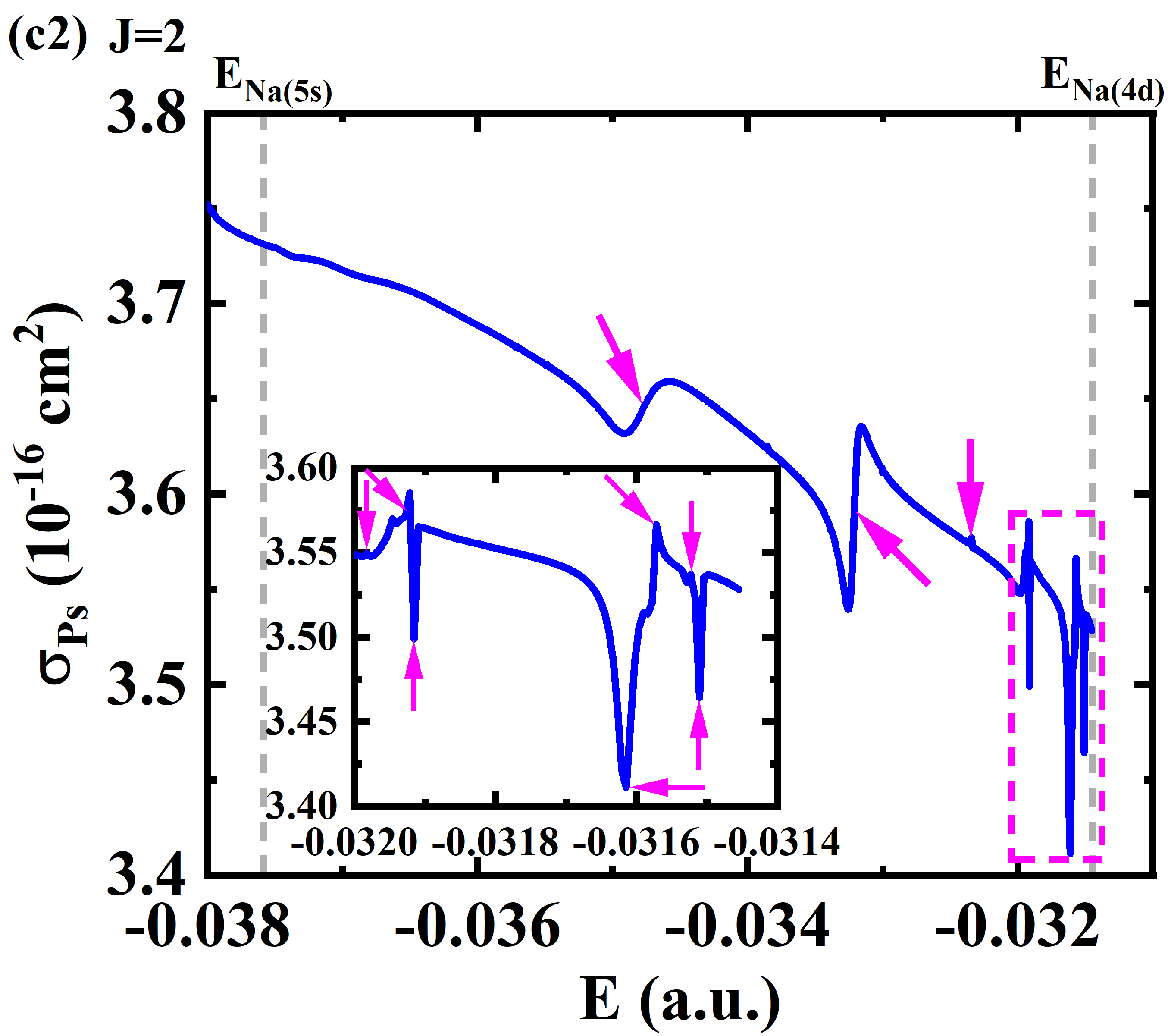}
		\label{fig11c2}
	}\\
	\renewcommand{\thesubfigure}{(d\arabic{subfigure})}
	\setcounter{subfigure}{0}
	\subfigure{
		\includegraphics[scale=0.28]{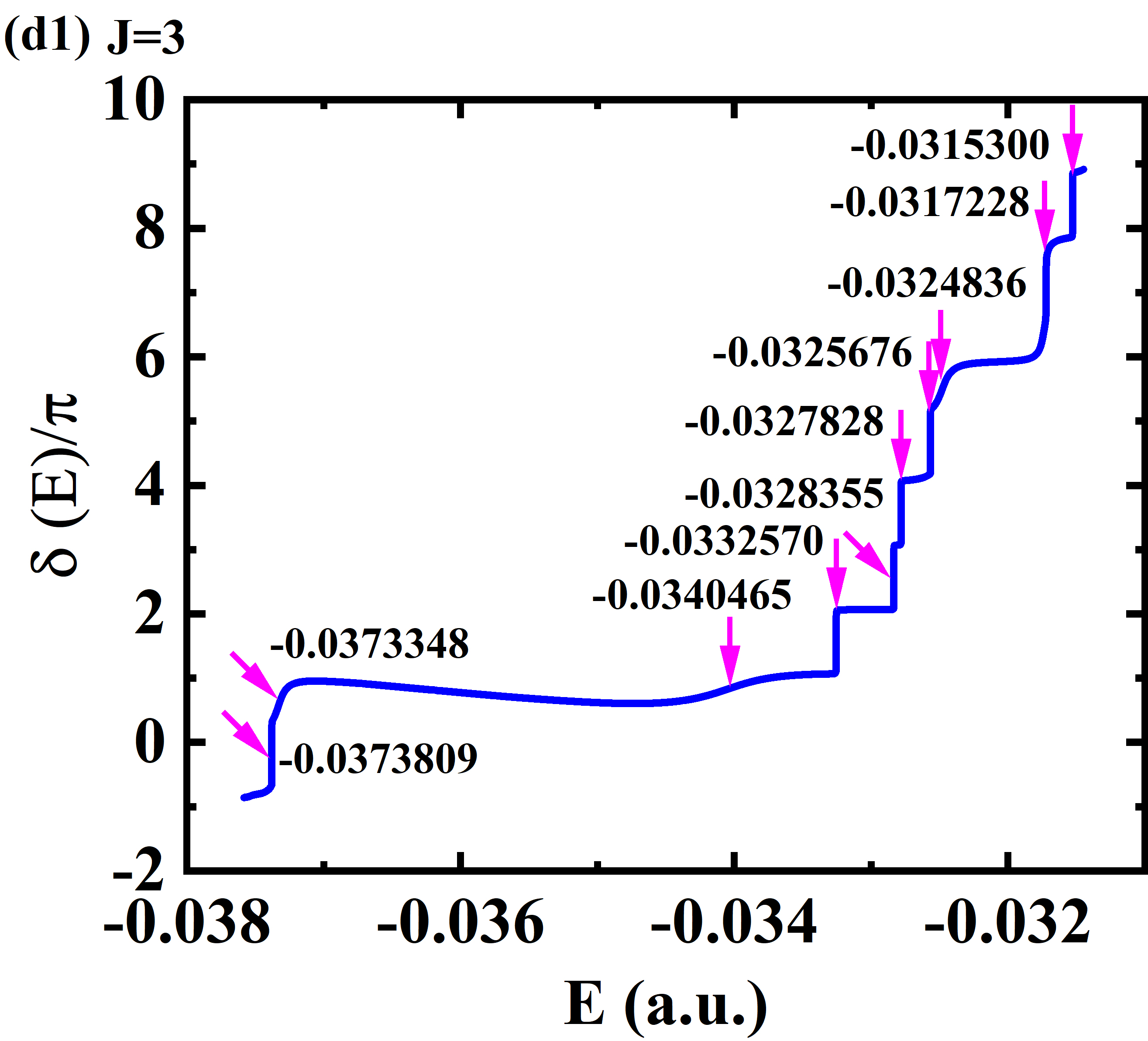}
		\label{fig11d1}
	}
	\subfigure{
		\includegraphics[scale=0.28]{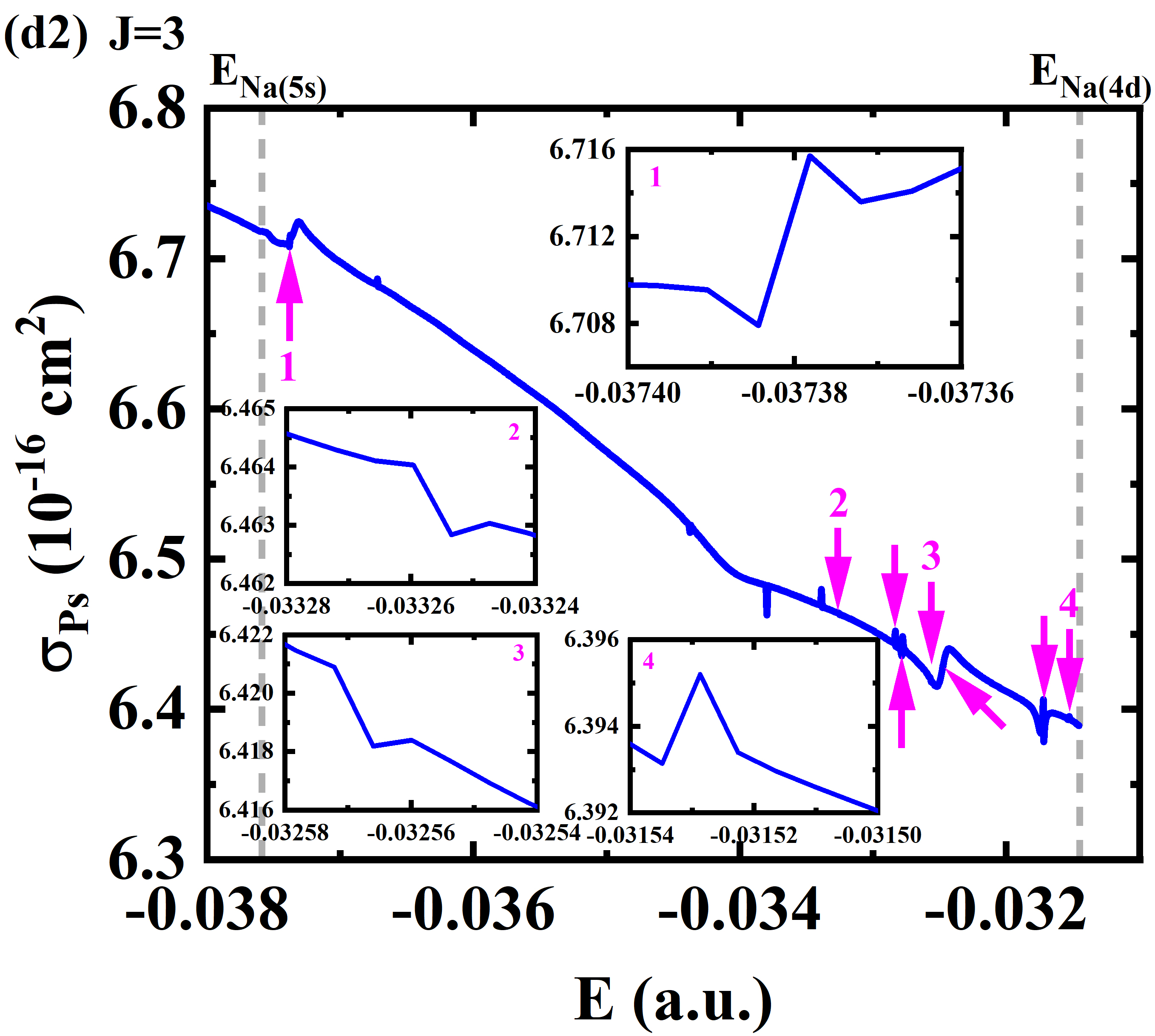}
		\label{fig11d2}
	}
	\caption{(Color online) (a1), (b1), (c1) and (d1)Eigenphase sums; (a2), (b2), (c2) and (d2) Ps-formation cross sections of the e$^{\scriptscriptstyle+}$-Na system with $J=0-3$ near the Na($4d$) threshold. Arrows indicate the resonance positions.}
\end{figure*}

	Figure~\ref{fig11a1}-\ref{fig11d1} presents the eigenphase sum spectra in the energy region near the Na($4d$) threshold for the e$^{\scriptscriptstyle+}$-Na system with total angular momentum $J=0-4$. The eigenphase sums increase by approximately $\pi$, indicating the presence of resonances, with their positions marked by arrows. The Ps($n=1,\,2$) formation cross sections also exhibit the corresponding resonance structures (Figs.
	~\ref{fig11a2}-\ref{fig11d2}). From the figure, we can see that most of these resonance series consist of narrow resonances whose corresponding structures in the scattering cross sections are extremely weak and become visible only upon magnification. As evident from the eigenphase sum spectra, the resonances become increasingly dense as the energy approaches the Na($4d$) threshold. The identified resonances are summarized in Table~\ref{t4}, together with the energy ratios of successive resonances. It is clear that, near the threshold, the ratios deviate from the expected values, which can be attributed to strong interchannel coupling effects in this energy region.
	
	\begin{table*}[ht]
		\caption{\label{t4} Lowest five partial-wave resonance energies $E_{R}$, widths $\Gamma$, and successive energy ratios for $J=0-4$ in the e$^{\scriptscriptstyle +}$-Na system near the Na($4d$) threshold. Here, x[y] denotes $10^{-y}$.}
		\renewcommand{\arraystretch}{1.3}
		\begin{ruledtabular}
			\begin{tabular}{cccccccc}						
				\multicolumn{1}{c}{$E_{R}$}&\multicolumn{1}{c}{$\Gamma$}&
				\multicolumn{1}{c}{$E_{\nu}/E_{\nu+1}$}&
				\multicolumn{1}{c}{Present}&\multicolumn{1}{c}{$E_{R}$}&\multicolumn{1}{c}{$\Gamma$}&
				\multicolumn{1}{c}{$E_{\nu}/E_{\nu+1}$}&
				\multicolumn{1}{c}{Present}\\
				\cline{1-4} \cline{5-8}
				\multicolumn{4}{c}{$S$ wave}&\multicolumn{4}{c}{$F$ wave}\\				
				-0.0368551&1.22[6]&$1/2$ &1.77 &-0.0373809&2.44[7]&$1/2$ &1.01 \\
				-0.0344996&9.45[5]&$2/3$ &1.15 &-0.0373348&9.58[5]&$2/3$ &2.27 \\
				-0.0341125&1.82[6]&$3/4$ &3.56 &-0.0340465&6.96[4]&$3/4$ &1.44 \\
				-0.0321962&1.46[4]&$4/5$ &2.55 &-0.0332570&5.81[7]&$4/5$ &1.30 \\
				-0.0317415&4.15[5]&$5/6$ &6.11 &-0.0328355&2.61[7]&$5/6$ &1.04 \\
				-0.0314968&2.73[6]&      &     &-0.0327828&2.50[7]&$6/7$ &1.19 \\
				\multicolumn{4}{c}{$P$ wave}&-0.0325676&2.17[7]&$7/8$ &1.08 \\
				-0.0375056&4.97[7]&$1/2$ &1.23 &-0.0324836&1.07[4]&$8/9$ &3.78 \\
				-0.0363906&1.91[7]&$2/3$ &1.29 &-0.0317228&1.20[6]&$9/10$&3.38 \\
				-0.0356480&2.71[7]&$3/4$ &1.58 &-0.0315300&2.25[7]&&\\
				-0.0352685&1.33[4]&$4/5$ &1.47 &\multicolumn{4}{c}{$G$ wave}\\
				-0.0341090&2.58[7]&$5/6$ &1.06 &-0.0370239&6.92[7]&$1/2$ &1.13 \\
				-0.0340498&2.23[7]&$6/7$ &2.27 &-0.0363993&4.16[4]&$2/3$ &1.32 \\
				-0.0339491&9.08[5]&$7/8$ &3.91 &-0.0351911&6.12[7]&$3/4$ &1.35 \\
				-0.0325945&4.42[7]&$8/9$ &7.25 &-0.0342184&2.40[7]&$4/5$ &1.12 \\
				-0.0320880&6.35[5]&$9/10$ &4.14 &-0.0339215&2.56[7]&$5/6$ &1.45 \\
				-0.0316070&2.89[6]&$10/11$ &1.32 &-0.0331579&1.24[6]&$6/7$ &1.77 \\
				-0.0316034&1.49[6]&$11/12$ &2.31 &-0.0324165&2.05[7]&$7/8$ &1.10 \\
				-0.0315691&1.08[5]&$12/13$ &1.80 &-0.0323271&2.61[7]&$8/9$ &1.64 \\
				-0.0315158&1.30[5]&       &      &-0.0319851&2.42[7]&$9/10$ &1.24 \\
				\multicolumn{4}{c}{$D$ wave}&-0.0318811&1.16[4]&$10/11$ &23.30\\
				-0.0354907&2.26[7]&$1/2$ &1.19& -0.0314675&3.05[5]&&\\
				-0.0348334&4.08[4]&$2/3$ &1.91 &&&&\\
				-0.0332223&9.62[5]&$3/4$ &1.14 &&&&\\
				-0.0330041&2.00[7]&$4/5$ &1.74 &&&&\\
				-0.0323440&1.97[7]&$5/6$ &1.68 &&&&\\
				-0.0319823&2.28[7]&$6/7$ &1.13 &&&&\\
				-0.0319206&2.61[7]&$7/8$ &1.01 &&&&\\
				-0.0319153&5.15[7]&$8/9$ &2.81 &&&&\\
				-0.0316151&2.14[5]&$9/10$ &1.31 &&&&\\
				-0.0315761&6.70[6]&$10/11$ &1.62 &&&&\\
				-0.0315274&6.22[7]&$11/12$ &1.25 &&&&\\
				-0.0315115&3.16[6]&&&&&&\\									
			\end{tabular}
		\end{ruledtabular}
	\end{table*}
	
\begin{figure}[htbp]
		\centering
		\includegraphics[scale=0.35]{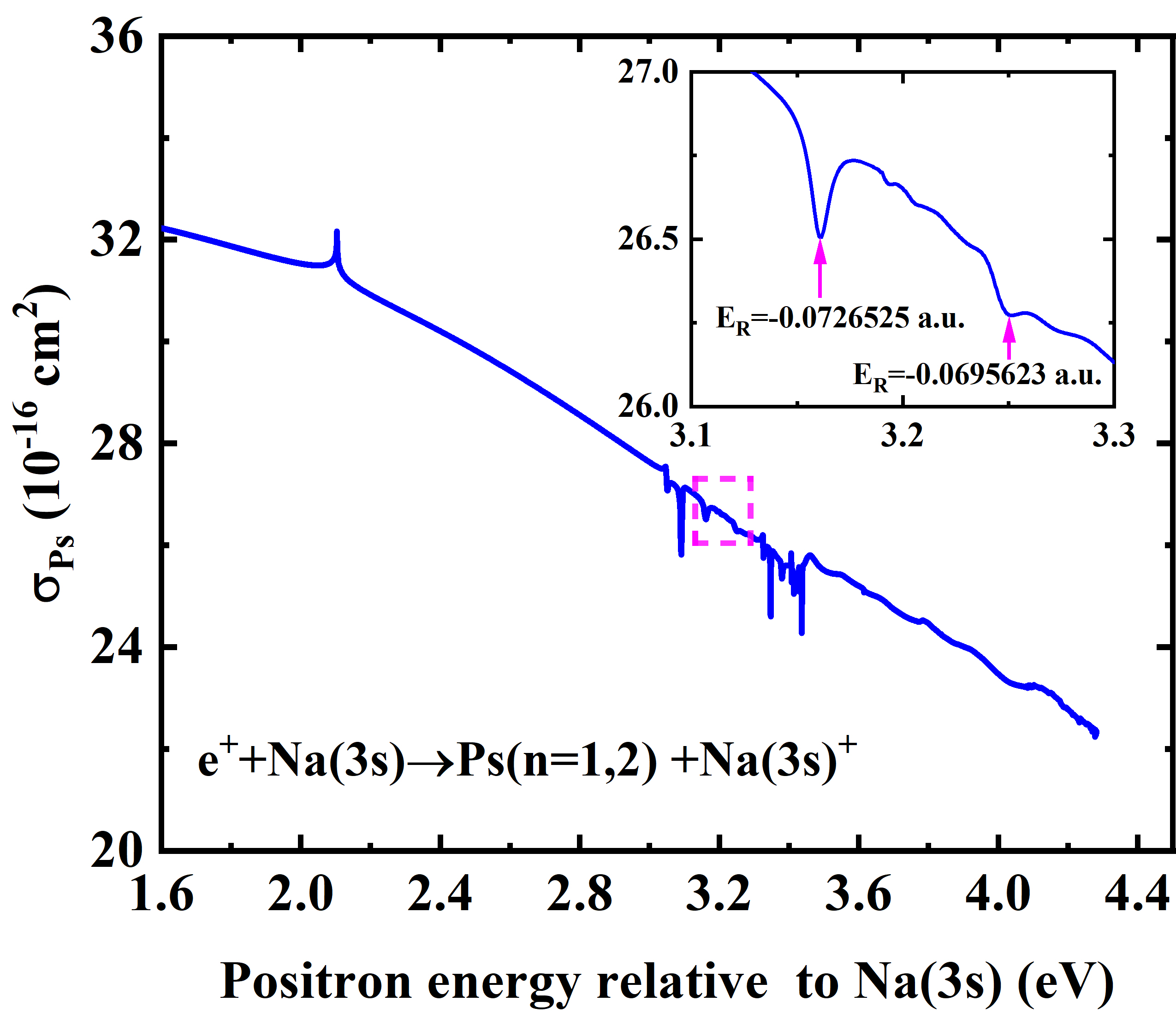}
		\caption{(Color online) Structural features of the total Ps-formation cross section in e$^{\scriptscriptstyle+}$-Na system. }
		\label{fig12}
	\end{figure}

	Figure~\ref{fig12} shows the structural features of the total Ps-formation cross section for the positron-sodium system. The structure observed near 2.1~eV originates from a shape resonance close to the Na($3p$) threshold. This is a narrow resonance with a width of about 0.27~meV, which is too small to be resolved by current positron-beam experiments. In addition, two broader resonances with widths on the order of 10~meV are identified. The structure appearing near 3.1~eV, with a width of approximately 11~meV, is attributed to an $D$-wave resonance lying just below the Na($4s$) threshold. The structure appearing near 3.2~eV, with a width of approximately 19~meV, is attributed to an $F$-wave resonance associated with the Ps($n=2$) threshold.
	
	\section{Summary}
	\label{sec:Summary}
	In this work, we carried out a comprehensive investigation of resonances in positron-sodium scattering using the $R$-matrix propagation method formulated in hyperspherical coordinates. The interaction between the sodium core and its valence electron was modeled using analytical model potentials. To gain a deeper understanding of the resonant structures, we examined both the energy dependence of the eigenphases of the multichannel $S$-matrix and the eigenvalues of the corresponding time-delay matrix. Several resonant states of debated character were identified, and their behavior was analyzed through phase-variation studies, the associated structures in the calculated cross sections, and the characteristic patterns observed in the stability plots. Our calculations show how the near threshold resonances appear in cross sections and explain why they show tiny phase variations.
	
	Our calculations successfully reproduced previously reported resonances and extended the understanding of the e$^{\scriptscriptstyle+}$-Na system to higher partial waves up to $J=4$. We found a relatively broad resonance localized at about 3.2~eV, with a width of approximately 19~meV, originating from an $F$-wave resonance below the Ps($n=2$) threshold. Modern trap-based positron beams can currently achieve a total energy resolution of about 40~meV\;\cite{Gribakin2010Sep,sullivanpositron2008}. With further improvements in energy resolution, this relatively broad resonance could become observable in positron-sodium scattering experiments through its signature in the Ps-formation cross section.
	
	Moreover, two distinct series of dipole resonances were identified: one converging toward the Ps($n=2$) threshold, arising from the interaction between the excited Ps atom and Na$^{\scriptscriptstyle+}$; and another, a sequence of quasi-dipole resonances, resulting from the near degeneracy of the Na($4d$) and Na($4f$) states in the e$^{\scriptscriptstyle+}$-Na system, which accumulate geometrically toward the Na($4d$) threshold. The energy ratios and fitted scaling parameters of these series exhibit universal scaling behavior consistent with dipole-interaction theory.
	
	Our findings provide a detailed characterization of resonance structures in the e$^{\scriptscriptstyle+}$-Na system, offering valuable information for future theoretical studies and potential experimental verification of positron-atom resonance phenomena.
	
	\begin{acknowledgments}
		We thank Bao-Chun Yang for helpful discussions. Hui-Li Han was supported by the National Natural Science Foundation of China under Grant No. 11874391. Ting-Yun Shi was supported by the National Natural Science Foundation of China under Grant No.12274423. Li-Yan Tang was supported by the Pioneer Research Project for Basic and Interdisciplinary Frontiers of Chinese Academy of Sciences under Grants No. XDB0920101 and XDB0920100, and by the National Natural Science Foundation of China under Grants Nos. 12393821 and 12174402. All the calculations are done on the APM-Theoretical Computing Cluster(APM-TCC).
	\end{acknowledgments}

	\bibliographystyle{apsrev_title.bst}

\end{document}